\documentclass[aps,prb,reprint,groupedaddress, showpacs]{revtex4-1}
\usepackage{graphicx,graphics}
\usepackage[table]{xcolor}
\usepackage{epstopdf}
\usepackage{amsmath,amssymb,amsfonts}
\usepackage{latexsym,verbatim}
\usepackage{bm}
\usepackage{bbold}
\usepackage{color}
\usepackage{ulem}
\usepackage[breaklinks=false,colorlinks,citecolor=blue,linkcolor=blue,urlcolor=blue]{hyperref}
\usepackage[abs]{overpic}
\usepackage{textcomp} 
\usepackage{mathtools}
\usepackage{braket}
\usepackage{comment}
\usepackage{tcolorbox}
\usepackage{soul}    
\newcommand{\y}{\color{black}}

\begin{document}
 
\title{Vibrational response functions for multidimensional electronic spectroscopy \\ in the adiabatic regime: a coherent-state approach}

\author{Frank Ernesto Quintela Rodriguez}
\affiliation{Università di Modena e Reggio Emilia, I-41125 Modena, Italy}
\author{Filippo Troiani$^*$}
\affiliation{Centro S3, CNR-Istituto di Nanoscienze, I-41125 Modena, Italy}
\affiliation{\rm $^*$Author to whom correspondence should be addressed: filippo.troiani@nano.cnr.it}

\begin{abstract}
Multi-dimensional spectroscopy represents a particularly insightful tool for investigating the interplay of nuclear and electronic dynamics, which plays an important role in a number of photophysical processes and photochemical reactions. Here we present a coherent state representation of the vibronic dynamics and of the resulting response functions for the widely used linearly displaced {\y harmonic} oscillator model. Analytical expressions are initially derived for the case of third-order response functions in an $N$-level system, with ground state initialization of the oscillator (zero-temperature limit). The results are then generalized to the case of $M$-th order response functions, with arbitrary $M$. The formal derivation is translated into a simple recipe, whereby the explicit analytical expressions of the response functions can be derived directly from the Feynman diagrams. We further generalize to the whole set of initial coherent states, which form an overcomplete basis. This allows one in principle to derive the dependence of the response functions on arbitrary initial states of the vibrational modes and is here applied to the case of thermal states. Finally, a non-Hermitian Hamiltonian approach is used to include in the above expressions the effect of vibrational relaxation.
\end{abstract}

\date{\today}

\maketitle

\section{Introduction}
Ultrafast spectroscopy allows the investigation of dynamical processes occurring in the femtosecond regime in atomic, molecular, and solid-state systems \cite{Mukamel95a}.
In multidimensional coherent spectroscopy, the sample is excited by a series of laser pulses and its nonlinear optical response is resolved with respect to multiple frequencies \cite{Hamm11a}. The spectrum thus provides correlations between excitation and detection frequencies, and allows a dissection of the dynamics underlying complex phenomena of physical, chemical, and biological interest\cite{Schlau11a,Scholes2011,Smallwood18a,Rozzi18a,Maiuri2020,Herkert21a}.

Multidimensional coherent spectroscopy also represents a powerful tool for investigating the dynamical interplay between electronic and vibrational (nuclear) degrees of freedom, which plays an important role in a number of photophysical processes and photochemical reactions \cite{Luer2009,Rury2017,Thouin2019,DeSio16a,Pandya2018,Womick2011,Falke14a,Rozzi2013,Rafiq2021,Monahan2017,Schultz2021,Horstmann2020,Scholes2017,Collini2019,Collini2021}. 
In broadband femtosecond transient-absorption spectroscopy, impulsive transitions between electronic states induced by the laser pulses launch vibrational wave packets on excited states potential energy surfaces \cite{Caram2012}. This triggers the wave packet motion, which takes place between consecutive transitions and results in an oscillating modulation of the (nonlinear) polarization as a function of the waiting time(s). 

In general, the vibrational dynamics can be affected by a number of factors, such as anharmonic terms in the potential energy surface\cite{Park2000a,Arpin21a}, coupling between different modes\cite{Schultz22a,Yan1986a}, conical intersections\cite{Duan2016,krcmar}, differences between the curvatures of the displaced oscillators \cite{Fidler13a}, vibronic couplings\cite{Butkus14a,Collini09a,Engel2007,Chin2013,Prior10a,Rozzi2013,Christensson2012,Tiwari13a,Fuller2014a,Romero2014}. However, the dominant contribution in the electron-vibrational coupling is often represented in terms of the displaced-oscillator model, where one or more harmonic oscillators undergo an electronic-state specific displacement in the nuclear coordinates. In fact, this model accounts for complex spectral features, especially as the number of electronic levels and of exciting laser pulses increase\cite{Mukamel95a,Kumar01a,Egorova2007a,Mancal10a,Pollard1990a,Pollard1992a,Butkus12a,Cina2016,Le21a,Turner2020}.

Here we compute the vibrational component of the response functions for a system with an arbitrary number ($N$) of electronic levels and for an arbitrary order ($M$) in the interaction with the field, {\y under the assumption that the optically-induced transitions satisfy the Franck-Condon principle \cite{fcp}}. This is done by {\y fully} exploiting the potentialities of the coherent state {\y picture} \cite{Scully97a,Gerry04a}, {\y within the linearly displaced harmonic oscillator model. Here, a vibrational state initialized in the ground state (or in any other coherent state) always preserves its coherent state character. More specifically, its dynamics consists in rotations around the minima of the potential energy surfaces, alternated by vertical transitions between different surfaces, induced by the electric field. For each pathway ({\it i.e.} sequence of optically-induced transitions between electronic states), the vibrational state is thus fully captured by a complex number that defines the coherent state, and by an additional real number, which accounts for an overall phase factor. The analytic expressions of these numbers are here computed for arbitrary number of energy levels and of interactions with the field. From these we derive the overlaps between the vibrational states corresponding to different pathways. These, according to the Fanck-Condon principle, account for the vibrational modulation of the electronic response function, which is observed in multidimensional electronic spectroscopy.}

This procedure is followed in detailed for the case of two-dimensional spectra ($M=3$, which represents the lowest-order nonlinear contribution in noncentrosymmetric systems) in systems with arbitrary number $N$ of electronic levels. This allows us to extends the long known results for two-level systems, on the one hand by accounting for the possibility of multiple pathways in similar contributions (ground state bleaching and stimulated emission) and, on the other hand, by including phenomena that cannot occur for $N=2$ (excited state absorption and coherences between ground and doubly excited states\cite{Kim2009}). The approach is then applied in a more compact form to directly derive the overlap between coherent states in the $M$-dimensional case, with arbitrary $M$. We show that the formal derivation of the response functions can be translated in a simple recipe and thus directly derived from the double-sided Feynman diagrams, which provide an intuitive representation of the relevant pathways. 

The above picture doesn't change significantly if the system is initialized in a generic coherent state, rather than in the ground state. In fact, the effect of such an initialization can be naturally incorporated in this approach, and is shown to consist in a phase factor, which we explicitly express in terms of the Hamiltonian parameters and of the waiting times. Given that the coherent states form an overcomplete basis, this expression in turn allows one to describe the full dependence of the response functions on the initial state of the vibrational mode, passing through its coherent state representation \cite{Scully97a}. Following such procedure in the case of thermal vibrational states, we recover the results known for the case of third-order response functions in two-level systems and show how they generalize if $M>3$ and/or $N>2$. 

Interactions with the environment can affect the dynamics of electronic degrees of freedom in different ways \cite{Mukamel95a}. However, decoherence also involves the dynamics of the strongly coupled (underdamped) vibrational modes, and this shows up in the response functions. Vibrational relaxation is easily incorporated in the present approach, because phonon emission (annihilation) preserves the coherent character of the state. In the presence of relaxation, the trajectory of the wave packet describes sequences of spirals, rather than circles, around the potential minima, and the modulus of the coherent state tends to decrease. These effects are formally included in the equations for the general case, and are discussed in some detail for the third-order response functions. Finally, the generalization of the above results to multiple modes is conceptually straightforward and is formally outlined in the final part of the paper.

The paper is organized as follows. In Sec. II we introduce the model and the connection between response function and the dynamics of the coherent vibrational states. Sections III and IV are devoted respectively to the derivation of all the relevant contributions in the third-order response function, and of their spectral components. In Sec. V a more compact derivation is presented, which allows one to derive  nonlinear contributions of arbitrary order. We additionally provide a recipe for inferring the expressions of these response functions directly from the Feynman diagrams. In Sec. VI we generalize the above findings to the case of a generic state, and apply this result to the case of thermal states. Section VII contains the expressions of the response functions in the presence of vibrational relaxation. Finally, we draw our conclusions in Sec. VIII. Further details on the relevant vibrational states and on the formal demonstrations are provided in the appendices. 

\section{Displaced oscillator model \\ and linear polarization}

\begin{figure}[h]
\centering
\includegraphics[width=0.4\textwidth]{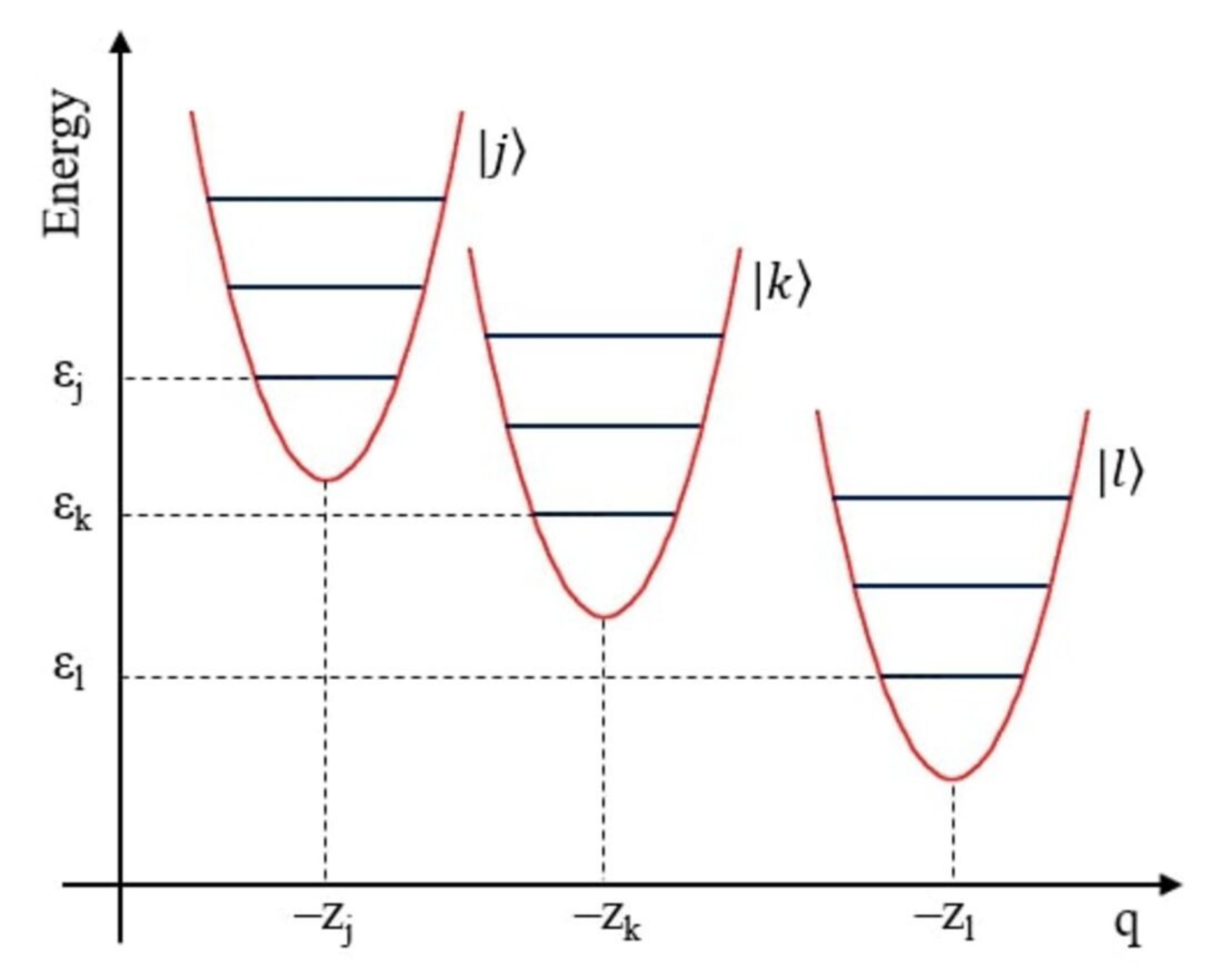}
\caption{Schematic representation of the displaced harmonic oscillator model. Each vibrational mode is modeled in terms of an harmonic oscillator, whose origin in the position and energy plane depends on the electronic eigenstates (here $|j\rangle$, $|k\rangle$, and $|l\rangle$). The parameters $z_{\chi = j,k,l}$ and $\epsilon_{\chi = j,k,l}$ are the ones entering the Hamiltonian in Eq. (\ref{eq:ham}).} \label{fig:dho}
\end{figure}

The results derived in the present paper are based on the {\y linearly} displaced harmonic oscillator model\cite{Mukamel95a}. This doesn't include nonadiabatic effects, i.e. coherent mixing of electronic and vibrational degrees of freedom in the eigenstates. More specifically, it assumes that the dependence of the vibrational modes on the electronic state can be reduced to that of their equilibrium positions.
The corresponding Hamiltonian reads:
\begin{eqnarray}\label{eq:ham}
    H &=& \sum_{j=0}^{N-1} |j \rangle\langle j| \otimes [ \epsilon_j + \hbar\omega_v (a^\dagger + z_j)(a + z_j)] \nonumber\\
    &\equiv& \sum_{j=0}^{N-1} |j \rangle\langle j| \otimes ( \epsilon_j + H_{v,j}) ,
\end{eqnarray}
where the $|j\rangle$ represent the $N$ eigenstates, with eigenvalues $\epsilon_j$ ($\epsilon_0=0$), of the electronic Hamiltonian. For each of these eigenstates, the harmonic oscillator that represents the vibrational mode undergoes a displacement {\y by }${\y -}z_j\in\mathbb{R}$ along the $X=\frac{1}{2}\langle a^\dagger + a \rangle$ (or $q$) axis with respect to the ground state position ($z_0=0$). The minimum of the oscillator's potential is thus shifted to $-z_j$ (see Fig. \ref{fig:dho}). The eigenstates of each $H_{v,j}$ are given by the displaced number states $|-z_{\y j},n\rangle = D(-z_{\y j}) \, |n\rangle $ (see Appendix \ref{app:A}).

The polarization of a system that is optically excited by a sequence of laser pulses can be derived from its response function. Within the perturbative approach in the light-matter interaction, the $M$-th order contribution to the induced polarization is related to an alternate sequences of $M$ instantaneous transitions between electronic states, induced by the electric field, and to time intervals (waiting times $t_1,\dots,t_M$), during which the system state undergoes a free evolution induced by its Hamiltonian $H$. The time evolution of the vibrational state is thus given by an Hamiltonian that is piecewise constant, coinciding with a given $H_{v,j}$ during each waiting time, but possibly changing to any other $H_{v,k}$ as a consequence of a transition $|j\rangle \longrightarrow |k\rangle$ between electronic states\cite{Mukamel95a,Hamm11a}.

\begin{figure}[h]
\centering
\includegraphics[width=0.35\textwidth]{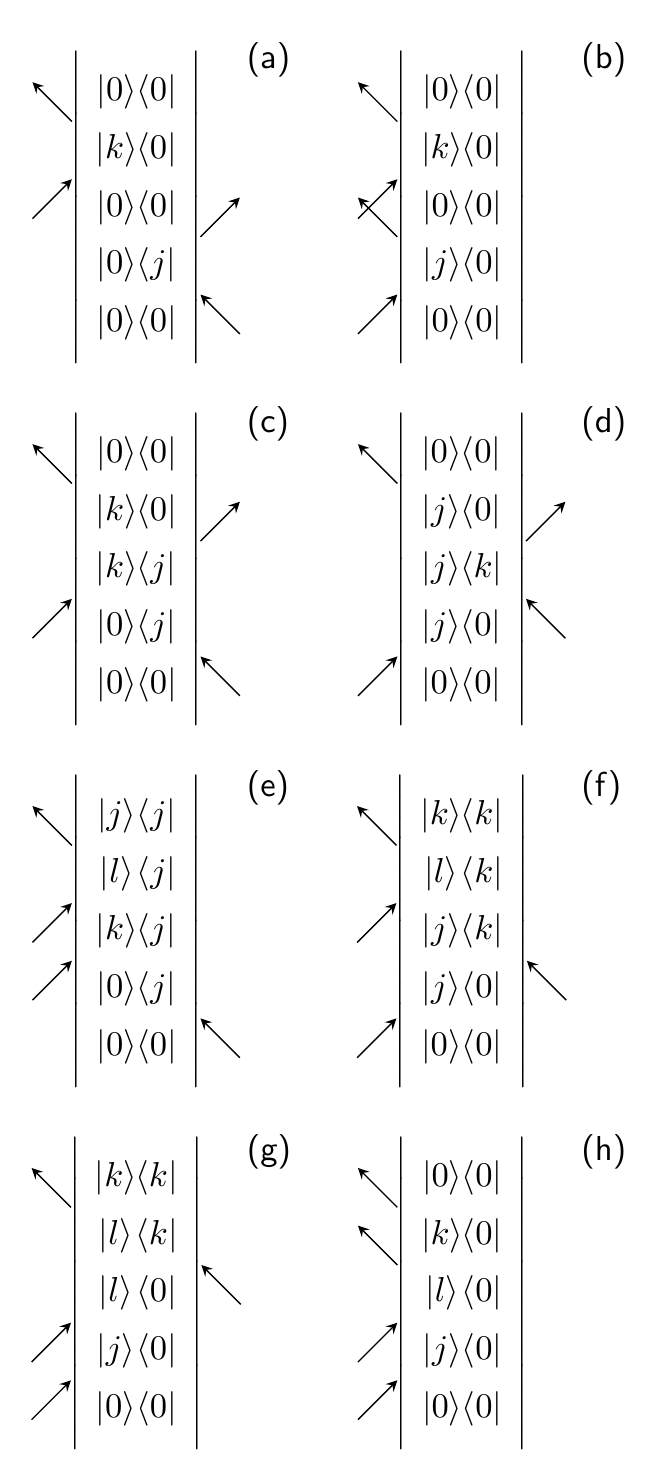}
\caption{Double-sided Feynman diagrams that are relevant to the third-order response function. In particular, (a) and (b) correspond respectively to $R_{2,jk}^{(v,3)}$ and $R_{5,jk}^{(v,3)}$ (ground-state bleaching process); (c) and (d) represent $R_{1,jk}^{(v,3)}$ and $R_{4,jk}^{(v,3)}$ (stimulated emission);  (e) and (f) correspond to $R_{3,jkl}^{(v,3)}$ and $R_{6,jkl}^{(v,3)}$ (excited-state absorption); (g) and (h) finally represent $R^{(v,3)}_{7,jkl}$ and $R^{(v,3)}_{8,jkl}$ (double quantum coherence). \label{fig:dsfd2}}
\end{figure}

The observable of interest is the polarization $P$, which is defined as the expectation value of the electric dipole operator $\mu$. This, in turn, can be expressed as an integral function of the response function $R$, and identified with the product of $R$ and the electric field in the semi-impulsive limit. 
In any case, the different contributions in the response function are modulated by the vibrational dynamics. This can be clearly seen in the case of the first-order contribution, which reads: 
\begin{align}
    R^{(1)} (t_1) = (i/\hbar) \sum_{j=1}^{N-1} |\mu_{j0}|^2\, e^{-i\epsilon_j t_1/\hbar} \,\langle 0 | \phi_j\rangle \equiv \sum_{j=1}^{N-1} R^{(1)}_j (t_1),
\end{align}
where $\mu_{j0}=\langle j | \mu | 0 \rangle$ are the matrix elements between the initial ground state and the excited electronic states. These matrix elements also contain the dependence of the overall response function on the polarization of the laser pulses, which doesn't directly affect the dynamics of the vibrational degrees of freedom. Each excited-state specific contribution can thus be written as the product of an electronic and a vibrational factor 
\begin{align}
 R^{(1)}_j (t_1) = R^{(1,e)}_j (t_1) \, R^{(1,v)}_j (t_1) ,
\end{align}
where the latter coincides with the overlap $\langle 0 | \phi_j\rangle$ between the vibrational state correlated with the excited state $|j\rangle$ and the one correlated with the electronic ground state $|0\rangle$. These are both coherent vibrational states, resulting from the application to the initial vacuum state of the time evolution operators $e^{-i H_{v,j} t_1}$ and  $e^{-i H_{v,0} t_1}$, respectively. 
While it is obvious that $e^{-i H_{v,0} t_1}|0\rangle = |0\rangle$, the expression $|\phi_j\rangle$ can be derived by reducing the displaced-oscillator Hamiltonians $H_{v,j}$ to the undisplaced one, $H_{v,0}$, through the displacement operators $\mathcal{D}(\pm z_j)$ (the definitions and basic properties of the coherent states and of the displacement operators are reported in Appendix \ref{app:A}). As a result, one has that
\begin{align}
|\phi_j\rangle &= \mathcal{D}(-z_j)\, e^{-i\omega_v a^\dagger a t_1}\, \mathcal{D}(z_j) |0\rangle \nonumber\\
& = e^{-iz_j^2\sin(\omega_v t_1)} |z_j(e^{-i\omega_v t_1}-1)\rangle .
\end{align}
From this, and from the fact that $\langle 0 | z_j(e^{-i\omega_v t_1}-1)\rangle = e^{-z_j^2[1-\cos (\omega_v t_1)]}$, it follows that: 
\begin{align}
    R_j^{(v)} (t_1) & \equiv \langle 0 | \phi_j \rangle = e^{-z_j^2} \exp(z_j^2 e^{-i\omega_v t_1}).
\end{align}

This simple example illustrates the procedure that will be followed to model the modulation induced by the vibrational degrees of freedom in the response functions of arbitrary order $M$, starting from the case $M=3$: we derive the coherent state of the quantum oscillator as a function of the waiting times and for the relevant pathways, and write the vibrational component of the response function as an overlap between the coherent states corresponding to the two sides of the Feynman diagrams.

\section{Third-order contributions \\ in a multi-level system\label{sec:rfs}}


Within the perturbative approach, each contribution in the response function can be associated to a double-sided Feynman diagram, which specifies the action of the field on the terms of the density operator that eventually contribute to the polarization. At each interaction with the field, the electronic state in the ket or in the bra (left and right sides of the Feynman diagram) undergoes a transition. As in the case of the linear contribution, the modulation of the polarization induced by the vibrational degrees of freedom can be expressed in terms of an overlap between the final oscillator states of the ket and of the bra, $e^{ia_{ket}} | \phi_{3,ket} \rangle$ and $e^{ia_{bra}} | \phi_{3,bra} \rangle$. Such overlap defines in fact the vibrational component of the response function: 
\begin{equation}\label{eq:exp}
    R^{(v,3)} (t_1,t_2,t_3) = \langle \alpha_{3,bra} | \alpha_{3,ket} \rangle e^{i(a_{ket}-a_{bra})} \equiv e^{r} e^{i \varphi},
\end{equation}
whose amplitude depends on the distance between the two wave packets in the $(X,P)$ plane, being $r = -\frac{1}{2} | \alpha_{3,ket} - \alpha_{3,bra} |^2 $. As is apparent in the above equation, the phase $\varphi$ results both from the phase factors accumulated in the ket and in the bra, and from the overlap between the two coherent states:   
\begin{equation}\label{eq:ph}
\varphi = a_{ket}-a_{bra} + \text{Im} (\alpha_{3,ket} \,\alpha_{3,bra}^*).
\end{equation}
In the present and in the following sections, the dependence of the third-order response functions on the waiting times is expressed in a compact form through the quantities
$\Lambda_{p_1\,p_2\,p_3} \equiv (p_1 t_1+p_2 t_2+p_3 t_3)\,\omega_v$.

\begin{figure}[h]
\centering
\includegraphics[width=0.5\textwidth]{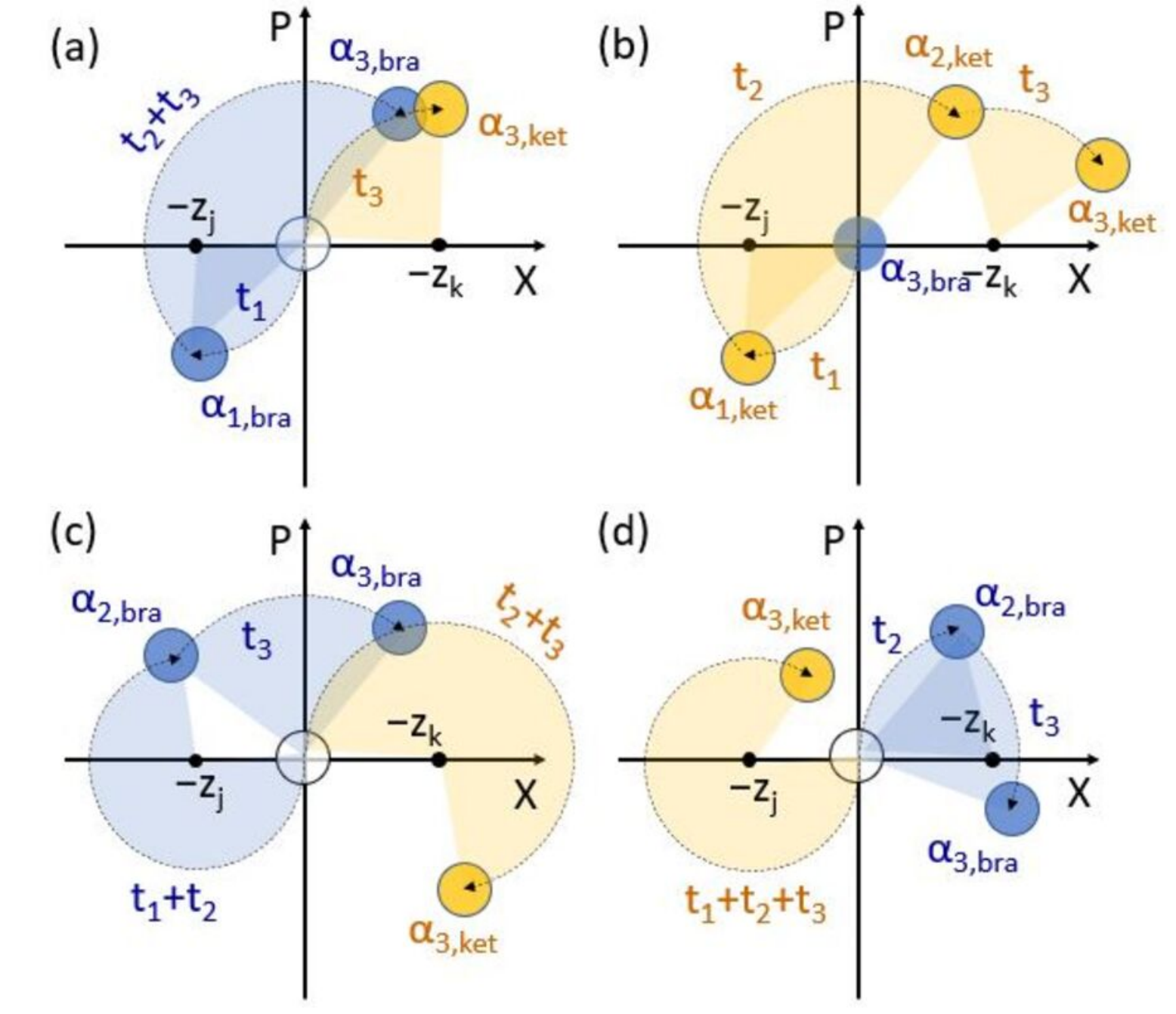}
\includegraphics[width=0.5\textwidth]{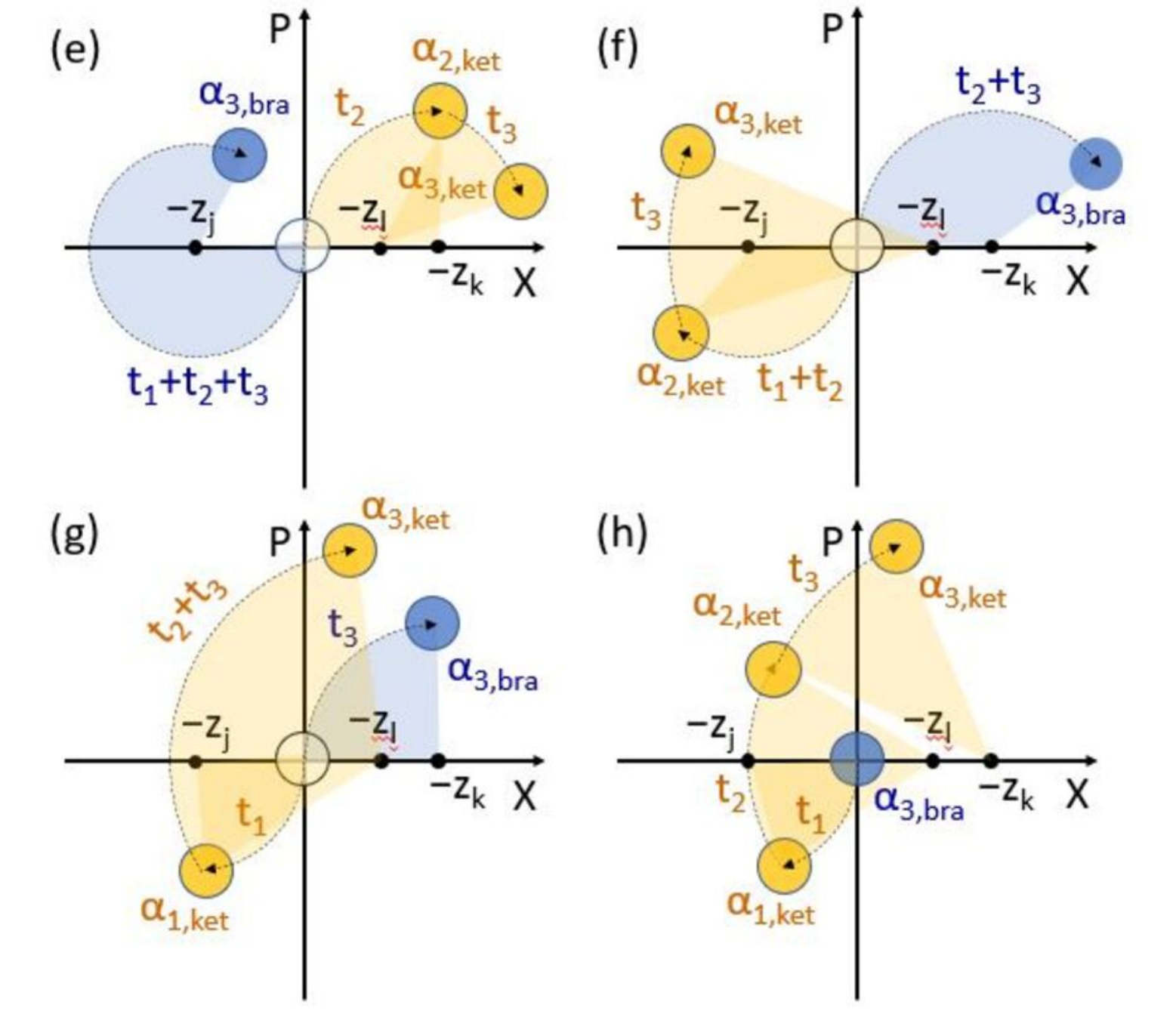}
\caption{Geometric representation of the time evolution in the $(X,P)$ plane of the vibrational wave packets corresponding to the third-order response functions. In particular: (a,b), (c,d), (e,f), (g,h) represent the rephasing and non-rephasing contributions respectively of the ground state bleaching, stimulated emission, excited state absorption, and double quantum coherence. \label{fig:te}}
\end{figure}

\subsection{Ground state bleaching}

The ground state bleaching is associated to those paths where both the ket and the bra are in the ground state during the second waiting time. It includes a rephasing and a non-rephasing contribution\cite{Mukamel95a,Hamm11a}. 

\paragraph{Rephasing contribution.}
The rephasing contributions correspond to all the sequences  
$ |0\rangle\langle 0| \longrightarrow |0 \rangle\langle j| \longrightarrow |0\rangle\langle 0| \longrightarrow |k\rangle\langle 0| $ [Fig. \ref{fig:dsfd2}(a)],
obtained for different combinations of the excited states $|j\rangle$ and $|k\rangle$. 
The electronic component of the response function, which oscillates with a positive (negative) frequency as a function of the first (third) waiting time, is given by:
\begin{align}\label{eq:2el}
    R^{(e,3)}_{2,jk} & = C_{2,jk} e^{i(\epsilon_{j}t_1-\epsilon_{k}t_3)/\hbar}.
\end{align}
Here, $\epsilon_{j/k}$ are the electronic energies ($\epsilon_0=0$) and $C_{2,jk}=(i/\hbar)^3 |\mu_{0j}\mu_{0k}|^2$\cite{Hamm11a}. 

As far as the vibrational degrees of freedom are concerned, the time evolution of the ket, leads, after the three waiting times, to the coherent state:
\begin{align}
    |\phi_{ket} \rangle & = e^{-iH_{v,k}t_3/\hbar} e^{-iH_{v,0}(t_1+t_2)/\hbar}|0\rangle \nonumber\\
    & = e^{-z_k^2\sin (\omega_v t_3)} | z_k (e^{-i\omega_v t_3}-1) \rangle .
\end{align}
The final position of the wave packet in the $(X,P)$ plane is thus obtained by applying a rotation by an angle $\omega_v t_3$ around the point $(0,-z_k)$, as shown in Fig. \ref{fig:te} (a). 
The vibrational state on the right side of the Feynman diagram (bra) undergoes a nontrivial evolution during all three waiting times, which eventually leads to the coherent state: 
\begin{align}
    |\phi_{bra} \rangle & = e^{-iH_{v,0}(t_2+t_3)/\hbar}e^{-iH_{v,j}t_1/\hbar}|0\rangle \nonumber\\
   & = e^{-z_j^2\sin (\omega_v t_1)}   | z_j (e^{-i\omega_v t_1}-1) e^{-i\omega_v (t_2+t_3)} \rangle .
\end{align}
This can be geometrically represented as a rotation by an angle $\omega_v t_1$ around $(0,-z_j)$, followed by a rotation by an angle $\omega_v (t_2+t_3)$ around the origin, as shown in Fig. \ref{fig:te} (a).
Further details on the intermediate ket and bra states, and on the expressions of $r$ and $\varphi$, are provided in Appendix \ref{app:B}.

The expression of the response function resulting from the above vibrational states reads: 
\begin{align}\label{eq:r2jk}
    R^{(v,3)}_{2,jk} & = \exp[ -(z_j^2+z_k^2) + z_j^2e^{i\Lambda_{100}} + z_k^2e^{-i\Lambda_{001}} \nonumber\\
    & + z_j z_k (-e^{i\Lambda_{010}}+e^{i\Lambda_{011}}+e^{i\Lambda_{110}}-e^{i\Lambda_{111}})],
\end{align}
which represents the main result of the present paragraph. 
Overall, the rephasing component of the response function associated to the ground state bleaching is obtained by combining the electronic and vibrational components for each pair $(j,k)$, and summing over all the possible combinations of such excited states:
\begin{align}\label{eq:r2}
    R_2^{(3)} = \sum_{j,k} R^{(e,3)}_{2,jk} \, R^{(v,3)}_{2,jk} .
\end{align}
The response function for the two-level system\cite{Mukamel95a} is obtained from the above expressions by setting $j=k=1$.

\paragraph{Non-rephasing contribution.}
The non-rephasing contributions correspond to the sequences
$ |0\rangle\langle 0| \longrightarrow |j \rangle\langle 0| \longrightarrow |0\rangle\langle 0| \longrightarrow |k\rangle\langle 0| $
[Fig. \ref{fig:dsfd2}(b)],
obtained for all the combinations of the excited states $|j\rangle$ and $|k\rangle$. 
The electronic component of the response function oscillates with a negative frequency as a function of both the first and the third waiting times:
\begin{align}\label{eq:5el}
    R^{(e,3)}_{5,jk} & = C_{5,jk} e^{-i(\epsilon_{j}t_1+\epsilon_{k}t_3)/\hbar},
\end{align}
where $C_{5,jk}=(i/\hbar)^3 |\mu_{0j}\mu_{0k}|^2$.

Both these excited states enter the expression of the final ket state, which is given by:
\begin{align}
    |\phi_{ket}\rangle & \!=\! e^{-iH_{v,k} t_3} e^{-iH_{v,0} t_2} e^{-iH_{v,j} t_1} |0\rangle \nonumber\\
    & \!=\! e^{ia_{ket}} | z_k (e^{-i\omega_v t_3}\!-\!1)\! +\! z_j (e^{-i\omega_v t_1}\!-\!1) e^{-i\omega_v (t_2+t_3)} \rangle .
\end{align}
The expression of $a_{ket}$, which coincides in this case with that of $\varphi$, reads:
\begin{align}
    a_{ket} & = - z_j^2 \sin\Lambda_{100} - z_k^2 \sin\Lambda_{001} \nonumber\\ & + z_j z_k (- \sin\Lambda_{010}+\sin\Lambda_{011}+\sin\Lambda_{110} -\sin\Lambda_{111}) .
\end{align}
The position of the wave packet in the $(X,P)$ plane is obtained by applying a sequence of three rotations: the first one by an angle $\omega_v t_1$, around $(0,-z_j)$; the second one by an angle $\omega_v t_2$ around the origin; the third one by an angle $\omega_v t_3$ around $(0,z_k)$ [Fig. \ref{fig:te} (b)].  
The bra, instead, doesn't undergo any time evolution, being the initial state $|0\rangle$ an eigenstate with zero eigenvalue of $H_{v,0}$. Further details on the intermediate ket states and on the quantity $r$ are provided in Appendix \ref{app:B}.

The above vibrational states result in the following expression of the response function:
\begin{align}\label{eq:r5jk}
    R_{5,jk}^{(v,3)} & = \exp[-(z_j^2+z_k^2) + z_j^2 e^{-i\Lambda_{100}} + z_k^2 e^{-i\Lambda_{001}} \nonumber\\
    & + z_j z_k (e^{-i\Lambda_{010}}-e^{-i\Lambda_{011}}-e^{-i\Lambda_{110}} +e^{-i\Lambda_{111}}) ].
\end{align}
Overall, the non-rephasing component of the response function associated to the ground state bleaching reads:
\begin{align}\label{eq:r5}
    R_5^{(3)} = \sum_{j,k} R^{(e,3)}_{5,jk} \, R^{(v,3)}_{5,jk} .
\end{align}
The response function for the two-level system\cite{Mukamel95a} is obtained from the above expressions by setting $j=k=1$.

\subsection{Stimulated emission}

The stimulated emission is associated to those paths where both the ket and the bra are in an excited state during the second waiting time, and the bra undergoes a deexcitation process at the end of such waiting time. It includes both a rephasing and a non-rephasing contribution\cite{Mukamel95a,Hamm11a}. 

\paragraph{Rephasing contribution.}
The rephasing contributions corresponds to the sequences
$ |0\rangle\langle 0| \longrightarrow |0 \rangle\langle j| \longrightarrow |k\rangle\langle j| \longrightarrow |k\rangle\langle 0| $ [Fig. \ref{fig:dsfd2}(c)].
For $j\neq k$, the term during the second waiting time is thus given by an electronic coherence between two different excited states, rather than by a population, as in the case of ground state bleaching. 
The electronic component of the response function is given by:
\begin{align}\label{eq:1el}
    R^{(e,3)}_{1,jk} & = C_{1,jk} e^{i[\epsilon_{j}(t_1+t_2)-\epsilon_{k}(t_2+t_3)]},
\end{align}
where $\epsilon_{j/k}$ are the energies of the electronic states $|j/k\rangle$ and $C_{1,jk}=(i/\hbar)^3 |\mu_{0j}\mu_{0k}|^2$\cite{Hamm11a}.

The ket undergoes a significant evolution during the second and third waiting times, resulting in:
\begin{align}
    |\phi_{ket}\rangle & = e^{-iH_{v,k}(t_2+t_3)/\hbar} e^{-iH_{v,0}t_1/\hbar} | 0 \rangle \nonumber\\
    & = e^{-z_k^2\sin [\omega_v (t_2+t_3)]} | z_k [e^{-i\omega_v (t_2+t_3)}-1] \rangle .
\end{align}
The final position of the wave packet is obtained by applying a rotation by an angle $\omega_v (t_2+t_3)$ around the point $(0,-z_k)$, as shown in Fig. \ref{fig:te} (c).
The final bra displays a dependence also on the first waiting time, being:
\begin{align}
    |\phi_{bra}\rangle & = e^{-iH_{v,0} t_3/\hbar} e^{-iH_{v,j}(t_1+t_2)/\hbar} | 0 \rangle \nonumber\\
    & = e^{-z_j^2\sin [\omega_v (t_1+t_2)]}  | z_j [e^{-i\omega_v (t_1+t_2)}-1]e^{-i\omega_v t_3} \rangle .
\end{align}
This geometrically results from a sequence of two rotations: one by an angle $\omega_v(t_1+t_2)$ around $(0,-z_j)$, and a second one by an angle $\omega_v t_3$ around the origin, as shown in Fig. \ref{fig:te} (c).
The expressions of the intermediate ket and bra states, as well as those of $r$ and $\varphi$, are reported in Appendix \ref{app:B}.

The response function corresponding to the above vibrational states is given by the following expression:
\begin{align}\label{eq:r1jk}
    R_{1,jk}^{(v,3)} & = \exp[-(z_j^2+z_k^2) + z_j^2 e^{i\Lambda_{110}} + z_k^2 e^{-i\Lambda_{011}} \nonumber\\
    & + z_j z_k (e^{i\Lambda_{001}}-e^{-i\Lambda_{010}}+e^{i\Lambda_{100}} -e^{i\Lambda_{111}}) ],
\end{align}
which represents the main result of the present paragraph.
The complete rephasing component of the response function associated to the stimulated emission thus reads:
\begin{align}\label{eq:r1}
    R_1^{(3)} = \sum_{j,k} R^{(e,3)}_{1,jk} \, R^{(v,3)}_{1,jk} .
\end{align}
By setting $j=k=1$ in the above expressions, one obtains the response function for the two-level system\cite{Mukamel95a}.

\paragraph{Non-rephasing contribution.}
The non-rephasing contribution corresponds to the sequence:
$ |0\rangle\langle 0| \longrightarrow |j \rangle\langle 0| \longrightarrow |j\rangle\langle k| \longrightarrow |j\rangle\langle 0| $ [Fig. \ref{fig:dsfd2}(d)].
As in the case of the rephasing contribution, one can have, during the second waiting time, a coherence between excited eletronic states. 
The electronic component of the response function is thus given by
\begin{align}\label{eq:4el}
    R^{(e,3)}_{4,jk} & = C_{4,jk} e^{i[\epsilon_{\y k} t_2-\epsilon_{\y j}(t_1+t_2+t_3)]/\hbar},
\end{align}
where $C_{4,jk}=(i/\hbar)^3 |\mu_{0j}\mu_{0k}|^2$\cite{Hamm11a}.

The time evolution of the ket after the three waiting times is given by
\begin{align}\label{eq:x}
    |\phi_{ket}\rangle & = e^{-iH_{v,j} (t_1+t_2+t_3)/\hbar} | 0 \rangle \nonumber\\
    & = e^{-z_j^2\sin [\omega_v (t_1+t_2+t_3)]} | z_j [e^{-i\omega_v (t_1+t_2+t_3)}-1] \rangle ,
\end{align}
geometrically corresponding to a rotation by an angle $\omega_v (t_1+t_2+t_3)$ around the point $(0,-z_j)$ [Fig. \ref{fig:te} (d)].
The vibrational state on the right side of the Feynman diagram reads instead
\begin{align}\label{eq:y}
    |\phi_{bra}\rangle & = e^{-iH_{v,0} t_3/\hbar} e^{-iH_{v,k} t_2/\hbar} e^{-iH_{v,0} t_1/\hbar} | 0 \rangle \nonumber\\
    & = e^{-z_k^2\sin (\omega_v t_2)} | z_k (e^{-i\omega_v t_2}-1)e^{-i\omega_v t_3} \rangle .
\end{align}
The position of the wave packet is obtained by performing a rotation by an angle $\omega_v t_2$ around $(0,-z_k)$, followed by a rotation by an angle $\omega_v t_3$ around the origin [Fig. \ref{fig:te} (d)].
The intermediate ket and bra states are reported in Appendix \ref{app:B}, together with the expressions of $r$ and $\varphi$.

The above dependence of the vibrational states on the waiting times leads to the response function:
\begin{align}\label{eq:r4jk}
    R_{4,jk}^{(v,3)} & = \exp[-(z_j^2+z_k^2) + z_j^2 e^{-i\Lambda_{111}} + z_k^2 e^{i\Lambda_{010}} \nonumber\\
    & + z_j z_k (e^{i\Lambda_{001}}+e^{-i\Lambda_{100}}-e^{i\Lambda_{011}}-e^{-i\Lambda_{110}}) ].
\end{align}
The overall expression of the response function results from the multiplication of the vibrational and the electronic contributions, summed over all the possible combinations of excited states $(j,k)$: 
\begin{align}\label{eq:r4}
R_4^{(3)} = \sum_{j,k} R^{(e,3)}_{4,jk} R^{(v)}_{4,jk} .
\end{align}
By setting $j=k=1$ in the above expressions, one obtains the response function for the two-level system\cite{Mukamel95a}.

\subsection{Excited state absorption}

The excited state absorption is associated to those paths where both the ket and the bra are in an excited state during the second waiting time, and the ket undergoes a further excitation process at the end of such period. It includes two non-equivalent set of contributions\cite{Mukamel95a,Hamm11a}. Both are absent in the response functions of two-level systems.

\paragraph{Rephasing contribution.} 
The first set of contributions corresponds to the sequences:
$ |0\rangle\langle 0| \longrightarrow |0 \rangle\langle j| \longrightarrow |k\rangle\langle j| \longrightarrow |l\rangle\langle j| $, where $l$ is a doubly excitated state [Fig. \ref{fig:dsfd2}(e)]. 
The electronic component of the response function, which oscillates with positive (negative) frequency as a function of the first (third) waiting time, is given by:
\begin{align}\label{eq:3el}
    R^{(e,3)}_{3,jkl} & = C_{3,jkl} e^{i[\epsilon_{j}(t_1+t_2+t_3)-\epsilon_{k}t_2-\epsilon_{l} t_3)]/\hbar},
\end{align}
where $C_{3,jk}=-(i/\hbar)^3 |\mu_{0j}|^2 \mu_{k0}\mu_{l k}$\cite{Hamm11a}.

The dependence of the ket on the three waiting times is given by the following equation:
\begin{align}
    |\phi_{ket}\rangle & = e^{-iH_{v,l}t_3/\hbar} e^{-iH_{v,k}t_2/\hbar}  e^{-iH_{v,0}t_1/\hbar}  |0\rangle  = e^{ia_{ket}} \nonumber\\ 
    &| z_{l} (e^{-i\omega_v t_3}-1) + z_k (e^{-i\omega_v t_2}-1) e^{-i\omega_v t_3} \rangle ,
\end{align}
where the phase factor is 
\begin{align}
    a_{ket} & = - z_k z_{kl} \sin (\omega_v t_2) - z_{l}z_k\sin [\omega_v (t_2+t_3)] \nonumber\\
    & - z_{l} z_{l k}\sin (\omega_v t_3) .
\end{align}
Here and in the following, we make use of the terms $z_{l k}\equiv z_{l}-z_k$. 
The corresponding position of the wave packet in the $(X,P)$ plane is obtained by applying to the point $(0,0)$ a rotation by an angle $\omega_v t_2$ around $(0,-z_k)$, followed by a rotation by $\omega_v t_3$ around $(0,-z_{l})$ [Fig. \ref{fig:te} (e)]. 
The expression of the bra reads instead:
\begin{align}
    |\phi_{bra}\rangle & = e^{-iH_{v,j}(t_1+t_2+t_3)/\hbar} |0\rangle  = e^{-z_j^2\sin [\omega_v (t_1+t_2+t_3)]} \nonumber\\ 
    &| z_j [e^{-i\omega_v (t_1+t_2+t_3)}-1] \rangle ,
\end{align}
corresponding to a rotation by an angle $\omega_v (t_1+t_2+t_3)$ around $(0,-z_j)$ [Fig. \ref{fig:te} (e)]. The intermediate ket and bra states, as well as the expressions of $r$ and $\varphi$, are reported in Appendix \ref{app:B}.

From the above equations, it follows that the response function reads:
\begin{align}\label{eq:re3jkl}
    R^{(v,3)}_{3,jkl} & = \exp\{-[z_j^2+z_{l}^2+z_k^2-z_{l}(z_j+z_k)]\nonumber\\
    & + z_{l k} z_{l j} e^{-i\Lambda_{001}} + z_k z_{kl} e^{-i\Lambda_{010}}  + z_j z_k e^{i\Lambda_{100}} \nonumber\\
    &+ z_k z_{l j} e^{-i\Lambda_{011}} - z_j z_{kl} e^{i\Lambda_{110}} - z_j z_{l j} e^{i\Lambda_{111}}\} ,
\end{align}
which is the main result of the present paragraph.
Overall, the response function associated to the rephasing part of the excited state absorption reads:
\begin{align}
R_3^{(3)} = \sum_{j,k,l} R^{(e,3)}_{3,jkl} R^{(v,3)}_{3,jkl} .
\end{align}

\paragraph{Non-rephasing contribution.}
The second set of contributions corresponds to the sequences:
$ |0\rangle\langle 0| \longrightarrow |j \rangle\langle 0| \longrightarrow |j\rangle\langle k| \longrightarrow |l\rangle\langle k| $
[Fig. \ref{fig:dsfd2}(f)].
The electronic part of the response function reads:
\begin{align}\label{eq:6el}
    R^{(e,3)}_{6,jkl} & = C_{6,jkl} e^{i[\epsilon_{k}(t_2+t_3)-\epsilon_{j}(t_1+t_2)-\epsilon_{l} t_3)]/\hbar}.
\end{align}
This oscillates with a negative frequency as a function of both $t_1$ and $t_3$, with $C_{6,jk}=-(i/\hbar)^3 |\mu_{0k}|^2 \mu_{j0}\mu_{l j}$\cite{Hamm11a}.

The time evolution of the ket is given by
\begin{align}
    |\phi_{ket}\rangle & = e^{-iH_{v,l}t_3/\hbar} e^{-iH_{v,j}(t_1+t_2)/\hbar} |0\rangle  = e^{ia_{ket}} \nonumber\\
    &| z_{l} (e^{-i\omega_v t_3}-1) + z_j [e^{-i\omega_v (t_1+t_2)}-1] e^{-i\omega_v t_3} \rangle,
\end{align}
where the phase factor reads
\begin{align}
    a_{ket} &= -z_j^2\sin [\omega_v(t_1+t_2)] -z_{l}^2 \sin(\omega_v t_3) +z_j z_{l} \{\sin (\omega_v t_3)\nonumber\\
    &-\sin [\omega_v(t_1+t_2+t_3)]+\sin [\omega_v(t_1+t_2)]\}.
\end{align}
The final position of the wave packet in the $(X,P)$ plane is obtained by applying a rotation by an angle $\omega_v(t_1+t_2)$ around $(0,-z_j)$, followed by a rotation by $\omega_v t_3$ around $(0,-z_{l})$ [Fig. \ref{fig:te} (f)].
The state of the bra reads 
\begin{align}
    |\phi_{bra}\rangle & = e^{-iH_{v,k}(t_2+t_3)/\hbar} e^{-iH_{v,0}t_1/\hbar} |0\rangle \nonumber\\
    & = e^{-z_k^2\sin [\omega_v (t_2+t_3)]}  | z_k [e^{-i\omega_v (t_2+t_3)}-1] \rangle ,
\end{align}
geometrically resulting from a single rotation, by $\omega_v (t_2+t_3)$, around $(0,-z_k)$ [Fig. \ref{fig:te} (f)].
The expressions of $r$, $\varphi$, and of the intermediate vibrational states are reported in Appendix \ref{app:B}.

These vibrational states result in a response function of the form:
\begin{align}\label{eq:re6jkl}
    R^{(v,3)}_{6,jkl} & = \exp\{-[z_j^2+z_k^2+z_{l}^2-z_{l}(z_j+z_k)] \nonumber\\ 
    & + z_{l k} z_{l j} e^{-i\Lambda_{001}} + z_k z_{l j} e^{i\Lambda_{010}}  + z_j z_k e^{-i\Lambda_{100}} \nonumber\\
    & - z_k z_{l k} e^{i\Lambda_{011}} + z_j z_{jl} e^{-i\Lambda_{110}} + z_j z_{l k} e^{-i\Lambda_{111}}\} ,
\end{align}
which represents the main result of the present paragraph.
Overall, the response function associated to the non-rephasing part of the excited state absorption reads:
\begin{align}
R_6^{(3)} = \sum_{j,k,l} R^{(e,3)}_{6,jkl} R^{(v,3)}_{6,jkl} .
\end{align}

\subsection{Double quantum coherence}

We finally consider the pathways that involve coherences between the ground and a doubly excited state\cite{Kim2009}.  
These include two non-equivalent kind of contributions, both of which are absent in the case of two-level systems.

\paragraph{First contribution.}
The first kind of contributions correspond to the sequence:
$ |0\rangle\langle 0| \longrightarrow |j \rangle\langle 0| \longrightarrow |{l}\rangle\langle 0| \longrightarrow |{l}\rangle\langle k| $ [Fig. \ref{fig:dsfd2}(g)]. The electronic component of the response function, which oscillates with a positive (negative) frequency as a function of the first (third) waiting time, is given by 
\begin{align}\label{eq:7el}
    R^{(e,3)}_{7,jkl} & = C_{7,jkl} e^{-i[\epsilon_{j}t_1+\epsilon_{l}(t_2+t_3)-\epsilon_k t_3]/\hbar},
\end{align}
where
$C_{7,jk}=-(i/\hbar)^3 |\mu_{0k}|^2 \mu_{j0}\mu_{l j}$.

The ket state after the three waiting times is given by:
\begin{align}
    |\phi_{ket} \rangle & =  e^{-iH_{v,l}(t_2+t_3)/\hbar}  e^{-iH_{v,j}t_1/\hbar} |0\rangle  = e^{ia_{ket}}  \nonumber\\ 
    &| z_{l} [e^{-i\omega_v (t_2+t_3)}-1] + z_j (e^{-i\omega_v t_1}-1) e^{-i\omega_v (t_2+t_3)} \rangle ,
\end{align}
where the phase factor reads
\begin{align}
    a_{ket} & = -z_j^2 \sin (\omega_v t_1) - z_{l} (z_{l}-z_j) \sin [\omega_v (t_2+t_3)]\nonumber\\ & -z_{l} z_j \sin [\omega_v (t_1+t_2+t_3)]+z_{l} z_j \sin (\omega_v t_1) .
\end{align}
The final position of the wave packet is determined by a sequence of two rotations, respectively by and angle $\omega_v t_1$ around $(0,-z_j)$ and by an angle $\omega_v (t_2+t_3)$ around $(0,-z_{l})$, as shown in Fig. \ref{fig:te} (g).
The time evolution of the bra is given by:
\begin{align}
    |\phi_{bra}\rangle & = e^{-iH_{v,\y k}t_3/\hbar}  e^{-iH_{v,0}(t_1+t_2)/\hbar} |0\rangle \nonumber \nonumber\\
    & = e^{-z_k^2\sin (\omega_v t_3)} |z_k (e^{-i\omega_v t_3}-1)\rangle \nonumber ,
\end{align}
which geometrically corresponds to a single rotation, by angle $\omega_v t_3$ around $(0,-z_k)$, as shown in Fig. \ref{fig:te} (g).
The quantities $r$ and $\varphi$, and the intermediate ket and bra states are given in Appendix \ref{app:B}.

Therefore, the vibrational component of the response function reads:
\begin{align}\label{eq:re7jkl}
    R_{7,jkl}^{(v,3)} & = \exp\{-[z_j^2+z_{l}^2+z_k^2-z_{l} (z_j+z_k)] \nonumber\\
    &  + z_k z_{kl} e^{i\Lambda_{001}} + z_k z_{l j} e^{-i\Lambda_{010}}  - z_j z_{l j} e^{-i\Lambda_{100}}  \nonumber\\ 
    & + z_{l j} z_{l k} e^{-i\Lambda_{011}} +z_j z_k e^{-i\Lambda_{110}}  + z_j z_{l k} e^{-i\Lambda_{111}}\} . 
\end{align}
The overall response function is obtained by multiplying the above expression by the electronic contribution, and summing over all the possible combinations of excited states $(j,k,l)$:
\begin{align}
R_7^{(3)} = \sum_{j,k,l} R^{(e,3)}_{7,jkl} R^{(v,3)}_{7,jkl} .
\end{align}

\paragraph{Second contribution.}
The second kind of contributions related to the double quantum coherence correspond to sequences:
$ |0\rangle\langle 0| \longrightarrow |j \rangle\langle 0| \longrightarrow |{l}\rangle\langle 0| \longrightarrow |k\rangle\langle 0| $ [Fig. \ref{fig:dsfd2}(h)].
The electronic component of the response function is given by:
\begin{align}\label{eq:8el}
    R^{(e,3)}_{8,jkl} & = C_{8,jkl} e^{-i(\epsilon_{j}t_1+\epsilon_{l}t_2+\epsilon_k t_3)/\hbar}.
\end{align}
This oscillates with a negative frequency as a function of both the first and third waiting times, 
where
$C_{8,jk}=(i/\hbar)^3 \mu_{j0} \mu_{l j} \mu_{l k} \mu_{k0}$.

The time evolution of the ket is given by
\begin{align}
    |\phi_{ket} \rangle & =  e^{-iH_{v,k}t_3/\hbar}  e^{-iH_{v,l}t_2/\hbar} e^{-iH_{v,j}t_1/\hbar} |0\rangle \nonumber\\ 
    & = e^{ia_{ket}} | -z_k + e^{-i\omega_vt_3} (z_k-z_{l}) \nonumber\\
    & + e^{-i\omega_v(t_2+t_3)} (z_{l}-z_j) +z_j e^{-i\omega_v(t_1+t_2+t_3)} \rangle .
\end{align}
The expression of $a_{ket}$, which here coincides with $\varphi$, is given by:
\begin{align}
    a_{ket} & = -z_k z_{kl}\sin\Lambda_{001}-z_{l k} z_{l j} \sin\Lambda_{010} - z_j z_{jl} \sin\Lambda_{100} \nonumber\\ &   
      -z_k z_{l j} \sin\Lambda_{011} - z_j z_{l k} \sin\Lambda_{110} - z_j z_k \sin\Lambda_{111} .
\end{align}

This geometrically corresponds to a sequence of three rotations: the first one by an angle $\omega_v t_1$ around $(0,-z_j)$; the second one by an angle $\omega_v t_2$ around $(0,-z_{l})$; the third one by an angle $\omega_v t_3$ around $(0,-z_k)$ [Fig. \ref{fig:te} (h)].
The bra doesn't undergo any time evolution: $|\phi_{bra}\rangle = |0\rangle \nonumber$.

Combining together the above equations, one obtains the expression of the vibrational component of the response function, which reads:
\begin{align}\label{eq:re8jkl}
    R_{8,jkl}^{(v,3)} & = \exp\{-[z_j^2+z_{l}^2+z_k^2- z_{l}(z_j +z_k)] \nonumber\\
    & +z_k z_{kl} e^{-i\Lambda_{001}} + z_{l k} z_{l j} e^{-i\Lambda_{010}} + z_j z_{jl} e^{-i\Lambda_{100}} \nonumber\\ & 
    + z_k z_{l j} e^{-i\Lambda_{011}} + z_j z_{l k} e^{-i\Lambda_{110}} + z_j z_k e^{-i\Lambda_{111}}\}.  
\end{align}
The overall response function is obtained by multiplying the above function by the electronic component, and summing over all the possible combination of excited states: 
\begin{align}
R_8^{(3)} = \sum_{j,k,l} R^{(e,3)}_{8,jkl} R^{(v,3)}_{8,jkl} .
\end{align}

\section{Spectral components in the third-order response functions}

The response functions derived in the previous Section give rise to an infinite number of spectral components, each identified by the frequency of the oscillations as a function of the three waiting times. In the present Section, we outline a route for explicitly deriving the weight of each individual spectral component.

The third-order response functions are exponential functions of $\Lambda_\chi$, with $\chi = p_1\,p_2\,p_3 = 001,010,100,011,110,111$. Expanding the exponential in Taylor series, the generic response function $R^{(v,3)}$ can be expressed as follows: 
\begin{align}\label{eq:30}
    R^{(v,3)} & = e^{-h({\bf z})} \prod_\chi \sum_{n_\chi=0}^\infty \frac{(s_{\chi} z_\chi z_\chi' e^{s_{\chi}' i\Lambda_\chi})^{n_\chi}}{n_\chi!}\nonumber\\
    & = e^{-h({\bf z})} \sum_{p_1,p_2,p_3=-\infty}^{+\infty} C_{p_1\,p_2\,p_3} e^{i\Lambda_{p_1\,p_2\,p_3}},
\end{align}
where $\Lambda_{p_1\,p_2\,p_3} \equiv (p_1 t_1+p_2 t_2+p_3 t_3)\,\omega_v$, the signs $s_{\chi},s_{\chi}'=\pm 1$, the displacements $z_\chi$ and $z_\chi'$, and the function $h({\bf z})$ (which includes all the non-oscillating terms in the exponent $f$), all depend on the specific response function under consideration. The weight of each spectral component is determined by the corresponding coefficient $C_{p_1,p_2,p_3}$, which is in turn is given by the sum of all the terms in the first line of Eq. (\ref{eq:30}) that fulfil the conditions:
\begin{align}
    p_1 & = s_{100}'n_{100}+s_{110}'n_{110}+s_{111}'n_{111} \label{eq:32} \\
    p_2 & = s_{010}'n_{010}+s_{011}'n_{011}+s_{110}'n_{110}+s_{111}'n_{111} \label{eq:31}\\
    p_3 & = s_{001}'n_{001}+s_{011}'n_{011}+s_{111}'n_{111} \label{eq:33} .
\end{align}
These equations can be used in order to express three of the exponents, for example $n_{100}$, $n_{010}$, and $n_{001}$, as a function of the other three. If, in addition, one is specifically interested in the terms of order $2q$ in the displacements, the number of independent exponents $n_\chi$ is further reduced by the condition
\begin{align}
    q = n_{100} + n_{010} + n_{001} + n_{110} + n_{011} + n_{111} , 
\end{align}
which can be used to reduce to two the number of independent exponents in Eq. (\ref{eq:30}). (The exponential $e^{-h({\bf z})}$ is not expanded in Taylor series, for its exact value can be easily derived from the knowledge of the displacements, and the function doesn't contribute to the values of the frequencies considered in the spectral decomposition.)

In the semi-impulsive limit (i.e. for laser pulses of infinitesimally short duration) the response function can be directly related to the observed spectra\cite{Hamm11a}. The amplitude and phase of each peak, and other spectral features, can thus be obtained from the above expressions. In particular, the amplitude of a peak centered in the point $(\omega_{1,p_1},\omega_{3,p_3})$ of the $(\omega_1,\omega_3)$ plane is given by the following function of $t_2$:
\begin{align}
    A_{p_1,p_3}(t_2) = e^{-h({\bf z})} \sum_{p_2=-\infty}^{+\infty} C_{p_1\,p_2\,p_3} e^{i\,p_2\,\omega_v\,t_2} ,
\end{align}
where $\omega_{1,0}$ and $\omega_{1,p_1}=\omega_{1,0}-p_1\,\omega_v$ are the frequencies of the zero-phonon line and of their replicas, respectively (analogously for $\omega_{3,p_3}$).

In the following, we derive analytical expressions for the coefficients $C_{p_1\,p_2\,p_3}$ corresponding to each pathway (i.e. set of involved electronic states $|j\rangle$, $|k\rangle$, and $|l\rangle$). This allows one to determine their values for each particular physical system, given the corresponding set of displacements $z_j$, $z_k$, and $z_l$.

\begin{figure}[h]
\centering
\includegraphics[width=0.45\textwidth]{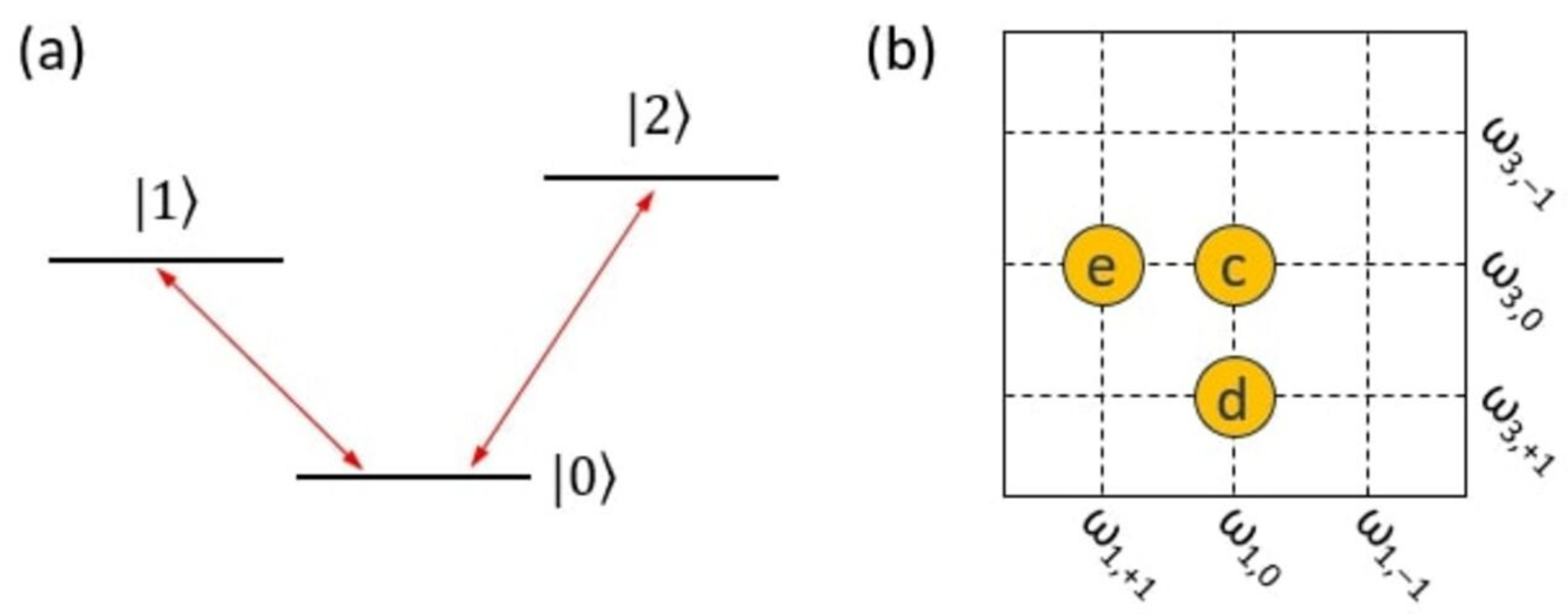}
\includegraphics[width=0.5\textwidth]{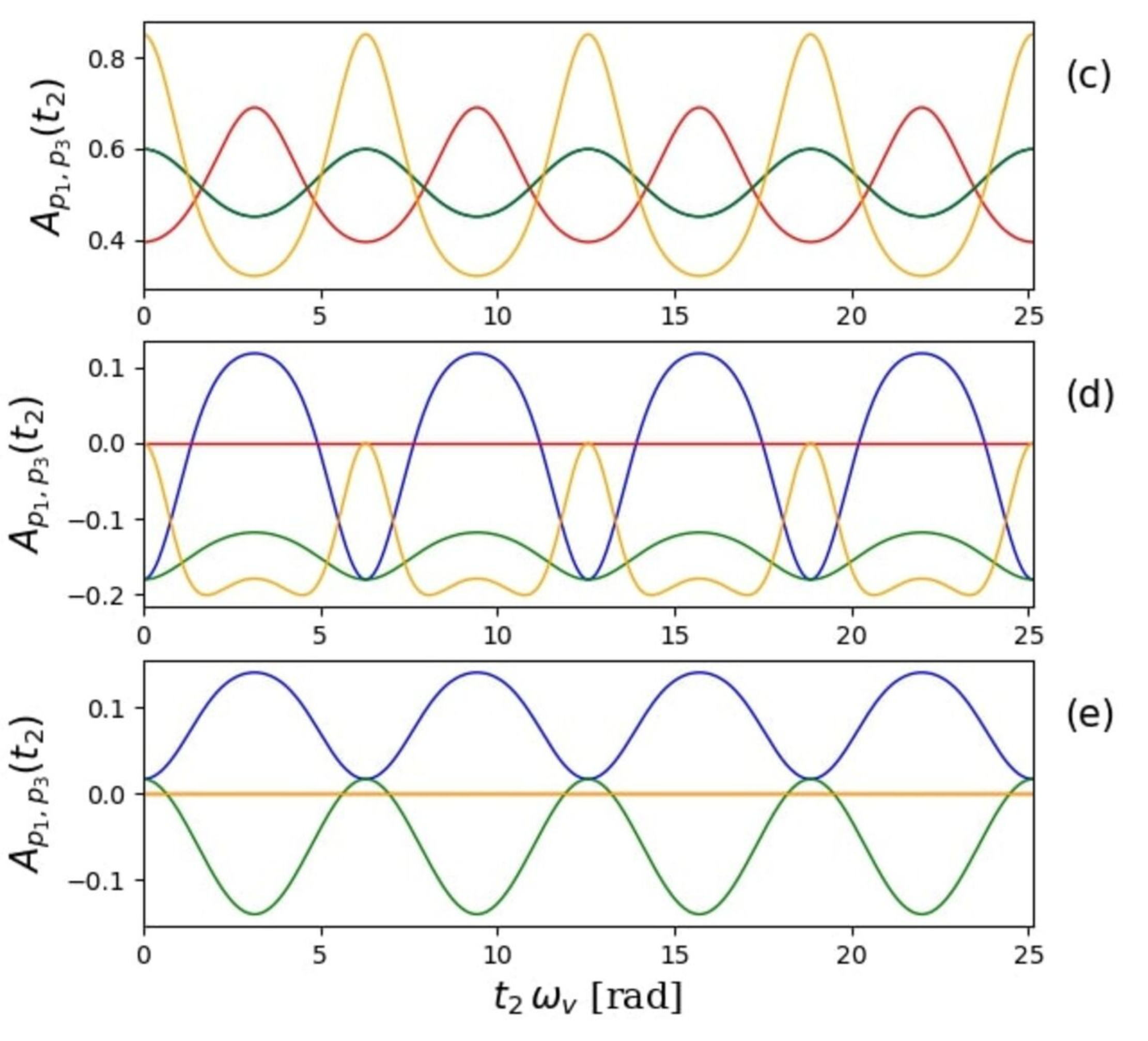}
\caption{Dependence on the waiting time $t_2$ of three representative peaks in the 2D map of a system characterized by a three-level $V$ scheme (a). The panels (c-e) report the (real part of) the amplitudes for the peaks highlighted in panel (b). Different colors of the solid lines correspond to different contributions: ground state bleaching, rephasing (blue) and non-rephasing (red); stimulated emission, rephasing (green) and non-rephasing (orange). The values of the displacements are: $z_1=0.4$ and $z_2=-0.7$.}\label{fig:sc1}
\end{figure}

\subsection{Ground state bleaching}
We start by considering the response functions related to ground state bleaching.
\paragraph{Rephasing contribution.}
From Eq. (\ref{eq:r2jk}), it follows that $s_{001}'=-1$, while $s_\chi'=+1$ in all other cases. 
The weight of the $q$-th order contribution is given by the sum
\begin{align}\label{eq:80}
    C_{p_1,p_2,p_3}^{(q)} & = \sum_{n_{110},n_{111}=0}^q
    \frac{z_j^{2n_{100}}}{n_{100}!}
    \frac{z_k^{2n_{001}}}{n_{001}!}\frac{(-z_j z_k)^{n_{010}}}{n_{010}!}\nonumber\\
    &
    \frac{(z_j z_k)^{n_{110}}}{n_{110}!}
    \frac{(z_j z_k)^{n_{011}}}{n_{011}!}
    \frac{(-z_j z_k)^{n_{111}}}{n_{111}!},
\end{align}
where 
$n_{100} = p_1 - n_{110} - n_{111}$, 
$n_{001} = q - p_1 -p_2 + n_{110} + n_{111} $,
$n_{010} = p_1 +2p_2-p_3-q-2n_{110}-n_{111}$, 
and
$n_{011} = n_{110} + q - p_1 -p_2 +p_3 $.
The terms that actually contribute to the sum in Eq. (\ref{eq:80}) correspond to the values of the independent exponents $n_{110}$ and $n_{111}$ such that all the other ones, resulting from the above relations, are non-negative. 
The exponent in the prefactor is given by $h({\bf z}) = z_j^2+z_k^2$.

\subparagraph{Example.} We consider as an example the case of a three-level $V$ system, for $j=1$ and $k=2$, where the zero-phonon peak corresponds to $(\omega_{1,0},\omega_{3,0})=(-\epsilon_1/\hbar,\epsilon_2/\hbar)$ (see Eq. \ref{eq:2el}). The phonon replicas are found at excitation and detection frequencies $\omega_{1,p_1}=-\epsilon_1/\hbar-p_1\omega_v$ and $\omega_{3,p_3}=\epsilon_2/\hbar-p_3\omega_v$, respectively. Their amplitude and phase, as a function of $t_2$, are given by:
\begin{align}\label{eq:x01}
    A_{p_1,p_3}(t_2) & = e^{-(z_1^2+z_2^2)} \sum_{n_{011},n_{110},n_{111}=0}^\infty
    \frac{(z_1 z_2)^{n_{110}+n_{011}}}{n_{110}!\,n_{011}!}
    \nonumber\\
    & \frac{z_2^{2n_{001}}}{n_{001}!} \frac{z_1^{2n_{100}}}{n_{100}!}
    \sum_{p_2=-\infty}^{+\infty} e^{i\,p_2\,\omega_v\,t_2} 
    \frac{(-z_1 z_2)^{n_{010}+n_{111}}}{n_{010}!\,n_{111}!},
\end{align}
where 
$n_{100} = p_1 - n_{110} - n_{111}$, 
$n_{001} = -p_3+n_{011} + n_{111} $,
$n_{010} = p_2-n_{011}-n_{110}-n_{111}$. 
The amplitude of the peak is thus identically zero for $p_1<0$, because in this case the first of the three equations above cannot be fulfilled. 
The function $A_{p_1,p_3}(t_2)$ is plotted in Fig. \ref{fig:sc1} (blue lines), for specific values of the displacements.
The other possible pathways, $j=k=1$ and $j=k=2$, give rise to contributions that are formally identical to the ones that are obtained for a two-level system: they can be obtained from Eq. (\ref{eq:x01}) by replacing respectively $z_2$ with $z_1$ or vice versa.

\paragraph{Non-rephasing contribution.}
From Eq. (\ref{eq:r5jk}) it follows that $s_\chi'=-1$ in all cases. 
The weight of the $q$-th order contribution is given by the sum
\begin{align}\label{eq:81}
    C_{p_1,p_2,p_3}^{(q)} & = \sum_{n_{110},n_{111}=0}^q
    \frac{z_j^{2n_{100}}}{n_{100}!}
    \frac{z_k^{2n_{001}}}{n_{001}!}\frac{(z_j z_k)^{n_{010}}}{n_{010}!}\nonumber\\
    &
    \frac{(-z_j z_k)^{n_{110}}}{n_{110}!}
    \frac{(-z_j z_k)^{n_{011}}}{n_{011}!}
    \frac{(z_j z_k)^{n_{111}}}{n_{111}!},
\end{align}
where 
$n_{100} = - p_1 - n_{110} - n_{111}$, 
$n_{001} = q - p_1 -p_2 - n_{110} - n_{111} $,
$n_{010} = -p_1 -2p_2+p_3+q-2n_{110}-n_{111}$, 
and
$n_{011} = n_{110} - q + p_1 +p_2 -p_3 $.
The sum in Eq. (\ref{eq:81}) actually involves only the values of the independent exponents $n_{110}$ and $n_{111}$ for which all the other ones, resulting from the above relations, are non-negative.
The exponent in the prefactor is given by $h({\bf z}) = z_j^2+z_k^2$.

\subparagraph{Example.} In the case of a three-level $V$ system, for $j=1$ and $k=2$, the zero-phonon peak corresponds to $(\omega_{1,0},\omega_{3,0})=(\epsilon_1/\hbar,\epsilon_2/\hbar)$ (see Eq. \ref{eq:5el}). The phonon replicas are found at excitation and detection frequencies $\omega_{1,p_1}=\epsilon_1/\hbar-p_1\omega_v$ and $\omega_{3,p_3}=\epsilon_2/\hbar-p_3\omega_v$, respectively. Their amplitude and phase, as a function of $t_2$, are given by:
\begin{align}\label{eq:x02}
    A_{p_1,p_3}(t_2) & = e^{-(z_1^2+z_2^2)} \sum_{n_{011},n_{110},n_{111}=0}^\infty
    \frac{(-z_1 z_2)^{n_{110}+n_{011}}}{n_{110}!\,n_{011}!} \nonumber\\
    &  
    \frac{z_1^{2n_{100}}}{n_{100}!}
    \frac{z_2^{2n_{001}}}{n_{001}!}
    \sum_{p_2=-\infty}^{+\infty} e^{i\,p_2\,\omega_v \,t_2} 
    \frac{(z_1 z_2)^{n_{010}+n_{111}}}{n_{010}!\,n_{111}!},
\end{align}
where 
$n_{100} = - p_1 - n_{110} - n_{111}$, 
$n_{001} = - p_3 - n_{011} - n_{111}  $,
$n_{010} = - p_2-n_{011}-n_{110}-n_{111}$. The amplitude of the peak is thus identically zero for $p_1>0$ or $p_3>0$. 
The function $A_{p_1,p_3}(t_2)$ is plotted in Fig. \ref{fig:sc1} (red lines), for specific values of the displacements.
The other possible pathways, $j=k=1$ and $j=k=2$, give rise to contributions that are formally identical to the ones that are obtained for a two-level system. They can be obtained from Eq. (\ref{eq:x02}) by replacing respectively $z_2$ with $z_1$ or vice versa.

\subsection{Stimulated emission}
The second set of response functions we consider are the ones related to stimulated emission.
\paragraph{Rephasing contribution.}
As can be seen from Eq. (\ref{eq:r1jk}), $s_{010}'=s_{011}'=-1$ and $s_\chi'=+1$ in all other cases, while $h({\bf z})=z_j^2+z_k^2$.
The weight of the $q$-th order contribution is given by the sum
\begin{align}\label{eq:82}
    C_{p_1,p_2,p_3}^{(q)} & = \sum_{n_{110},n_{111}=0}^q
    \frac{(z_j z_k)^{n_{100}}}{n_{100}!}
    \frac{(z_j z_k)^{n_{001}}}{n_{001}!}
    \frac{(-z_j z_k)^{n_{010}}}{n_{010}!}\nonumber\\
    &
    \frac{z_j^{2n_{110}}}{n_{110}!}
    \frac{z_k^{2n_{011}}}{n_{011}!}
    \frac{(-z_j z_k)^{n_{111}}}{n_{111}!},
\end{align}
where 
$n_{100} = p_1 - n_{110} - n_{111}$, 
$n_{001} = - q + p_1 + p_2 - n_{110} - n_{111} $,
$n_{010} = -p_1 -2p_2+p_3+q+2n_{110}+n_{111}$, 
and
$n_{011} = -n_{110} - q + p_1 +p_2 -p_3 $.
The terms that actually contribute to the sum in Eq. (\ref{eq:82}) correspond to the values of the independent exponents $n_{110}$ and $n_{111}$ such that all the other ones, resulting from the above relations, are non-negative.

\subparagraph{Example.} In the case of a three-level $V$ system, for $j=1$ and $k=2$, the zero-phonon peak corresponds to $(\omega_{1,0},\omega_{3,0})=(-\epsilon_1/\hbar,\epsilon_2/\hbar)$ (see Eq. \ref{eq:1el}). This contribution involves a coherence between the two exited states, $|1\rangle$ and $|2\rangle$, during the second waiting time. 
The phonon replicas are found at excitation and detection frequencies $\omega_{1,p_1}=-\epsilon_1/\hbar-p_1\omega_v$ and $\omega_{3,p_3}=\epsilon_2/\hbar-p_3\omega_v$, respectively. Their amplitude and phase, as a function of $t_2$, are given by:
\begin{align}\label{eq:x03}
    A_{p_1,p_3}(t_2) & = e^{-(z_1^2+z_2^2)} \sum_{n_{011},n_{110},n_{111}=0}^\infty
    \frac{(z_1 z_2)^{n_{100}+n_{001}}}{n_{100}!\,n_{001}!} \nonumber\\
    &  \frac{z_1^{2n_{110}}}{n_{110}!}
    \frac{z_2^{2n_{011}}}{n_{011}!}
    \sum_{p_2=-\infty}^{+\infty} e^{i\,p_2\,\omega_v \,t_2} 
    \frac{(-z_1 z_2)^{n_{010}+n_{111}}}{n_{010}!\,n_{111}!},
\end{align}
where 
$n_{100} = p_1 - n_{110} - n_{111}$, 
$n_{001} = p_3+n_{011} - n_{111} $,
$n_{010} = - p_2-n_{011}+n_{110}+n_{111}$. The amplitude of the peak is thus identically zero for $p_1<0$. The function $A_{p_1,p_3}(t_2)$ is plotted in Fig. \ref{fig:sc1} (green lines), for specific values of the displacements. The other possible pathways, $j=k=1$ and $j=k=2$, give rise to contributions that are formally identical to the ones that are obtained for a two-level system. They can be obtained from Eq. (\ref{eq:x03}) by replacing respectively $z_2$ with $z_1$ or vice versa.

\paragraph{Non-rephasing contribution.}
From Eq. (\ref{eq:r4jk}) it follows that $s_{100}'=s_{110}'=s_{111}'=-1$, while the sign is positive in the other cases, while $h({\bf z})=z_j^2+z_k^2$.
The weight of the $q$-th order contribution is given by the sum
\begin{align}\label{eq:84}
    C_{p_1,p_2,p_3}^{(q)} & = \sum_{n_{110},n_{111}=0}^q
    \frac{(z_j z_k)^{n_{100}}}{n_{100}!}
    \frac{(z_j z_k)^{n_{001}}}{n_{001}!}
    \frac{z_k^{2n_{010}}}{n_{010}!}\nonumber\\
    &
    \frac{(-z_j z_k)^{n_{110}}}{n_{110}!}
    \frac{(-z_j z_k)^{n_{011}}}{n_{011}!}
    \frac{z_j^{2n_{111}}}{n_{111}!},
\end{align}
where 
$n_{100} = -p_1 - n_{110} - n_{111}$, 
$n_{001} = -q + p_1 +p_2 + n_{110} + n_{111} $,
$n_{010} = p_1 +2p_2-p_3-q+2n_{110}+n_{111}$, 
and
$n_{011} = -n_{110} + q - p_1 -p_2 +p_3 $.
The terms that actually contribute to the sum in Eq. (\ref{eq:84}) correspond to the values of the independent exponents $n_{110}$ and $n_{111}$ such that all the other ones, resulting from the above relations, are non-negative.

\subparagraph{Example.} In the case of a three-level $V$ system, for $j=1$ and $k=2$, the zero-phonon peak corresponds to $(\omega_{1,0},\omega_{3,0})=(\epsilon_1/\hbar,\epsilon_1/\hbar)$ (see Eq. \ref{eq:4el}).
This contribution involves a coherence between the two exited states, $|1\rangle$ and $|2\rangle$, during the second waiting time. 
The phonon replicas are found at excitation and detection frequencies $\omega_{1,p_1}=\epsilon_1/\hbar-p_1\omega_v$ and $\omega_{3,p_3}=\epsilon_1/\hbar-p_3\omega_v$, respectively. Their amplitude and phase, as a function of $t_2$, are given by:
\begin{align}\label{eq:x04}
    A_{p_1,p_3}(t_2) & = e^{-(z_1^2+z_2^2)} \sum_{n_{011},n_{110},n_{111}=0}^\infty
    \frac{(z_1 z_2)^{n_{100}+n_{001}}}{n_{100}!\,n_{001}!}
     \nonumber\\
    &   
    \frac{z_1^{2n_{111}}}{n_{111}!}
    \frac{(-z_1 z_2)^{n_{110}+n_{011}}}{n_{110}!\,n_{011}!}
    \sum_{p_2=-\infty}^{+\infty} e^{i\,p_2\,\omega_v \,t_2} \frac{z_2^{2n_{010}}}{n_{010}!},
\end{align}
where 
$n_{100} = - p_1 - n_{110} - n_{111}$, 
$n_{001} = p_3 - n_{011} + n_{111} $,
$n_{010} = p_2-n_{011}+n_{110}+n_{111}$. The amplitude of the peak is thus identically zero for $p_1>0$. The function $A_{p_1,p_3}(t_2)$ is plotted in Fig. \ref{fig:sc1} (orange lines), for specific values of the displacements.
The other possible pathways, $j=k=1$ and $j=k=2$, give rise to contributions that are formally identical to the ones that are obtained for a two-level system. They can be obtained from Eq. (\ref{eq:x04}) by replacing respectively $z_2$ with $z_1$ or vice versa.

\begin{figure}[h]
\centering
\includegraphics[width=0.45\textwidth]{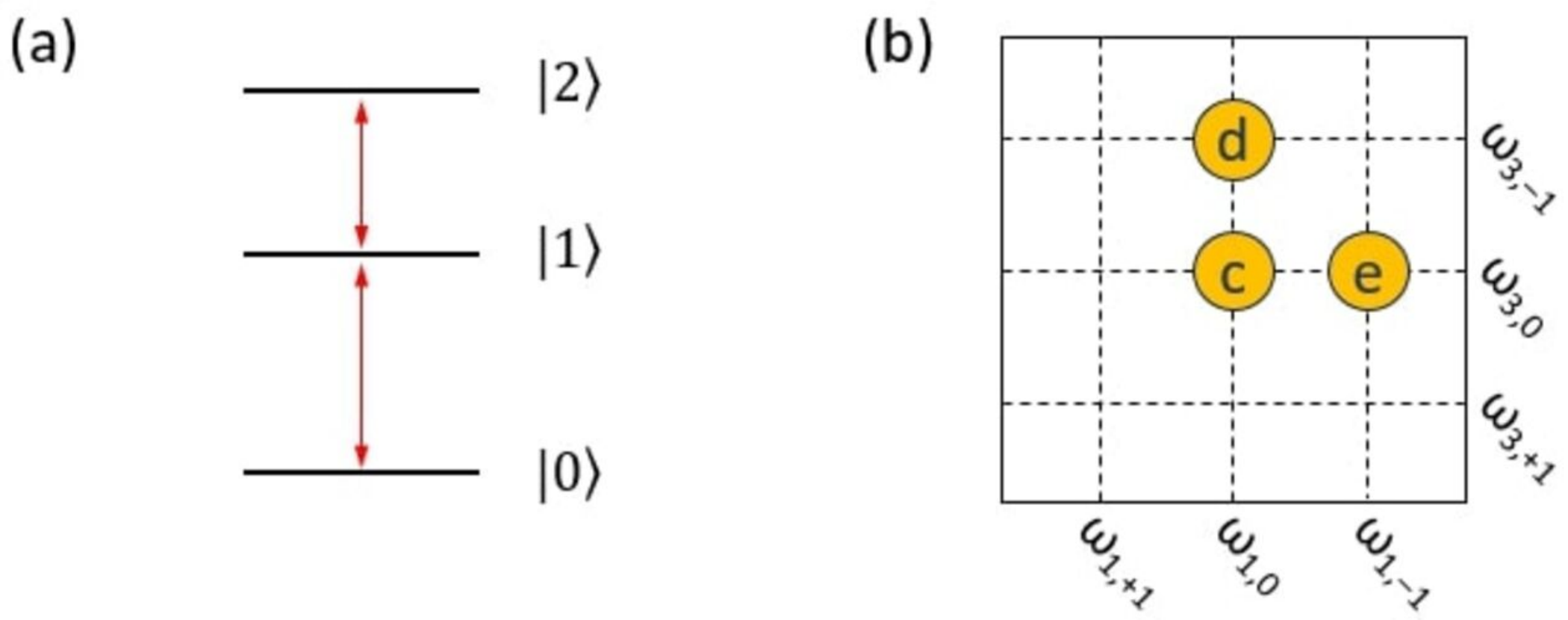}
\includegraphics[width=0.5\textwidth]{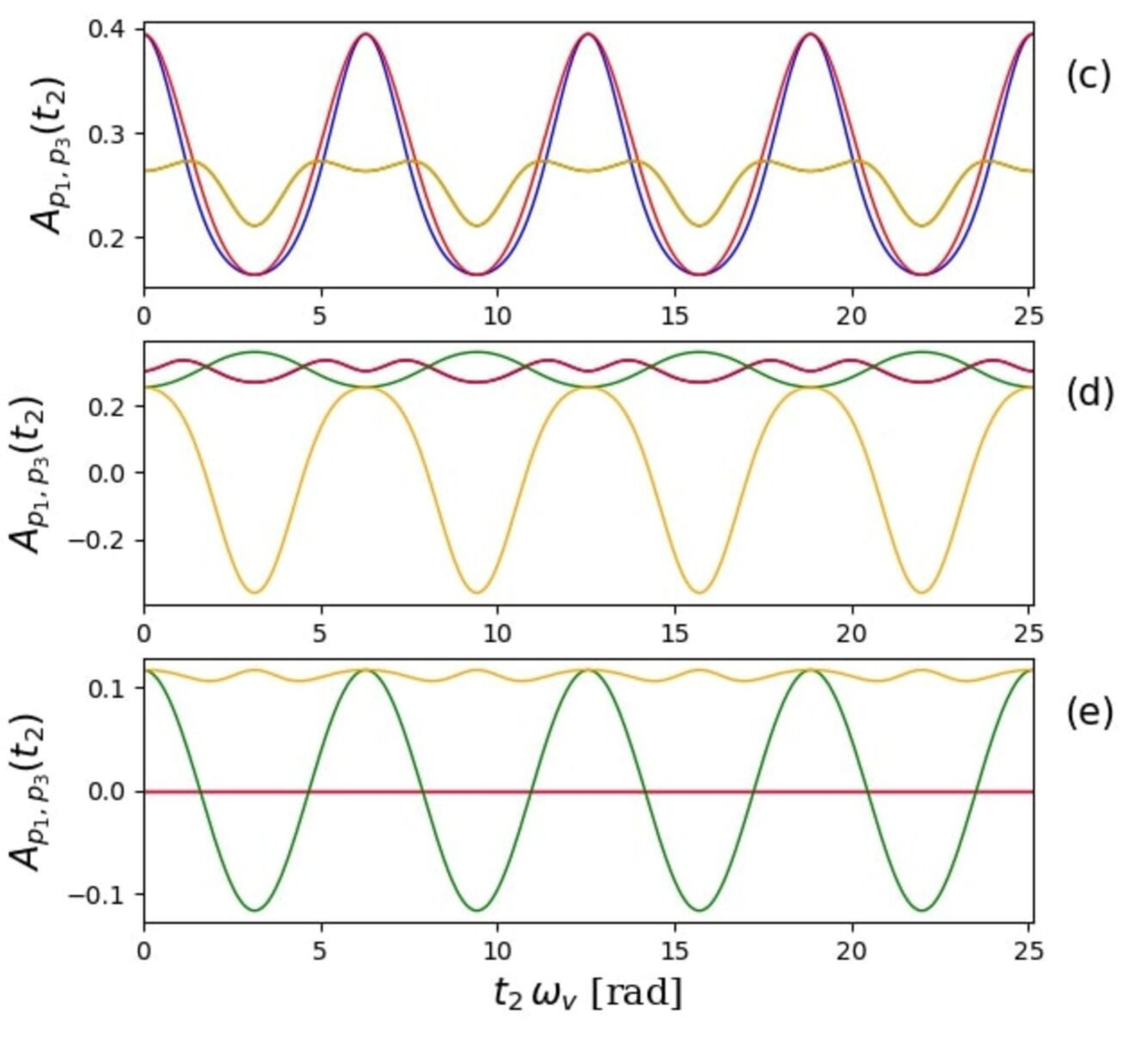}
\caption{Dependence on the waiting time $t_2$ of three representative peaks in the 2D map of a system characterized by a three-level $\Xi$ scheme (a). The panels (c-e) report the (real part of) the amplitudes for the peaks highlighted in panel (b). Different colors of the solid lines correspond to different contributions: excited state absorption, rephasing (blue) and non-rephasing (red); double quantum coherence, rephasing (green) and non-rephasing (orange). The values of the displacements are: $z_1=0.4$ and $z_2=-0.7$.}
\end{figure}

\subsection{Excited state absorption}

The third response function is related to excited-state absorption processes.
\paragraph{Rephasing contribution.}
As can be seen from Eq. (\ref{eq:re3jkl}), $s_{010}'=s_{001}'=s_{011}'=-1$ and $s_\chi'=+1$ in all other cases, while $h({\bf z})=z_j^2+z_k^2+z_{l}^2 -z_{l} (z_j+z_k)$.
The weight of the $q$-th order contribution is given by the sum
\begin{align}\label{eq:85}
    C_{p_1,p_2,p_3}^{(q)} & = \sum_{n_{110},n_{111}=0}^q
    \frac{(z_j z_k)^{n_{100}}}{n_{100}!}
    \frac{(z_{lj} z_{l k})^{n_{001}}}{n_{001}!}
    \frac{(z_k z_{kl})^{n_{010}}}{n_{010}!}\nonumber\\
    &
    \frac{(-z_j z_{kl})^{n_{110}}}{n_{110}!}
    \frac{( z_k z_{l j})^{n_{011}}}{n_{011}!}
    \frac{(-z_j z_{l j})^{n_{111}}}{n_{111}!},
\end{align}
where 
$n_{100} = p_1 - n_{110} - n_{111}$, 
$n_{001} = q - p_1 - p_2 + n_{110} + n_{111} $,
$n_{010} = -p_1 -2p_2+p_3+q+2n_{110}+n_{111}$, 
and
$n_{011} = -n_{110} - q + p_1 +p_2 -p_3 $.
The relevant contributions in the sum in Eq. (\ref{eq:85}) correspond to the values of the independent exponents $n_{110}$ and $n_{111}$ such that all the other ones, resulting from the above relations, are non-negative.

\subparagraph{Example.} We consider as an example the case of a three-level $\Xi$ system, for $j=k=1$ and $l=2$, where the zero-phonon peak corresponds to $(\omega_{1,0},\omega_{3,0})=[-\epsilon_1/\hbar,(\epsilon_2-\epsilon_1)/\hbar]$ (see Eq. \ref{eq:3el}). The phonon replicas are found at excitation and detection frequencies $\omega_{1,p_1}=-\epsilon_1/\hbar-p_1\omega_v$ and $\omega_{3,p_3}=(\epsilon_2-\epsilon_1)/\hbar-p_3\omega_v$, respectively. Their amplitude and phase, as a function of $t_2$, are given by:
\begin{align}
    A_{p_1,p_3}(t_2) & = e^{-h({\bf z})} \sum_{n_{011},n_{110},n_{111}=0}^\infty
    \frac{( z_1 z_{21})^{n_{110}+n_{011}}}{n_{110}!\,n_{011}!}
    \nonumber\\ &
    \frac{z_1^{2n_{100}}}{n_{100}!}
    \frac{z_{21}^{2n_{001}}}{n_{001}!}
    \sum_{p_2=-\infty}^{+\infty} e^{i\,p_2\,\omega_v \,t_2} 
    \frac{(-z_1 z_{21})^{n_{010}+n_{111}}}{n_{010}!\,n_{111}!},
\end{align}
where 
$n_{100} = p_1 - n_{110} - n_{111}$, 
$n_{001} = -p_3-n_{011} + n_{111} $,
$n_{010} = -p_2-n_{011}+n_{110}+n_{111}$, 
and $h({\bf z})=2z_1(z_1-z_2)+z_2^2$. The amplitude of the peak is thus identically zero for $p_1<0$.

\paragraph{Non-rephasing contribution.}
As results from Eq. (\ref{eq:re6jkl}), $s_{010}'=s_{011}'=+1$ and $s_\chi'=-1$ in all other cases, while $h({\bf z})=z_j^2+z_k^2+z_{l}^2 -z_{l} (z_j+z_k)$.
The weight of the $q$-th order contribution is given by the sum
\begin{align}\label{eq:86}
    C_{p_1,p_2,p_3}^{(q)} & = \sum_{n_{110},n_{111}=0}^q
    \frac{(z_j z_k)^{n_{100}}}{n_{100}!}
    \frac{(z_{l k} z_{l j})^{n_{001}}}{n_{001}!}
    \frac{(z_k z_{l j})^{n_{010}}}{n_{010}!}\nonumber\\
    &
    \frac{(z_j z_{jl})^{n_{110}}}{n_{110}!}
    \frac{(-z_k z_{l k})^{n_{011}}}{n_{011}!}
    \frac{(z_j z_{l k})^{n_{111}}}{n_{111}!},
\end{align}
where 
$n_{100} = - p_1 - n_{110} - n_{111}$, 
$n_{001} = q - p_1 - p_2 - n_{110} - n_{111} $,
$n_{010} = p_1 +2p_2-p_3-q+2n_{110}+n_{111}$, 
and
$n_{011} = -n_{110} + q - p_1 -p_2 +p_3 $.
The terms in Eq. (\ref{eq:86}) that matter correspond to the values of the independent exponents $n_{110}$ and $n_{111}$ such that all the other ones, resulting from the above relations, are non-negative.

\subparagraph{Example.} In the case of a three-level $\Xi$ system, for $j=k=1$ and $l=2$, the zero-phonon peak corresponds to $(\omega_{1,0},\omega_{3,0})=[\epsilon_1/\hbar,(\epsilon_2-\epsilon_1)/\hbar]$ (see Eq. \ref{eq:3el}). The phonon replicas are found at excitation and detection frequencies $\omega_{1,p_1}=\epsilon_1/\hbar-p_1\omega_v$ and $\omega_{3,p_3}=(\epsilon_2-\epsilon_1)/\hbar-p_3\omega_v$, respectively. Their amplitude and phase, as a function of $t_2$, are given by:
\begin{align}
    A_{p_1,p_3}(t_2) & = e^{-h({\bf z})} \sum_{n_{011},n_{110},n_{111}=0}^\infty
    \frac{(-z_1 z_{21})^{n_{110}+n_{011}}}{n_{110}!\,n_{011}!}
    \nonumber\\ &
    \frac{z_1^{2n_{100}}}{n_{100}!}
    \frac{z_{21}^{2n_{001}}}{n_{001}!}
    \sum_{p_2=-\infty}^{+\infty} e^{i\,p_2\,\omega_v \,t_2} 
    \frac{(z_1 z_{21})^{n_{010}+n_{111}}}{n_{010}!\,n_{111}!},
\end{align}
where 
$n_{100} = - p_1 - n_{110} - n_{111}$, 
$n_{001} = -p_3 + n_{011} - n_{111} $,
$n_{010} = p_2 - n_{011} + n_{110} + n_{111}$, 
and $h({\bf z})=2z_1(z_1-z_2)+z_2^2$. The amplitude of the peak is thus identically zero for $p_1>0$.

\subsection{Double quantum coherence}

Finally, we address the spectral contributions related to double quantum coherences.
\paragraph{Rephasing contribution.}
As results from Eq. (\ref{eq:re7jkl}), $s_{001}'=+1$ and $s_\chi'=-1$ in all other cases. 
The weight of the $q$-th order contribution is given by the sum
\begin{align}\label{eq:87}
    C_{p_1,p_2,p_3}^{(q)} & = \sum_{n_{110},n_{111}=0}^q
    \frac{(-z_j  z_{l j})^{n_{100}}}{n_{100}!}
    \frac{(z_k  z_{kl})^{n_{001}}}{n_{001}!}
    \frac{(z_k  z_{l j})^{n_{010}}}{n_{010}!}\nonumber\\
    &
    \frac{(z_j z_k)^{n_{110}}}{n_{110}!}
    \frac{( z_{l j}  z_{l k})^{n_{011}}}{n_{011}!}
    \frac{(z_j  z_{l k})^{n_{111}}}{n_{111}!},
\end{align}
where 
$n_{100} = -p_1 - n_{110} - n_{111}$, 
$n_{001} = - q + p_1 + p_2 + n_{110} + n_{111} $,
$n_{010} = -p_1 -2p_2+p_3+q-2n_{110}-n_{111}$, 
and
$n_{011} = n_{110} - q + p_1 + p_2 - p_3 $.
The relevant terms in Eq. (\ref{eq:87}) correspond to the values of the independent exponents $n_{110}$ and $n_{111}$ such that all the other ones, resulting from the above relations, are non-negative.
The exponent in the prefactor is given by $h({\bf z})=z_j^2+z_k^2+z_{l}^2 -z_{l} (z_j+z_k)$.

\subparagraph{Example.} In the case of a three-level $\Xi$ system, for $j=k=1$ and $l=2$, the zero-phonon peak corresponds to $(\omega_{1,0},\omega_{3,0})=[\epsilon_1/\hbar,(\epsilon_2-\epsilon_1)/\hbar]$ (see Eq. \ref{eq:7el}). The phonon replicas are found at excitation and detection frequencies $\omega_{1,p_1}=\epsilon_1/\hbar-p_1\omega_v$ and $\omega_{3,p_3}=(\epsilon_2-\epsilon_1)/\hbar-p_3\omega_v$, respectively. Their amplitude and phase, as a function of $t_2$, are given by:
\begin{align}
    A_{p_1,p_3}(t_2) & = e^{-h({\bf z})} \sum_{n_{011},n_{110},n_{111}=0}^\infty
    \frac{(-z_1 z_{21})^{n_{100}+n_{001}}}{n_{100}!\,n_{001}!}
    \nonumber\\ &
    \frac{z_1^{2n_{110}}}{n_{110}!}
    \frac{z_{21}^{2n_{011}}}{n_{011}!}
    \sum_{p_2=-\infty}^{+\infty} e^{i\,p_2\,\omega_v \,t_2} 
    \frac{(z_1 z_{21})^{n_{010}+n_{111}}}{n_{010}!\,n_{111}!},
\end{align}
where 
$n_{100} = - p_1 - n_{110} - n_{111}$, 
$n_{001} = p_3 + n_{011} + n_{111} $,
$n_{010} = - p_2 - n_{011} - n_{110} - n_{111}$, 
and $h({\bf z})=2z_1(z_1-z_2)+z_2^2$. The amplitude of the peak is thus identically zero for $p_1>0$.

\paragraph{Non-rephasing contribution.}
As can be seen from Eq. (\ref{eq:re8jkl}), $s_\chi'=-1$ for all the cases. 
The weight of the $q$-th order contribution is given by the sum
\begin{align}\label{eq:88}
    C_{p_1,p_2,p_3}^{(q)} & = \sum_{n_{110},n_{111}=0}^q
    \frac{(z_j  z_{jl})^{n_{100}}}{n_{100}!}
    \frac{(z_k  z_{kl})^{n_{001}}}{n_{001}!}
    \frac{( z_{l k}  z_{l j})^{n_{010}}}{n_{010}!}\nonumber\\
    &
    \frac{(z_j  z_{l k})^{n_{110}}}{n_{110}!}
    \frac{(z_k  z_{l j})^{n_{011}}}{n_{011}!}
    \frac{(z_j z_k)^{n_{111}}}{n_{111}!},
\end{align}
where 
$n_{100} = -p_1 - n_{110} - n_{111}$, 
$n_{001} = q - p_1 - p_2 - n_{110} - n_{111} $,
$n_{010} = -p_1 -2p_2+p_3+q-2n_{110}-n_{111}$, 
and
$n_{011} = n_{110} - q + p_1 +p_2 -p_3 $.
The actual contributions in the sum in Eq. (\ref{eq:88}) correspond to the values of the independent exponents $n_{110}$ and $n_{111}$ such that all the other ones, resulting from the above relations, are non-negative.
The exponent in the prefactor is given by $h({\bf z})=z_j^2+z_k^2+z_{l}^2 -z_{l} (z_j+z_k)$.

\subparagraph{Example.} In the case of a three-level $\Xi$ system, for $j=k=1$ and $l=2$, the zero-phonon peak corresponds to $(\omega_{1,0},\omega_{3,0})=(\epsilon_1/\hbar,\epsilon_1/\hbar)$ (see Eq. \ref{eq:8el}). The phonon replicas are found at excitation and detection frequencies $\omega_{1,p_1}=\epsilon_1/\hbar-p_1\omega_v$ and $\omega_{3,p_3}=\epsilon_1/\hbar-p_3\omega_v$, respectively. Their amplitude and phase, as a function of $t_2$, are given by:
\begin{align}
    A_{p_1,p_3}(t_2) & = e^{-h({\bf z})} \sum_{n_{011},n_{110},n_{111}=0}^\infty
    \frac{(z_1 z_{21})^{n_{110}+n_{011}}}{n_{110}!\,n_{011}!}
    \nonumber\\ &
    \frac{z_1^{2n_{111}}}{n_{111}!}
    \frac{(-z_1 z_{21})^{n_{100}+n_{001}}}{n_{100}!\,n_{001}!}
    \sum_{p_2=-\infty}^{+\infty} e^{i\,p_2 \,\omega_v\,t_2} 
    \frac{z_{21}^{2n_{010}}}{n_{010}!},
\end{align}
where 
$n_{100} = - p_1 - n_{110} - n_{111}$, 
$n_{001} = -p_3 - n_{011} - n_{111} $,
$n_{010} = - p_2 - n_{011} - n_{110} - n_{111}$, 
and $h({\bf z})=2z_1(z_1-z_2)+z_2^2$. The amplitude of the peak is thus identically zero for $p_1>0$ or $p_3>0$.

\section{Higher-order and multi-mode generalizations\label{sec:gen}}

Hereafter, we derive the expressions of the response functions corresponding to $M$-th order in the interaction with the field, with arbitrary $M$. The generalization to the case of multiple vibrational modes is also briefly discussed.

\subsection{Higher-order nonlinear contributions}

So far, third-order response functions have been derived from the expressions of the time-dependent vibrational states for the ket and the bra. This approach allows one to develop a clear physical picture, where the vibrational component of the response function corresponds to the overlap between the coherent states that correspond to the left and right sides of the Feynman diagrams. In order to derive higher-order response functions, it is however convenient to follow a slightly different approach, which simplifies the calculations (Appendix \ref{app:C}). 

The final result, consisting in the expression of the vibrational component of the response function in terms of the waiting times and of the displacements $z_k$, reads:
\begin{align}\label{eq:65}
   R^{(v,M)} & =  \exp [f(t_1,\dots,t_M)] = \nonumber\\
   & \exp\!\left[\sum_{k=1}^M\!\sum_{l=1}^{M-k+1}\!\! z_{j_{l-1},j_l} z_{j_{l+k-1},j_{l+k}}\! \left(\! 1\!-\!\prod_{p=l}^{l+k-1} v_p\!\right)\!\right].
\end{align}
Each the $v_p$, as well as the above products of consecutive $v_p$ functions, take the form $\exp (i\,s\, \omega_v\,\sum_j t_j) $, where the sum in the exponent is performed on variable numbers of consecutive waiting times, each one corresponding to a time interval between consecutive interactions with the field on the left ($s=-1$) or on the right ($s=+1$) side of the diagram. More specifically, the sums in the exponents that define $v_1$ ($v_M$) include all the waiting times between the first and second interactions of the bra (ket) with the field, those in $v_2$ ($v_{M-1}$) include the times between the second and third interactions; and so on.

From the above equation it follows that the response function $R^{(v,M)}$ corresponding to a given pathway can in practice be directly derived from the Feynman diagram, by adopting the following recipe for composing the exponent $f(t_1,\dots,t_M)$. In particular, this includes:
\begin{itemize}
    \item $M(M+1)/2$ terms $\chi_{mn}\equiv 1-e^{-i\,\omega_v \sum_{j=m}^n t_j}$ or $\chi_{mn}^*$, where the sum includes from 1 to $M$ terms, and runs over all combinations of consecutive waiting times. The sign in the exponent is assigned as follows: if the arrows at the beginning and at the end of the considered time interval are both on the right (left) side, then the sign is positive and the function is $\chi_{mn}^*$ (the sign is negative and the function is $\chi_{mn}$); if the two arrows are on opposite sides, then the sign is positive or negative, depending on whether the earliest interaction is on the right or on the left.
    \item Each of these oscillating terms is multiplied by $ z_{jk} z_{k'j'} $, where $j$ and $k$ ($j'$ and $k'$) specify the electronic states before and after the first (second) delimiting arrow. Here we refer to an order that goes from the bra at the bottom right corner to the ket at the bottom left corner of the double-sided Feynman diagrams, proceeding counterclockwise.
\end{itemize}

\begin{figure}[h]
\centering
\includegraphics[width=0.4\textwidth]{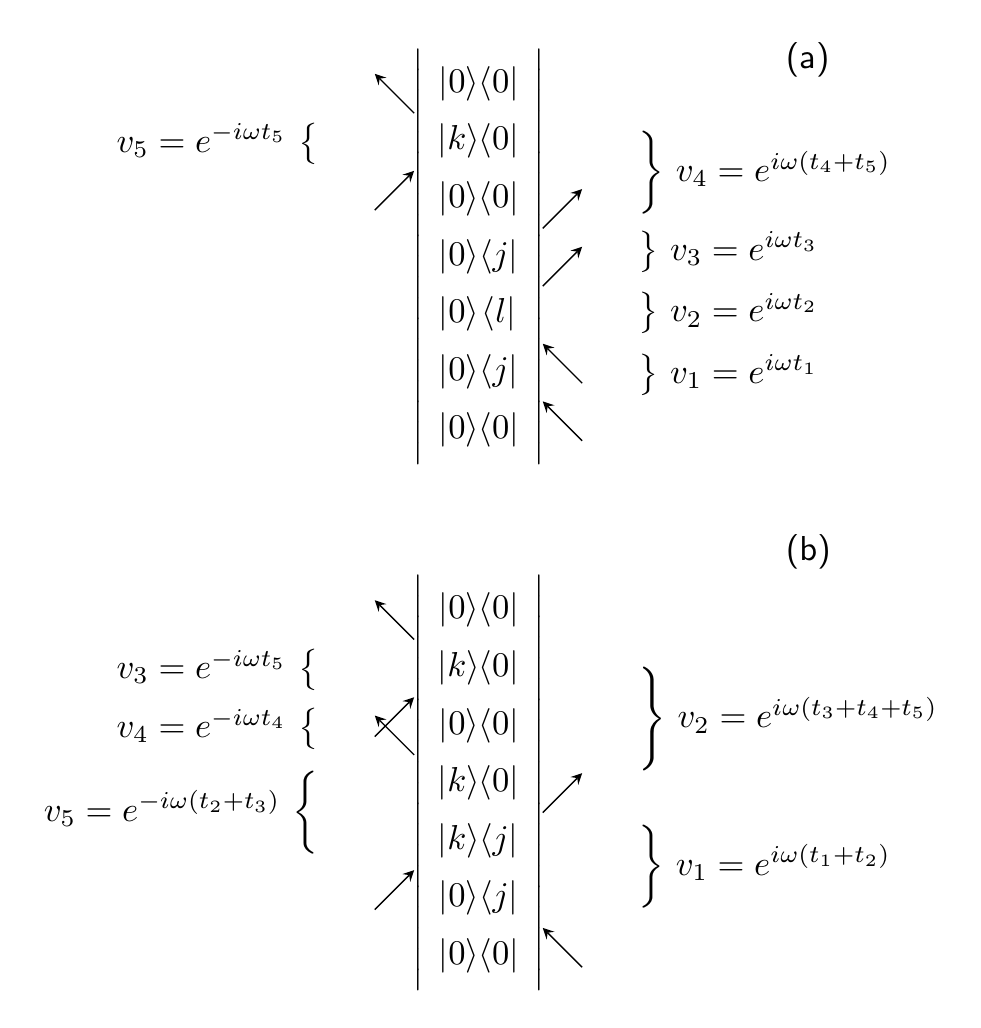}
\caption{Two representative examples of a double-sided Feynman diagram corresponding to a fifth-order ($M=5$) response function, to which we refer in the paragraphs entitled First example (a) and Second Example (b).\label{fig:n5}}
\end{figure}

{\y It can be readily verified} that, by applying the above recipe to the third-order response functions, one recovers all the expressions derived in Section III. In order to further illustrate the recipe, we apply it to two representative examples of a fifth-order response function. The identification of the oscillating terms in $f$ with products of the $v_k$ operators will be included in order to clear the connection with the formal derivation of the response functions, given in Appendix \ref{app:C}, but is not necessary in order to apply the recipe and can be disregarded by the uninterested reader.

\paragraph{First example.} 
The considered pathway is defined by the Feynman diagram in Fig. \ref{fig:n5}(a). Following step by step the above procedure, we derive the following composition of the function $f$ in the exponent [Eq. (\ref{eq:65})]:
\begin{itemize}
    \item The oscillating terms $1-\chi_{mn}$, which result from the products of the $v_k$ functions, can be directly identified with: $v_1=e^{i\omega_v t_1}$, $v_2=e^{i\omega_v t_2}$, $v_3=e^{i\omega_v t_3}$, $v_4v_5=e^{i\omega_v t_4}$, $v_5=e^{-i\omega_v t_5}$ (one waiting time); $v_1v_2=e^{i\omega_v t_{12}}$, $v_2v_3=e^{i\omega_v t_{23}}$, $v_3v_4v_5=e^{i\omega_v t_{34}}$, $v_4=e^{i\omega_v t_{45}}$ (two waiting times); $v_1v_2v_3=e^{i\omega_v t_{13}}$, $v_2v_3v_4v_5=e^{i\omega_v t_{24}}$, $v_3v_4=e^{i\omega_v t_{35}}$ (three waiting times); $v_1v_2v_3v_4v_5=e^{i\omega_v t_{14}}$, $v_2v_3v_4=e^{i\omega_v t_{25}}$ (four waiting times); $v_1v_2v_3v_4=e^{i\omega_v t_{15}}$ (five waiting times). Here, we have adopted the convention: $t_{ij}\equiv t_i+t_{i+1}+\dots+t_{j-1}+t_j$.
    \item The functions $\chi_{mn}$ are multiplied respectively by the prefactors: $z_{0j} z_{jl}$, $z_{jl} z_{lj}$, $z_{lj} z_{j0}$, $z_{j0} z_{\y 0k}$, $z_{0k} z_{k0}$; $z_{0j} z_{lj}$, $z_{jl} z_{j0}$,  $z_{lj} z_{k0}$, $z_{j0} z_{0k}$; $z_{0j} z_{j0}$, $z_{jl} z_{k0}$, $z_{lj} z_{0k}$; $z_{0j} z_{k0}$, $z_{jl} z_{0k}$; $z_{0j} z_{0k}$.
\end{itemize}
\begin{figure}[h]
\centering
\includegraphics[width=0.35\textwidth]{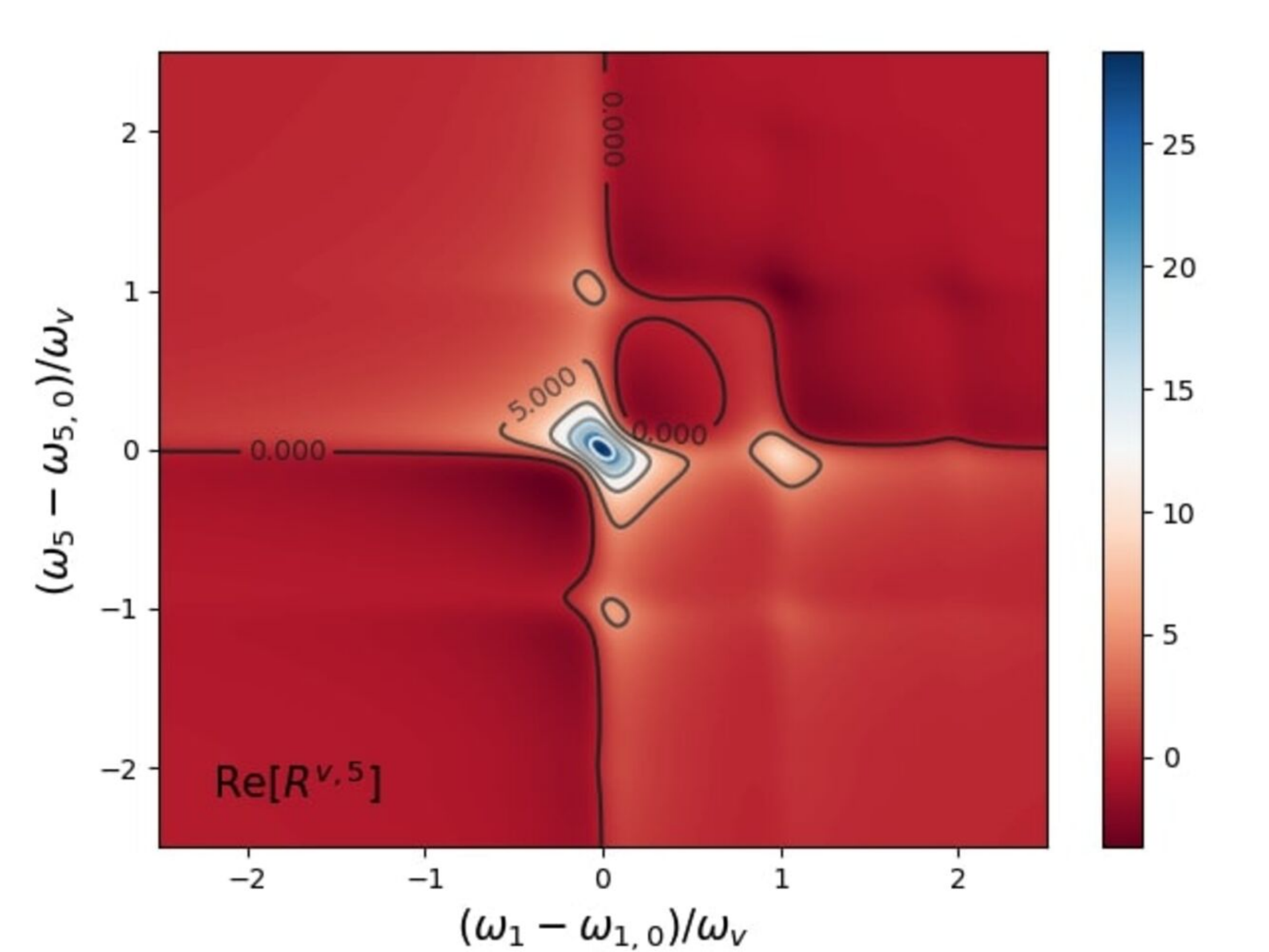}
\includegraphics[width=0.35\textwidth]{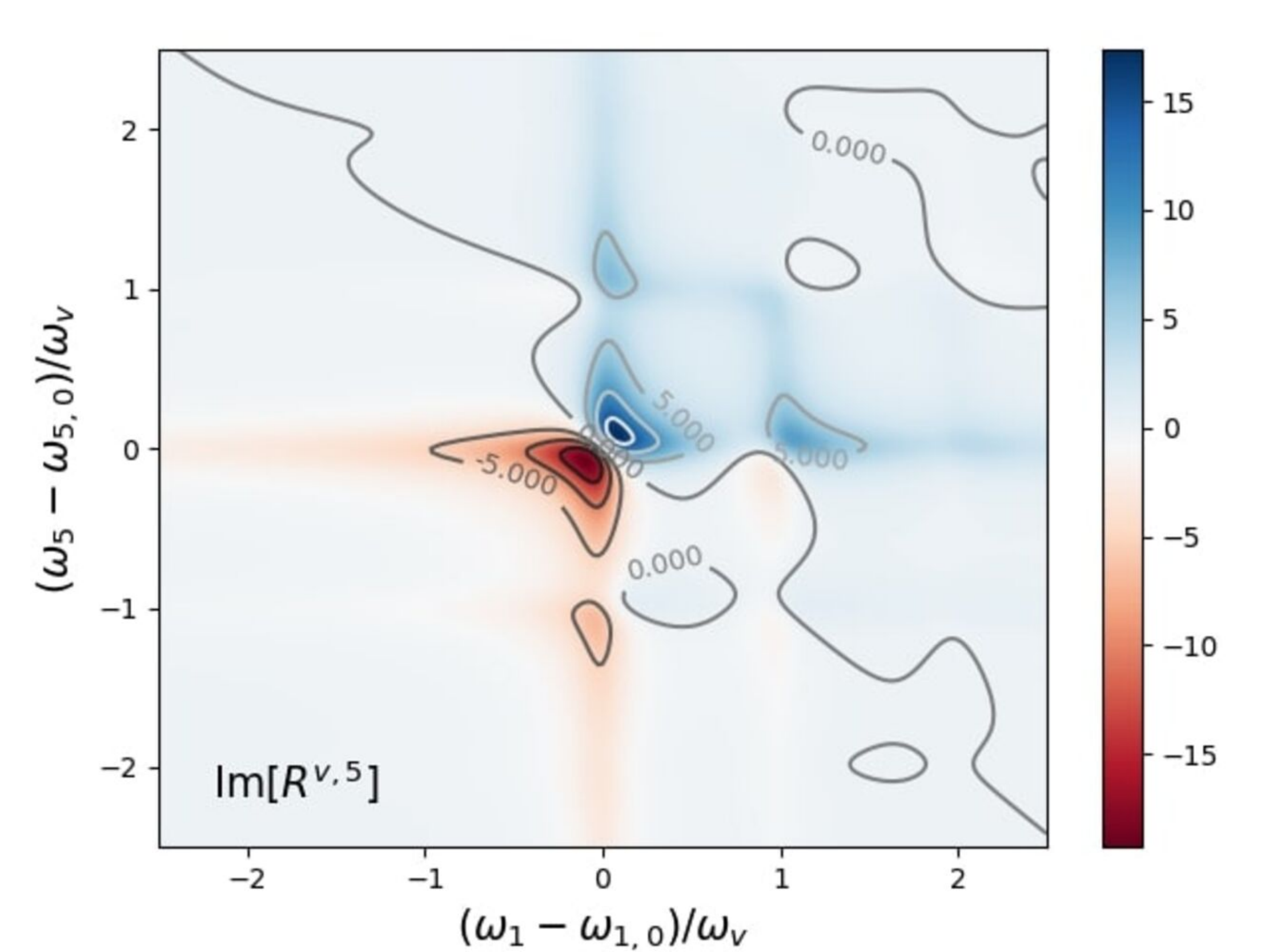}
\includegraphics[width=0.35\textwidth]{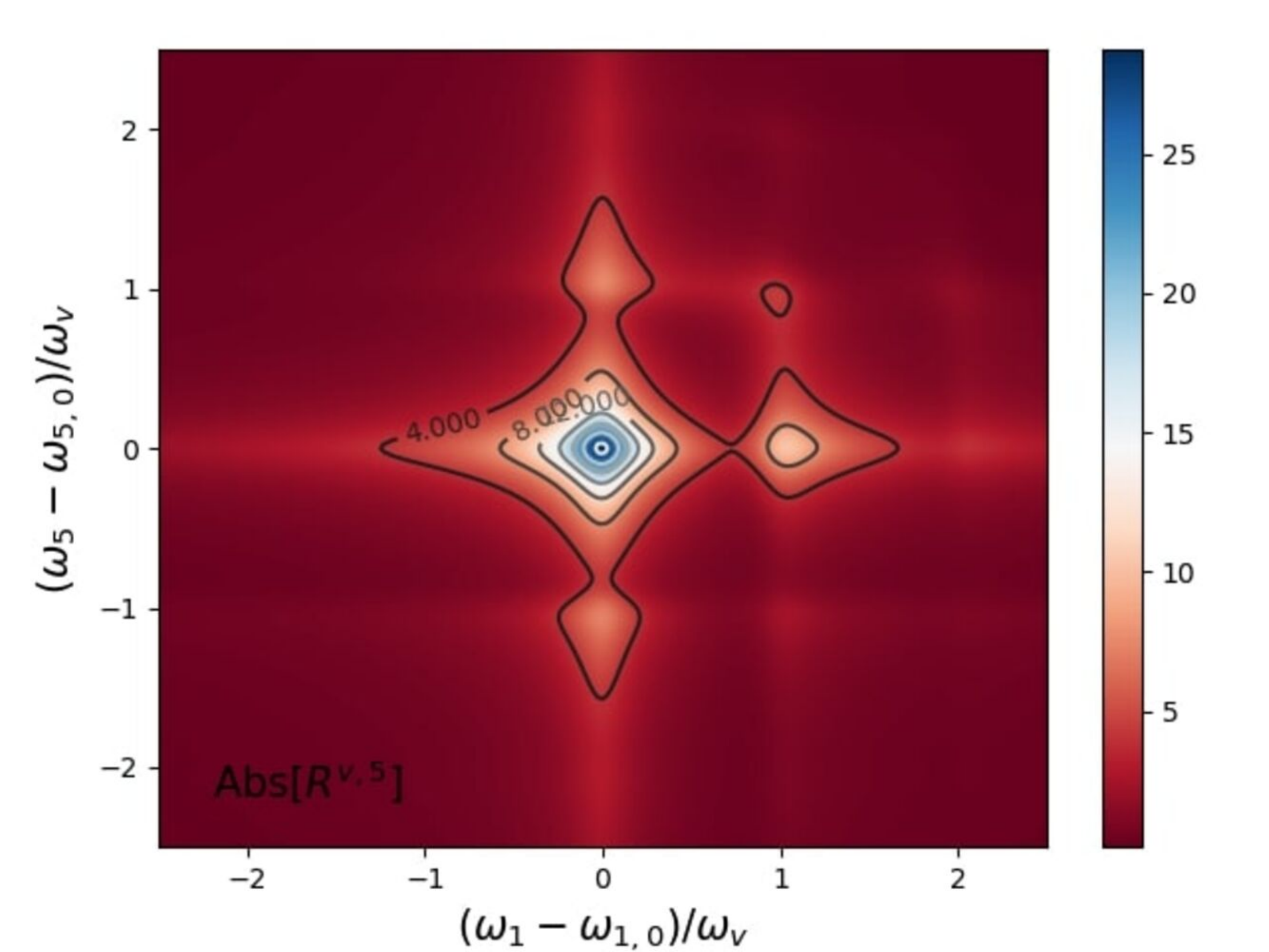}
\caption{Fifth-order response function $R^{(v,5)}(\omega_1,\omega_5)$ of a system with a three-level $\Xi$ scheme, corresponding to the Fourier transform of the $R^{(v,5)}(t_1,t_5)$ reported in Eq. \ref{eq:rf1}. The values of the displacements are: $z_1=0.4$ and $z_2=-0.7$, the lines are broadened by  assuming for the electronic coherences a dephasing rate $\gamma/\omega_v=0.15$.}\label{fig:2dm1}
\end{figure}

\begin{figure}[h]
\centering
\includegraphics[width=0.35\textwidth]{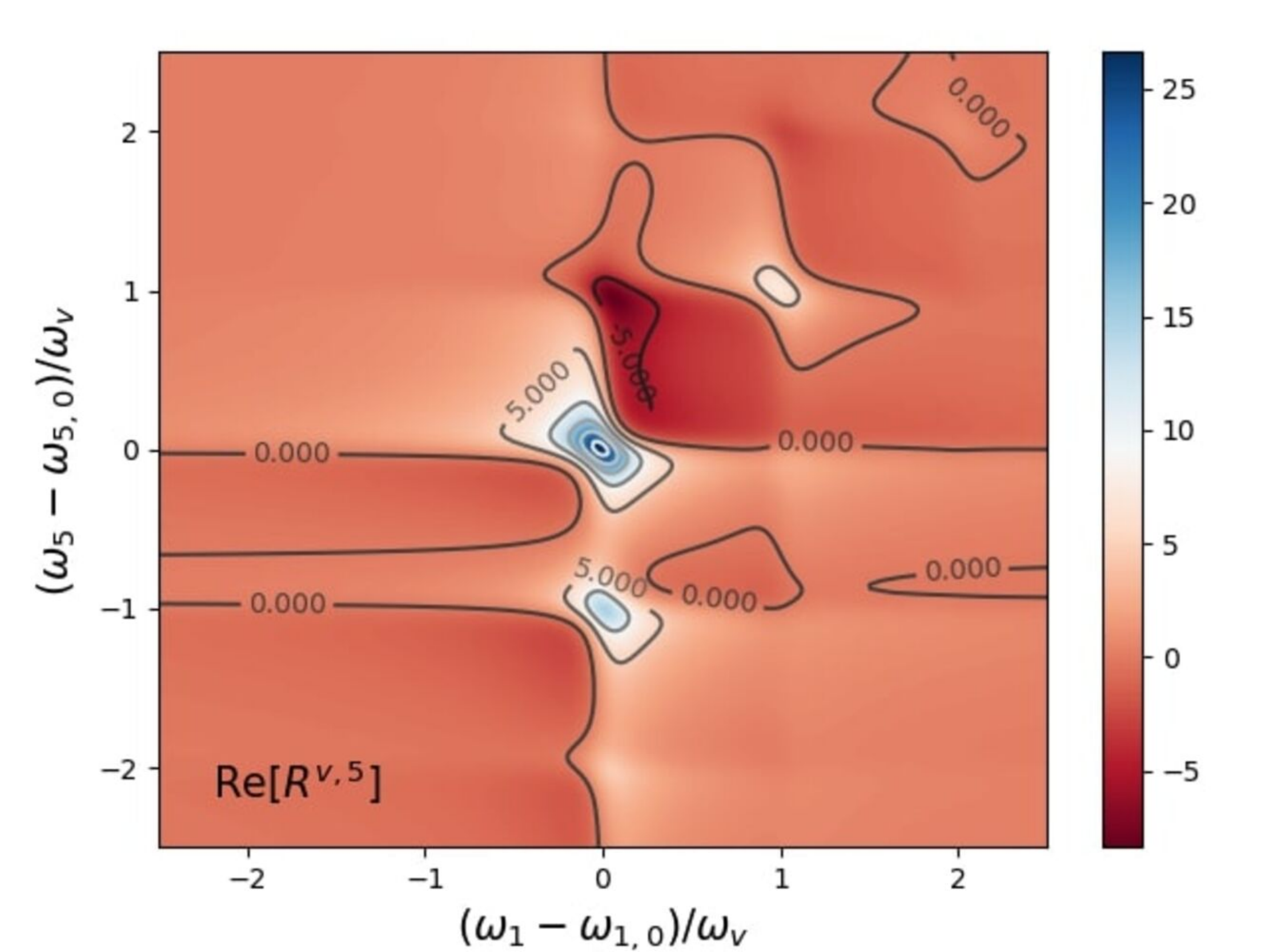}
\includegraphics[width=0.35\textwidth]{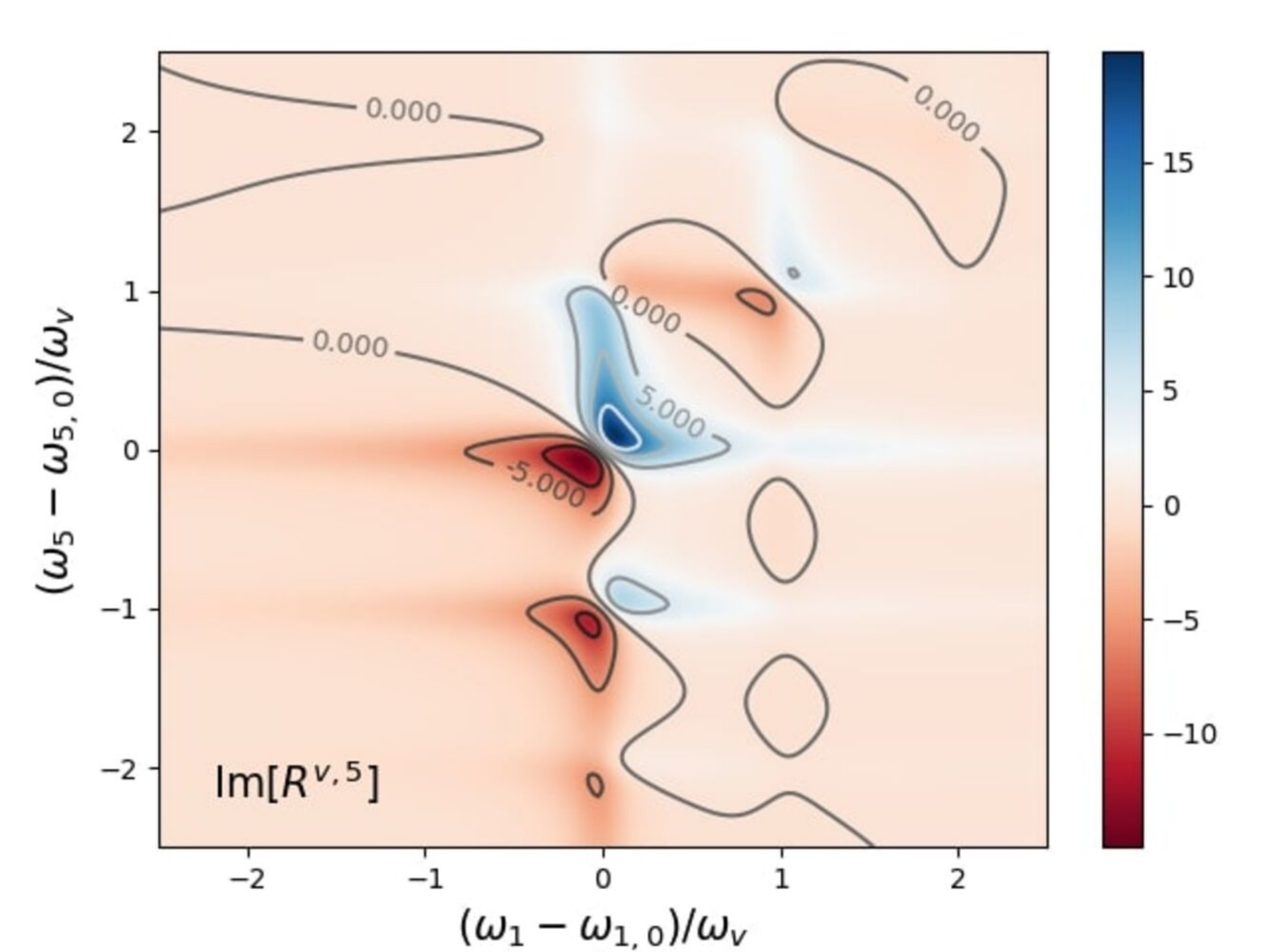}
\includegraphics[width=0.35\textwidth]{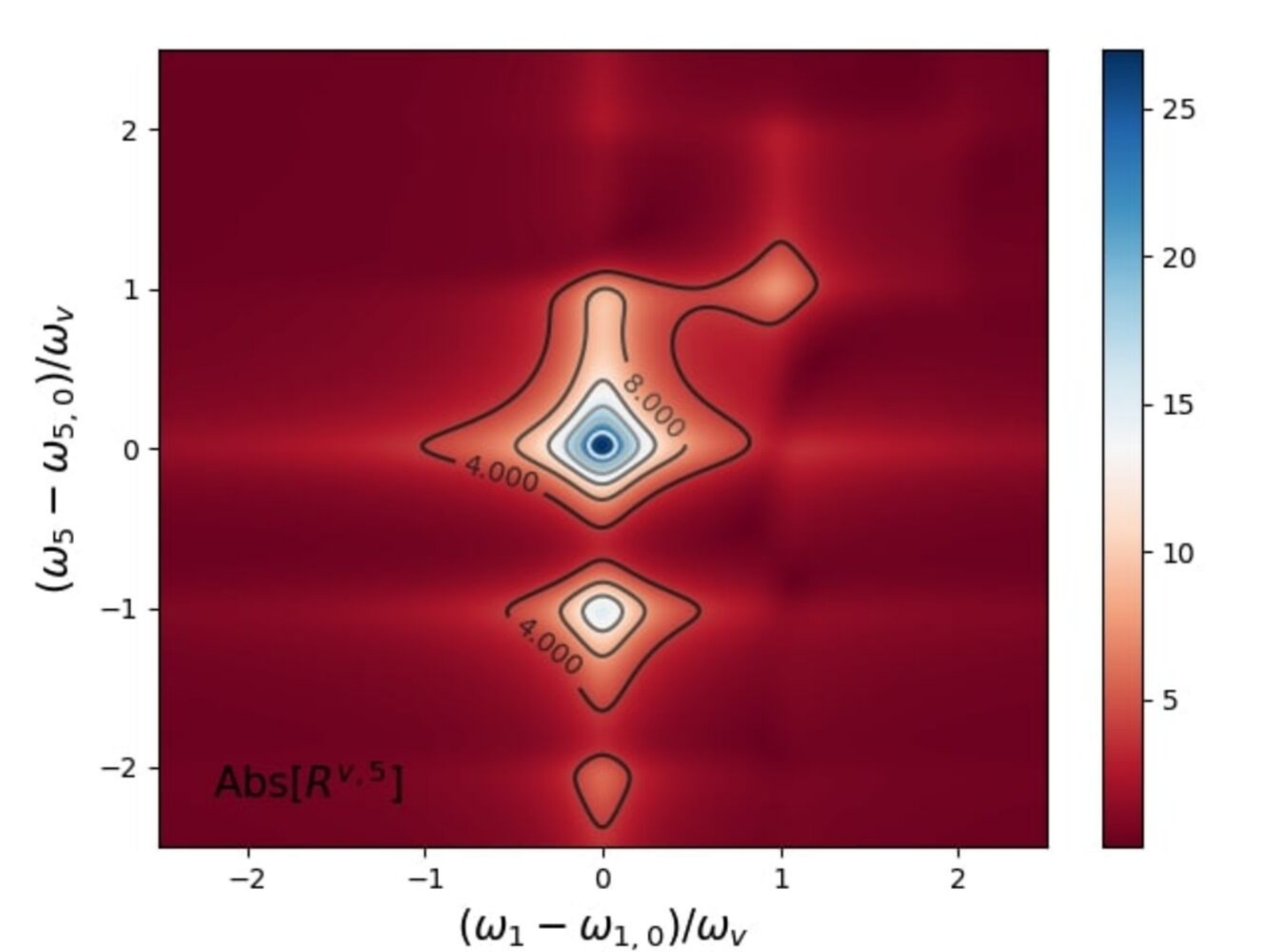}
\caption{Fifth-order response function $R^{(v,5)}(\omega_1,\omega_5)$ of a system with a three-level $V$ scheme, corresponding to the Fourier transform of the $R^{(v,5)}(t_1,t_5)$ reported in Eq. \ref{eq:rf2}. The values of the displacements are: $z_1=0.4$ and $z_2=-0.7$, the lines are broadened by  assuming for the electronic coherences a dephasing rate $\gamma/\omega_v=0.15$.}\label{fig:2dm2}
\end{figure}

\paragraph{Second example.} 
The pathway is here defined by the Feynman diagram in Fig. \ref{fig:n5}(b). The function $f$ in the exponent [Eq. (\ref{eq:65})] is composed as follows:
\begin{itemize}
    \item The oscillating terms $1-\chi_{mn}$, which result from the products of the $v_k$ functions, can be directly identified with: $v_1v_2v_3v_4v_5=e^{i\omega_v t_1}$, $v_2v_3v_4v_5=e^{-i\omega_v t_2}$, $v_2v_3v_4=e^{i\omega_v t_3}$, $v_4=e^{-i\omega_v t_4}$, $v_3=e^{-i\omega_v t_5}$ (one waiting time); $v_1=e^{i\omega_v t_{12}}$, $v_5=e^{-i\omega_v t_{23}}$, $v_2v_3=e^{i\omega_v t_{34}}$, $v_3v_4=e^{-i\omega_v t_{45}}$ (two waiting times); $v_1v_2v_3v_4=e^{i\omega_v t_{13}}$, $v_4v_5=e^{-i\omega_v t_{24}}$, $v_2=e^{i\omega_v t_{35}}$ (three waiting times); $v_1v_2v_3=e^{i\omega_v t_{14}}$, $v_3v_4v_5=e^{-i\omega_v t_{25}}$ (four waiting times); $v_1v_2=e^{i\omega_v t_{15}}$ (five waiting times).     
    \item The functions $\chi_{mn}$ are multiplied respectively by the prefactors: $z_{0j} z_{k0}$, $z_{j0} z_{k0}$, $z_{j0} z_{0k}$, $z_{k0} z_{0k}$, $z_{0k} z_{k0}$; $z_{0j} z_{j0}$, $z_{0k} z_{k0}$, $z_{j0} z_{k0}$, $z_{0k} z_{0k}$; $z_{0j} z_{0k}$, $z_{k0} z_{k0}$, $z_{j0} z_{0k}$; $z_{0j} z_{k0}$, $z_{0k} z_{k0}$; $z_{0j} z_{0k}$. 
    \item Applying this result to the case of a three-level $\Xi$ system {\y (which implies that $j=k=1$)}, for $t_2=t_3=t_4=0$, one obtains for the response function the expression
    \begin{align}\label{eq:rf1}
        R^{(v,5)}&=z_1^2 [2e^{i\omega_v t_1}+e^{i\omega_v t_5}-e^{i\omega_v (t_1+t_5)}\nonumber\\ &+e^{-i\omega_v t_5}-3].
    \end{align}
    Its Fourier transform with respect to $t_1$ and $t_5$ is plotted in Fig. \ref{fig:2dm1}.
    \item Applying this result to the case of a three-level $V$ system, for {\y $j=1$, $k=2$ and} $t_2=t_3=t_4=0$, one obtains for the response function the expression
    \begin{align}\label{eq:rf2}
    R^{(v,5)}&= z_1(z_1\!+\!z_2) e^{i\omega_v t_1} \!-\!(z_1^2\!+\!z_2^2\!+\!z_1z_2)\nonumber\\&\!+\!z_1 z_2 [e^{i\omega_v t_5}\!-\!e^{i\omega_v (t_1+t_5)}] \!+\! z_2^2  e^{-i\omega_v t_5} . 
    \end{align}
    Its Fourier transform with respect to $t_1$ and $t_5$ is plotted in Fig. \ref{fig:2dm2}. 
\end{itemize}

\subsection{Multimode case}

The generalization of the above results to the case of $B$ vibrational modes is straightforward. The Hamiltonian becomes 
\begin{eqnarray}\label{qe:mmdho}
    H &=& \sum_{j=0}^{N-1} |{\y j} \rangle\langle j| \otimes \left[ \epsilon_j + \sum_{\xi=1}^B\hbar\omega_{v,\xi} (a_\xi^\dagger + z_{j,\xi})(a_\xi + z_{j,\xi})\right] \nonumber\\
    &\equiv& \sum_{j=0}^{N-1} |{\y j} \rangle\langle j| \otimes \left( \epsilon_j + \sum_{\xi=1}^B H_{v,\xi,j}\right) .
\end{eqnarray}
The vibrational component of the response function has to be replaced by a product of terms such as the ones derived in the previous Sections, with mode-dependent frequencies $\omega_{v,\xi}$ and displacements $z_{j,\xi}$. 
The overall response functions, including electronic and vibrational degrees of freedom, are thus given by 
\begin{align}\label{eq:mmode}
    R_{p}^{(M)} &= \sum_{j_1,\dots,j_M} R^{(e,M)}_{p,j_1,\dots,j_M} \prod_{\xi=1}^B R^{(v,M,\xi)}_{p,j_1,\dots,j_M} ,
\end{align}
where $p$ specifies the kind of response function (in analogy to the classification in ground state bleaching, stimulated emission, excited state absorption, and double quantum coherence that has been considered for the case $M=3$) and $j_1,\dots,j_M$ are the involved electronic states, which specify the pathway. 

\section{Finite temperature case and application to the simulation \\ of a phonon bath}

So far, we have considered the zero-temperature limit, corresponding to vibrational modes initialized in the ground state of the undisplaced oscillator. In the present Section, we show how the response functions generalize to the case of an arbitrary initial state $|\alpha_0\rangle$. From the resulting expressions, one can in principle derive the response function for arbitrary initial states of the vibrational modes, by expressing the state of interest as a combination of coherent states, through the coherent state representation\cite{Scully97a}. This possibility will be exploited to derive the case of a thermal state.  Applying such result to the case of a phonon bath, we derive the expression of the line shape functions. The derivations of the following results are provided in the Appendices \ref{app:D}-\ref{app:Z}.

\subsection{Initialization to coherent and thermal states}

In order to derive the effect of such initialization, we refer to the expression of a response function in terms of an overlap between the vibrational states of the ket and of the bra:
\begin{equation}
    R^{(v,M)} = \langle \alpha_{M,bra} | \alpha_{M,ket} \rangle\, e^{i(a_{ket}-a_{bra})} \equiv e^{r} e^{i \varphi}.
\end{equation}
With respect to the $\alpha_0=0$ case, the final coherent states are defined by complex numbers $ \alpha_{M,ket} $ and $\alpha_{M,bra}$ that include the same, additional term $\beta\equiv\alpha_0 e^{-i\omega_v (t_1+\dots+t_M)}$. As a result, the modulus of the response function, given by an exponential function of $r=-\frac{1}{2}|\alpha_{M,ket}-\alpha_{M,bra}|^2$
is left unchanged. Therefore, the dependence of $R^{(v,M)}$ on $\alpha_0$ only concerns phase $\varphi$. 
In particular, the relation between the response functions in the zero-temperature limit ($\alpha_0=0$) considered so far and those at arbitrary $\alpha_0$ can be written in the compact form
\begin{align}
    R^{(v,M)}_{\alpha_0} & = e^{i\Delta\varphi}\,R^{(v,M)}_{\alpha_0=0} \nonumber\\
    & = \exp\{{\rm Re}(f)+i[{\rm Im} (f)+\Delta\varphi]\} ,
\end{align}
which reduces to $R^{(v,M)}=e^f$ [Eq. (\ref{eq:65})] for the standard initialization $\alpha_0=0$. The initial-state dependent change in the phase reads 
\begin{align}
    \Delta\varphi & = 2\sum_{j=1}^M (z_{b_j}-z_{k_j}) {\rm Im}[\alpha^*_0(e^{i\omega_v t_j}-1)\, e^{i\omega_v\sum_{k=1}^{j-1}t_k}] , 
\end{align}
being $|k_j\rangle$ and $|b_j\rangle$ are the electronic ket and bra states during the $j$-th waiting time.

Knowing the response function for any initial coherent states allows in principle to derive their expression for arbitrary initial vibrational states, passing through their coherent state representation\cite{Scully97a}. As a representative example, one can consider the case of thermal states, corresponding to $P(\alpha_0,\alpha_0^*) = \frac{1}{\pi\langle n \rangle} e^{-|\alpha_0|^2/\langle n \rangle} $. After averaging $R^{(v,M)}_{\alpha_0}$ in the phase space with such function, one obtains
\begin{align}\label{eq:thermal}
    R^{(v,M)}_T &= \int\,d^2\alpha\, \frac{R^{(v,M)}_{\alpha_0}}{\pi\langle n \rangle} e^{-|\alpha_0|^2/\langle n \rangle} \nonumber\\ 
    &= \exp[\coth(\hbar\omega_v/2k_BT){\rm Re}(f)+i{\rm Im} (f)],
\end{align}
which reduces to $R^{(v,M)}=e^f$ [Eq. (\ref{eq:65})] in the zero-temperature limit. In other words, the finite temperature response function is obtained from that at zero-temperature, derived in the previous Section, simply by multiplying by a factor $\coth(\hbar\omega_v/2k_BT)$ the real part in the exponent, while leaving the imaginary part unaffected. This {\y relation}, already known for the third-order response function of a two-level system\cite{Mukamel95a}, thus {\y has a wider validity}.

\subsection{Phonon bath and line shape function}

{\y A bath formed by a quasi-continuum of independent harmonic oscillators represents the prototypical model of environment, responsible for the decay of the oscillating features in multidimensional coherent spectroscopy and for the specific features of the observed line shapes \cite{Mukamel95a,Abramavicius2009a}. Here, the results obtained for thermal states [Eq. (\ref{eq:thermal})] and for the multimode case [Eq. (\ref{eq:mmode})] are combined together, in order to derive the expression of the line shape function for a generic multilevel system.}

{\y In line with the approach adopted so far, the $B$ harmonic oscillators that form the bath are assumed to be linearly coupled to the system [Eq. (\ref{qe:mmdho})].} In the limit where the discrete set ($\xi$) of modes is replaced by a continuum, the effect of the bath on the system can be fully characterized in terms of the density of the displacements as a function of the mode frequency $\omega$: $z_{jk,\xi}\, z_{j'k',\xi} \rightarrow s_{jk,j'k'}(\omega)\,d\omega$, along the lines of what has been done in the case of a two-level system \cite{Butkus2012b}. 
The resulting response function is obtained from that of the single mode case by replacing each function $z_{jk}\,z_{j'k'}\chi_{mn}=z_{jk}\,z_{j'k'}[1-\exp(i\omega_v t_{mn})]$ appearing in the exponent $f$ [Eq. \ref{eq:65}] with 
\begin{align}\label{eq:gf}
    g_{jk,k'j'} (t_{mn}) &\equiv \int\, d\omega\,s_{jkk'j'}(\omega)\, \{ \coth (\hbar\omega/2k_BT) \nonumber\\ & [1-\cos (\omega t_{mn})] + i \sin (\omega t_{mn})\} .
\end{align}
Further details on the derivation are provided in Appendix \ref{app:Z}.

Hereafter, we apply the above result to the case of the third-order response functions of multilevel systems (arbitrary $N$). The exponents of such functions read:
\begin{align}\label{eq:x1}
    f_{T;1,jk} &= g^*_{0j,k0} (t_1) + g_{j0,k0} (t_2) + g^*_{0j,k0} (t_3)  \nonumber\\ & + g^*_{0j,j0} (t_{12}) + g_{k0,0k} (t_{23}) + g^*_{0j,0k} (t_{13}) \\
    f_{T;2,jk} &= g^*_{0j,j0} (t_1) + g^*_{j0,k0} (t_2) + g_{0k,k0} (t_3)  \nonumber\\ &  + g^*_{0j,k0} (t_{12}) + g^*_{j0,0k} (t_{23}) + g^*_{0j,0k} (t_{13}) \\
    f_{T;3,jkl} &= g^*_{0j,k0} (t_1) + g_{k0,lk} (t_2) + g_{lk,jl} (t_3)  \nonumber\\ &  + g^*_{j0,kl} (t_{12}) + g_{0k,lj} (t_{23}) + g^*_{j0,lj} (t_{13}) \\
    f_{T;4,jk} &= g_{0j,k0} (t_1) + g^*_{0k,k0} (t_2) + g^*_{0j,k0} (t_3)  \nonumber\\ &  + g_{j0,k0} (t_{12}) + g^*_{j0,k0} (t_{23}) + g_{0j,j0} (t_{13}) \\
    f_{T;5,jk} &= g_{0j,j0} (t_1) + g_{0j,k0} (t_2) + g_{0k,k0} (t_3)  \nonumber\\ &  + g_{j0,k0} (t_{12}) + g_{j0,k0} (t_{23}) + g_{0j,k0} (t_{13}) \\
    f_{T;6,jkl} &= g_{0j,k0} (t_1) + g^*_{0k,lj} (t_2) + g_{lk,jl} (t_3)  \nonumber\\ &  + g_{j0,lj} (t_{12}) + g^*_{k0,lk} (t_{23}) + g_{0j,lk} (t_{13}) \\
    f_{T;7,jkl} &= g_{j0,lj} (t_1) + g_{0k,lj} (t_2) + g_{0k,kl}^* (t_3)  \nonumber\\ &  + g_{0j,k0} (t_{12}) + g_{jl,lk} (t_{23}) + g_{0j,lk} (t_{13}) \\
    f_{T;8,jkl} &= g_{0j,jl} (t_1) + g_{kl,lj} (t_2) + g_{0k,kl} (t_3) \nonumber\\ &   + g_{0j,lk} (t_{12}) + g_{0k,lj} (t_{23}) + g_{0j,k0} (t_{13}) ,\label{eq:x2}
\end{align}
where $f_{T;p,jk(l)}=\ln [R^{v,3}_{T;p,jk(l)}]$. 

If the {\y electronic degrees of freedom of the system are not only weakly coupled to a bath, but also strongly coupled to a number of high-frequency vibrational modes \cite{HOLSTEIN1959325,Cheng,PhysRevLett.105.050404,Womick2011,Zhu2011,HuJie,Christensson2012,Butkus2012b},
then the overall response function is given by a product of the ones reported Eqs. (\ref{eq:x1}-\ref{eq:x2}) and of those derived in the previous Section \ref{sec:rfs}. The extension of this to higher-order functions follows directly from Eq. (\ref{eq:gf}) and from the more general expression of the vibrational response function given in Eq. (\ref{eq:65}).}

\section{Effect of vibrational relaxation}

The results obtained in the previous Sections refer to the case of a coherent vibrational dynamics. However, the vibrational states can undergo relaxation, resulting in incoherent transitions between vibrational eigenstates. Formally the phonon emission process is represented as a transition:
\begin{equation}
    (a+z_k) |n,-z_k\rangle = \sqrt{n} |n-1,-z_k\rangle , 
\end{equation}
where $ |n,-z_k\rangle = \mathcal{D} (-z_k) |n\rangle $ is the displaced number state, eigenstate of $H_{v,k}$, and the index $k$ denotes the electronic state.

The effect of relaxation on the time evolution of the vibrational state can be simulated by including a non-Hermitian term in the Hamiltonian\cite{Gerry04a}, and, more specifically, by performing the following replacements:
\begin{equation}
    H_{v,k} \longrightarrow \tilde H_{v,k} = H_{v,k} -i \hbar\frac{\kappa}{2} (a^\dagger + z_k) (a+z_k) ,
\end{equation}
where $\kappa$ is the decay rate.
As a result, a coherent state evolves as
\begin{equation}\label{eq:10}
    e^{-i \tilde H_{v,k} /\hbar} |\alpha\rangle = f_k(t) \, e^{i \tilde g}| (\alpha +z_k) e^{-(\kappa/2+i\omega_v)t}-z_k\rangle ,
\end{equation}
where the phase $\tilde g$ is obtained by replacing $\omega_v$ with $\omega_v-i\kappa/2$ in the expression of $g$ (see Appendix \ref{app:B}), while the amplitude is given by the prefactor
\begin{equation}\label{eq:11}
    f_k(t,\alpha) = \exp\left[-\frac{|\alpha + z_k|^2}{2}\left(1-e^{-\kappa t}\right) \right].
\end{equation}

The effect of relaxation on the coherent state $|\alpha\rangle$ evolution is thus twofold. On the one hand [Eq. \ref{eq:10}], the wave packet no longer describes a circle of radius $|\alpha+z_k\rangle$, but rather a spiral, around the point $-z_k$ in the complex plane. On the other hand, the modulus of the state vector decreases as a function of time, asymptotically approaching a value that decreases exponentially with the distance between the initial state $|\alpha\rangle$ and the displaced origin $-z_k$ [Eq. \ref{eq:11}]. In particular, one has that $f_k(t,-z_k) {\y =} 1$.

The first effect can be accounted for by replacing the positive real parameter $\omega_v$ with the complex parameter $\tilde\omega_v\equiv\omega_v\pm i\kappa/2$ in the exponential functions $e^{\pm i\omega_v t}$. This leads to modified expressions of the third-order response functions, which can be written as functions of 
\begin{equation}\label{eq:lambda}
    \tilde\Lambda_{p_1 p_2 p_3} \equiv (\omega_v+i\kappa/2)(p_1 t_1 + p_2 t_2 + p_3 t_3 ) ,
\end{equation}
being $p_1,p_2,p_3 \ge 0$. In the presence of a minus sign in the exponent, the above $\tilde\Lambda$ has to be replaced by its complex conjugate, in order for the real part of the exponent to be negative. The generalization of Eq. (\ref{eq:lambda}) to the $M$-th order case, with $M>3$ is straightforward.

The second effect of the vibrational relaxation can be accounted for by multiplying each response function by an $F$ that is given by the product of functions $f_k(t_i,\alpha_{ket,i-1})$ (for the ket) and $f_k(t_i,\alpha_{bra,i-1})$ (for the bra), each one corresponding to the relevant electronic states $k$, waiting times $t_i$ and initial states $|\alpha_{i-1}\rangle$. Formally, 
\begin{equation}
    F = \prod_{i=1}^M f_{k_i} (t_i,\alpha_{ket,i-1})\, f_{b_i} (t_i,\alpha_{bra,i-1}),
\end{equation}
where $k_i$ and $b_i$ are the electronic states in the ket and bra, respectively, during the $i$-th waiting time, while $\alpha_{ket,i-1}$ and $\alpha_{bra,i-1}$ represent the vibrational states at the beginning of the same waiting time.
These factors $F$ tend to suppress the response functions and display a dependence on the waiting times that is both explicit and implicit, the latter one being included in the expression of $\alpha_{ket,i-1}$ and $\alpha_{bra,i-1}$. 

\subsection{Third-order response functions}
In the following we derive such factor for each of the four considered processes, omitting the expressions of $\alpha_{ket,i}$ and $\alpha_{bra,i}$, which are reported in Appendix \ref{app:B}. 

\paragraph{Ground state bleaching.}
The response function corresponding to the refocusing contribution is given by
\begin{align}
    \tilde R^{(v)}_{2,jk} & = F_2 \, \exp[ -(z_j^2+z_k^2) + z_j^2e^{i\tilde\Lambda_{100}} + z_k^2e^{-i\tilde\Lambda_{001}^*} \nonumber\\
    & + z_j z_k (-e^{i\tilde\Lambda_{010}}+e^{i\tilde\Lambda_{011}}+e^{i\tilde\Lambda_{110}}-e^{i\tilde\Lambda_{111}})].
\end{align}
The prefactor that accounts for the time dependence of the coherent state modulus is
\begin{align}
    F_2 = & f_k(t_3,\alpha_{ket,2}) \, f_j(t_1,\alpha_{bra,0})\, f_0(t_2+t_3,\alpha_{bra,1}) .
\end{align}

The response function corresponding to the non-refocusing contribution reads
\begin{align}
    \tilde R_{5,jk}^{(v)} & = F_5 \, \exp[-(z_j^2+z_k^2) + z_j^2 e^{-i\tilde\Lambda_{100}^*} + z_k^2 e^{-i\tilde\Lambda_{001}^*} \nonumber\\
    & + z_j z_k (e^{-i\tilde\Lambda_{010}^*}-e^{-i\tilde\Lambda_{011}^*}-e^{-i\tilde\Lambda_{110}^*} +e^{-i\Lambda_{111}}) ].
\end{align}
The prefactor that accounts for the time dependence of the coherent state modulus is
\begin{align}
    F_5 = & f_j(t_1,\alpha_{ket,0}) \, f_0(t_2,\alpha_{ket,1})\, f_k(t_3,\alpha_{ket,2}) .
\end{align}

\paragraph{Stimulated emission.} 
We start by considering the response function related to the refocusing contribution, which reads:
\begin{align}
    \tilde R_{1,jk}^{(v)} & = F_1\,\exp[-(z_j^2+z_k^2) + z_j^2 e^{i\tilde\Lambda_{110}} + z_k^2 e^{-i\tilde\Lambda_{011}^*} \nonumber\\
    & + z_j z_k (e^{i\tilde\Lambda_{001}}-e^{-i\tilde\Lambda_{010}^*}+e^{i\tilde\Lambda_{100}} -e^{i\tilde\Lambda_{111}}) ],
\end{align}
The prefactor $F_1$, which accounts for the time dependence of the coherent state modulus is
\begin{align}
    F_1 = & f_k(t_2+t_3,\alpha_{ket,1}) \, f_j(t_1+t_2,\alpha_{bra,0})\, f_0(t_3,\alpha_{ket,2}) .
\end{align}

The response function corresponding to the non-refocusing contribution to the stimulated emission is given by
\begin{align}
    \tilde R_{4,jk}^{(v)} & = F_4\, \exp[-(z_j^2+z_k^2) + z_j^2 e^{-i\tilde\Lambda_{111}^*} + z_k^2 e^{i\tilde\Lambda_{010}} \nonumber\\
    & + z_j z_k (e^{i\tilde\Lambda_{001}}+e^{-i\tilde\Lambda^*_{100}}-e^{i\tilde\Lambda_{011}} -e^{-i\tilde\Lambda^{\y *}_{110}}) ].
\end{align}
The prefactor $F_4$ is given by the product of two nontrivial contributions, namely 
\begin{align}
    F_4 = & f_j(t_1+t_2+t_3,\alpha_{ket,0}) \, f_k(t_2,\alpha_{bra,1}) .
\end{align}

\paragraph{Excited state absorption.}
The response function corresponding to the refocusing component of the excited state absorption is given by:
\begin{align}
    \tilde R^{(v)}_{3,jkl} & = F_3 \exp\{-[z_j^2+z_{l}^2+z_k^2-z_{l}(z_j+z_k)]\nonumber\\
    & + z_{l k} z_{l j} e^{-i\tilde \Lambda_{001}^*} + z_k z_{kl} e^{-i\tilde \Lambda_{010}^*}  + z_j z_k e^{i\tilde \Lambda_{100}} \nonumber\\
    &+ z_k z_{l j} e^{-i\tilde \Lambda_{011}^*} - z_j z_{kl} e^{i\tilde \Lambda_{110}} - z_j z_{l j} e^{i\tilde \Lambda_{111}}\} ,
\end{align}
The prefactor $F_3$, accounting for the decay in the coherent state modulus, is given by the product of three nontrivial terms: 
\begin{align}
    F_3 = & f_k(t_2,\alpha_{ket,1}) \, f_{l}(t_3,\alpha_{ket,2})\, f_j(t_1+t_2+t_3,\alpha_{bra,0}) .
\end{align}

The response function corresponding to the non-refocusing component of the excited state absorption reads:
\begin{align}
    \tilde R^{(v)}_{6,jkl} & = F_6 \exp\{-[z_j^2+z_k^2+z_{l}^2-z_{l}(z_j+z_k)] \nonumber\\ 
    & + z_{l k} z_{l j} e^{-i\tilde \Lambda_{001}^*} + z_k z_{l j} e^{i\tilde \Lambda_{010}}  + z_j z_k e^{-i\tilde \Lambda_{100}^*} \nonumber\\
    &- z_k z_{l k} e^{i\tilde \Lambda_{011}} + z_j z_{jl} e^{-i\tilde \Lambda_{110}^*} + z_j z_{l k} e^{-i\tilde \Lambda^{\y *}_{111}}\} .
\end{align}
The prefactor $F_6$ results from the product of three terms, namely 
\begin{align}
    F_6 = & f_j(t_1+t_2,\alpha_{ket,0}) \, f_{l}(t_3,\alpha_{ket,2})\, f_k(t_2+t_3,\alpha_{bra,1}) .
\end{align}

\paragraph{Double quantum coherence.}
The response function corresponding to the first component of the double quantum coherence reads:
\begin{align}
    \tilde R_{7,jkl}^{(v)} & = F_7 \exp\{-[z_j^2+z_{l}^2+z_k^2-z_{l} (z_j+z_k)] \nonumber\\
    &  + z_k z_{kl} e^{i\tilde \Lambda_{001}} + z_k z_{l j} e^{-i\tilde \Lambda_{010}^*}  {\y -} z_j z_{l j} e^{-i\tilde \Lambda_{100}^*}  \nonumber\\ 
    & + z_{l j} z_{l k} e^{-i\tilde \Lambda_{011}^*} +z_j z_k e^{-i\tilde \Lambda_{110}^*}  + z_j z_{l k} e^{-i\tilde \Lambda_{111}^*}\} . 
\end{align}
The prefactor $F_7$ is given by the following product of nontrivial terms: 
\begin{align}
    F_7 = & f_j(t_1,\alpha_{ket,0}) \, f_{l}(t_2+t_2,\alpha_{ket,1})\, f_k(t_3,\alpha_{bra,2}) .
\end{align}

Finally, the response function corresponding to the second component of the double quantum coherence is given by:
\begin{align}
    \tilde R_{8,jkl}^{(v)} & = F_8 \exp\{-[z_j^2+z_{l}^2+z_k^2- z_{l}(z_j +z_k)] \nonumber\\
    & + z_k z_{kl} e^{-i\tilde \Lambda_{001}^*} + z_{l k} z_{l j} e^{-i\tilde \Lambda_{010}^*} + z_j z_{jl} e^{-i\tilde \Lambda_{100}^*} \nonumber\\ &  + z_k z_{l j} e^{-i\tilde \Lambda_{011}^*} +z_j z_{l k} e^{-i\tilde \Lambda_{110}^*} + z_j z_k e^{-i\tilde \Lambda_{111}^*}\}.  
\end{align}
The prefactor $F_8$, accounting for  the decay in the coherent state modulus, reads:
\begin{align}
    F_8 = & f_j(t_1,\alpha_{ket,{\y 0}}) \, f_{l}(t_2,\alpha_{ket,1})\, f_k(t_3,\alpha_{bra,2}) .
\end{align}

\section{Conclusions}

In conclusion, we have developed a coherent state representation of the vibrational dynamics and of its effect on the nonlinear response functions, within the {\y linearly} displaced harmonic oscillator model. The underlying physical assumption is that nonadiabatic effects can be neglected and that the dependence of the vibrational modes on the electronic state can be reduced to that of their equilibrium positions. {\y Besides, the optically-induced transitions are assumed to satisfy the Franck-Condon principle.}

Within such model, and with no further assumption or approximation, a number of results have been derived. Crucial to the derivations is the fact that, within each pathway (i.e. any sequence of optically-induced transitions between electronic states) the vibrational state can always be described as a single coherent state. First, starting from the expressions of the path-dependent vibrational states, analytical expressions for the third-order response functions in $N$-level systems have been computed in the zero-temperature limit, where the oscillator is initialized in the ground state. Frome there, in order to highlight connections with observable quantities, we have derived within the semi-impulsive limit the amplitude of the peaks appearing in the 2D spectroscopy, as a function of the waiting time $t_2$ of the amplitude.

The expressions of the third-order response functions have been generalized to the case of $M$-th order case, with arbitrary $M$. The formal derivation has been translated into a simple recipe, which allows one to derive the vibrational component of the response functions directly from the Feynman diagrams, without performing any calculation. 

These results have been generalized to the case where the vibrational mode is initialized in a generic coherent state. This was shown to imply only a phase change in the response functions, with respect to the zero-temperature case. Besides, the expression of the response functions for arbitrary initial coherent states allows one in principle to derive the dependence on an arbitrary initial state of the vibrational mode. In fact, the coherent states form an overcomplete basis, in terms of which one can express any state of the mode, given its coherent state representation. Such possibility has been exploited in order to derive the response function for a thermal state (finite temperature) initialization.

The above result, combined with the straightforward extension to the multimode case, has been used to address the case of a system coupled to a phonon bath and to derive the effect of such coupling on the line width function. Also in this case, our approach {\y applies to an arbitrary order $M$ in the interaction with the field and to an arbitrary number $N$ of electronic levels required to model the system of interest.} 

Finally, the effect of vibrational relaxation on the vibrational response functions has been accounted for by means of a non-Hermitian Hamiltonian approach. This allows one to address the case where a few vibrational modes, strongly coupled to the electronic degrees of freedom, contribute to the coherent features in the multidimensional spectra but are themselves subject to a relaxation process.

\acknowledgements

This work has been supported by the European Union’s Horizon 2020 research and innovation programme under  the  Marie  Sklodowska-Curie  Grant  Agreement No. 812992.

\appendix

\section{Basics about coherent states \\ and displacement operators \label{app:A}}

We start by recalling some of the basic properties of the coherent states and of the displacement operators\cite{Scully97a,Gerry04a} that are used throughout the paper. 

A coherent state of a quantum harmonic oscillator, specified by the complex number $\alpha$, is given by the following linear superposition of Fock (number) states $|n\rangle$:
\begin{equation}
|\alpha\rangle = e^{-|\alpha|^2/2} \sum_{n=0}^\infty \frac{\alpha^n}{\sqrt{n!}}|n\rangle .
\end{equation}
From this it follows that the overlap between two coherent states $|\alpha\rangle$ and $|\beta\rangle$ is always finite, and is given by
\begin{equation}
    \langle \beta | \alpha \rangle = e^{-|\alpha-\beta|^2/2} e^{i{\rm Im} (\beta^*\alpha)} .
\end{equation}

The ground state of the harmonic oscillator corresponds to the number state $n=0$ and also to the coherent state $\alpha=0$. Any coherent state can be obtained from any other by applying the displacement operator, defined as
\begin{equation}
    \mathcal{D} (\alpha) = e^{\alpha a^\dagger - \alpha^* a} = e^{-|\alpha|^2/2} e^{\alpha a^\dagger} a^{-\alpha^* a}.
\end{equation}
In fact, one can show that such application leads to:
\begin{equation}
    \mathcal{D} (\alpha) |\beta\rangle = e^{i{\rm Im} (\beta^*\alpha)} |\alpha + \beta \rangle .
\end{equation}

The displacement operator can also be characterized by its action on the creation and annihilation operators, which is given by
\begin{align}
    \mathcal{D} (-\alpha) a^\dagger \mathcal{D} (\alpha) & = a^\dagger + \alpha^*,\ 
    \mathcal{D} (-\alpha) a \mathcal{D} (\alpha)  = a + \alpha ,
\end{align}
being $\mathcal{D} (-\alpha) = \mathcal{D}^\dagger (\alpha) = \mathcal{D}^{-1} (\alpha)$. From this it follows that any function of $a$ and $a^\dagger$ can be mapped onto the same function of $a+\alpha$ and $a^\dagger+\alpha^*$, as is the case for the time evolution operator of the free oscillator: 
\begin{align}
    \mathcal{D} (-\alpha) e^{-i\omega_v a^\dagger a} \mathcal{D} (\alpha) & = e^{-i\omega_v (a^\dagger + \alpha^*) (a+\alpha)} .
\end{align}
We finally remind that the free evolution of a coherent state can be represented as a rotation by an angle $\omega_v t$ around the origin of the $(X,P)\equiv\frac{1}{2}(\langle a + a^\dagger \rangle,\langle i(a^\dagger -a)\rangle)$ plane, being
\begin{equation}
    e^{-i \omega_v a^\dagger a t} |\alpha\rangle = |e^{-i\omega_v t}\alpha\rangle .
\end{equation}
Analogously, the evolution induced by a displaced-oscillator Hamiltonian geometrically corresponds, up to a phase factor (see Appendix \ref{app:B}), to a rotation around the point $[{\rm Re}(\alpha),{\rm Im}(\alpha)]$.

\section{Time evolution of the vibrational states entering the nonlinear response functions \label{app:B}}

The vibrational state in the ket and in the bra evolves, during each waiting time, under the effect of an Hamiltonians $H_{v,k}$, where $k$ denotes the electronic state. At each interaction with the field, the electronic state changes, either in the ket or in the bra. The overall evolution of the states on the two sides of the Feynman diagrams is thus induced by the alternate action of different time-evolution operators $e^{-i H_{v,k} t/\hbar}$. This can be reduced to $e^{-i H_{v,0} t/\hbar}$ by means of the displacement operators $\mathcal{D}(z_k) $, where $z_k$ is a real number (see Appendix A). As a result, one has that:
\begin{align}
e^{-i H_{v,k} t/\hbar} = \mathcal{D}(-z_k)\, e^{-i \omega_v a^\dagger a t} \, \mathcal{D}(z_k) .
\end{align}

When applied to a coherent state $|\alpha\rangle$, the above operator displaces the coherent state and generates a phase factor. The displacement results from a translation by $z_k$ in the $(X,P)$ plane, followed by a rotation by an angle $\omega_v t$ around the origin and by a translation by $-z_k$, induced respectively by the operators $\mathcal{D}(z_k)$, $e^{-i \omega_v a^\dagger a t}$ and $\mathcal{D}(-z_k)$ in the above equation. The phase factor is given by the sum of two contributions, $-z_k{\rm Im}(\alpha)$ and $z_k{\rm Im}[(\alpha+z_k)e^{-i\omega_v t}]$, induced respectively by $\mathcal{D}(z_k)$ and $\mathcal{D}(-z_k)$.
As a result, one has that 
\begin{align}
    e^{-i H_{v,k} t/\hbar} |\alpha\rangle & = e^{i g(\alpha,z_k,t)} |(\alpha\!+\!z_k)e^{-i\omega_v t}\!-\!z_k\rangle ,
\end{align}
where $g(t;z_k,\alpha)=z_k{\rm Im}[(\alpha+z_k)e^{-i\omega_v t}-\alpha]$ represents the overall phase.

Applying the above expression for the transformation of the vibrational states on the two sides of the Feynman diagrams, one obtains for the complex numbers that define the coherent states at the end of the $j$-th waiting time, given by:
\begin{align}
    |\phi_{j,\chi}\rangle = e^{i\sum_{k=1}^j a_{k,\chi} } |\alpha_{j,\chi}\rangle ,
\end{align}
where $\chi = ket, bra$, the complex numbers $\alpha_{j,\chi}$ specify the coherent states, and the real numbers $a_{j,\chi}$ represent the phases accumulated by the state within the $j$-th waiting time.
The coherent states corresponding to consecutive waiting times are related by the equations
\begin{align}
\alpha_{ket,j} & = (\alpha_{ket,j-1}+z_{k_j}) e^{-i\omega_v t_j} -z_{k_j} \\
\alpha_{bra,j} & = (\alpha_{bra,j-1}+z_{b_j}) e^{-i\omega_v t_j} -z_{b_j} .
\end{align}
The real numbers that define the phases satisfy the relations
\begin{align}
    a_{ket,j} =&  z_{k_j} {\rm Im} (\alpha_{ket,j}-\alpha_{ket,j-1}) \\
    a_{bra,j} =&  z_{b_j} {\rm Im} (\alpha_{bra,j}-\alpha_{bra,j-1}),
\end{align}
where $k_j$ and $b_j$ specify the electronic-state component of the ket and of the bra, respectively.
In the following, we derive the explicit expressions for these quantities that are relevant for the different contributions to the response functions, assuming $\alpha_{ket,0} = \alpha_{bra,0} = 0$. The case of a generic coherent state is considered in the final paragraph.

\paragraph{Ground state bleaching, rephasing term.}
In this case [Fig. \ref{fig:dsfd2}(a)], the sequence of electronic state in the ket is given by
$k_1=k_2=0$ and $k_3=k$. Therefore, the sequence of coherent states at the end of the three waiting times is specified by
$ \alpha_{ket,1}=\alpha_{ket,2}=0$ 
and
\begin{align}
    \alpha_{ket,3}=z_k(e^{-i\omega_v t_3}-1) .
\end{align}
The phases accumulated within each waiting times are $a_{ket,1}=a_{ket,2}=0 $ and
\begin{align}
a_{ket,3} = -z_k^2 \sin (\omega_v t_3) .
\end{align}

The sequence of electronic states in the bra is given by
$b_1=j$, $b_2=b_3=0$.
This results in the following sequence of coherent states:
\begin{align}
    \alpha_{bra,1}& =z_j(e^{-i\omega_v t_1}-1)\label{eq:A01}\\
    \alpha_{bra,2}& =z_j(e^{-i\omega_v t_1}-1)e^{-i\omega_v t_2}\\
    \alpha_{bra,3}& =z_j(e^{-i\omega_v t_1}-1)e^{-i\omega_v (t_2+t_3)} .
\end{align}
The phases accumulated within each waiting time by the vibrational state are:
\begin{align}
a_{bra,1} = -z_j^2 \sin (\omega_v t_1) \label{eq:A02}
\end{align}
and $a_{bra,2}=a_{bra,3}=0$.

The resulting expression of $r=-\frac{1}{2}|\alpha_{ket}-\alpha_{bra}|^2$, which determines the amplitude of the response function, is given by:
\begin{align}
    r & = z_j^2 (\cos\Lambda_{100}-1) + z_k^2 (\cos\Lambda_{001}-1) \nonumber\\
    & + z_j z_k (-\cos\Lambda_{010}+\cos\Lambda_{011}+\cos\Lambda_{110}-\cos\Lambda_{111}) .
\end{align}
Finally, the phase of the response function, $\varphi$, reads: 
\begin{align}
    \varphi & = z_j^2\sin\Lambda_{100} - z_k^2\sin\Lambda_{001} \nonumber\\
    & + z_j z_k (-\sin\Lambda_{010}+\sin\Lambda_{011}+\sin\Lambda_{110}-\sin\Lambda_{111}).
\end{align}

\paragraph{Ground state bleaching, non-rephasing term.}
In this case [Fig. \ref{fig:dsfd2}(b)], the electronic part of the ket evolves according to the sequence
$k_1=j$, $k_2=0$, and $k_3=k$.
The evolution of the vibrational part is thus given by:
\begin{align}
    \alpha_{ket,1}&=z_j(e^{-i\omega_v t_1}-1)\label{eq:A03}\\
    \alpha_{ket,2}&=z_j(e^{-i\omega_v t_1}-1)e^{-i\omega_v t_2}\\
    \alpha_{ket,3}&=z_j(e^{-i\omega_v t_1}-1)e^{-i\omega_v (t_2+t_3)} +z_k(e^{-i\omega_v t_3}-1) .
\end{align}
The phases accumulated at the end of the three waiting times are:
\begin{align}
    a_{ket,1}&=-z_j^2\sin(\omega_v t_1)\label{eq:A04}\\
    a_{ket,3}&=-z_k^2\sin(\omega_v t_3) - z_j z_k \{ \sin[\omega_v (t_1+t_2+t_3)] \nonumber\\ & 
    - \sin[\omega_v (t_2+t_3)] - \sin[\omega_v (t_1+t_2)] + \sin(\omega_v t_2)\},
\end{align}
while $a_{ket,2}=0$.

The electronic state in the bra doesn't undergo any evolution
($b_1=b_2=b_3=0$). 
Therefore, the vibrational state is also frozen
($\alpha_{bra,1}=\alpha_{bra,2}=\alpha_{bra,3}=0$),
and no phase is accumulated during the three waiting times
($a_{bra,1}=a_{bra,2}=a_{bra,3}=0$).

The above vibrational states correspond to the following expression of $r=-|\alpha_{ket}|^2/2$:
\begin{align}
    r & = z_j^2 (\cos\Lambda_{100}-1) + z_k^2 (\cos\Lambda_{001}-1) \nonumber\\ &+ z_j z_k ( \cos\Lambda_{010}-\cos\Lambda_{011}-\cos\Lambda_{110}+\cos\Lambda_{111}) .
\end{align}

\paragraph{Stimulated emission, rephasing term.}
In this case [Fig. \ref{fig:dsfd2}(c)], the sequence of electronic state in the ket is given by
$k_1=0$ and $k_2=k_3=k$. Therefore, the sequence of coherent states at the end of the three waiting times is specified by
$ \alpha_{ket,1}=0$ 
and
\begin{align}
    \alpha_{ket,2}& =z_k(e^{-i\omega_v t_2}-1) \label{eq:A05}\\
    \alpha_{ket,3}& =z_k[e^{-i\omega_v (t_2+t_3)}-1].
\end{align}
The phases accumulated within the three waiting times are $a_1=0$ and
\begin{align}
a_{ket,2} & = -z_k^2 \sin (\omega_v t_2) \label{eq:A06}\\
a_{ket,3} & = -z_k^2 \{\sin [\omega_v (t_2+t_3)]-\sin (\omega_v t_2)\},
\end{align}

The sequence of electronic states in the bra is given by
$b_1=b_2=j$ and $b_3=0$.
This results in the following sequence of coherent states:
\begin{align}
    \alpha_{bra,2}& =z_j[e^{-i\omega_v (t_1+t_2)}-1]\label{eq:A08}\\
    \alpha_{bra,3}& =z_j[e^{-i\omega_v (t_1+t_2)}-1]e^{-i\omega_v t_3} ,
\end{align}
while the expression of $\alpha_{bra,1}$ coincides with that given in Eq. (\ref{eq:A01}).
The phases accumulated within each waiting time by the vibrational state are given by:
\begin{align}
a_{bra,2} & = -z_j^2 \{\sin [\omega_v (t_1+t_2)]-\sin (\omega_v t_1)\} \label{eq:A10}
\end{align}
and $a_{bra,3}=0$, while $a_{bra,1}$ reads as in Eq. (\ref{eq:A02}).

From the above equations it follows that $r=-\frac{1}{2}|\alpha_{ket}-\alpha_{bra}|^2$ is given by:
\begin{align}
    r & =  z_j^2 (\cos\Lambda_{110}-1) + z_k^2 (\cos\Lambda_{011}-1) \nonumber\\
    & + z_j z_k (\cos\Lambda_{001}-\cos\Lambda_{010} + \cos\Lambda_{100}-\cos\Lambda_{111} )  .
\end{align}
The phase $\varphi$, resulting from the difference between the phase factors accumulated by the ket and the bra, and from the inner product $\langle\alpha_{bra}|\alpha_{ket}\rangle$, reads: 
\begin{align}
    \varphi & = z_j^2\sin\Lambda_{110} - z_k^2\sin\Lambda_{011} \nonumber\\
    & +z_j z_k (\sin\Lambda_{001}+\sin\Lambda_{010}+\sin\Lambda_{100}-\sin\Lambda_{111}).
\end{align}

\paragraph{Stimulated emission, non-rephasing term.}
In this case [Fig. \ref{fig:dsfd2}(d)], the electronic part of the ket evolves according to the sequence
$k_1=k_2=k_3=j$.
The evolution of the vibrational part is thus given by:
\begin{align}
    \alpha_{ket,2}&=z_j[e^{-i\omega_v (t_1+t_2)}-1]\label{eq:A11}\\
    \alpha_{ket,3}&=z_j[e^{-i\omega_v (t_1+t_2+t_3)}-1],
\end{align}
while the expression of $\alpha_{ket,1}$ coincides with that reported in Eq. (\ref{eq:A03}).
The phases accumulated at the end of the three waiting times are:
\begin{align}
    a_{ket,2}&= -z_j^2 \{\sin [\omega_v (t_1+t_2)]-\sin (\omega_v t_1)\} \label{eq:A12}\\
    a_{ket,3}&= -z_j^2 \{\sin [\omega_v (t_1+t_2+t_3)]-\sin [\omega_v (t_1+t_2)]\} ,
\end{align}
with $a_{ket,1}$ given by Eq. (\ref{eq:A04}). 

The electronic state in the bra undergoes the following evolution:
$b_2=k$, $b_1=b_3=0$. 
This results in the following sequence of coherent states
$\alpha_{bra,1}=0$ and
\begin{align}
    \alpha_{bra,2}& =z_k(e^{-i\omega_v t_2}-1)\\
    \alpha_{bra,3}& = z_k(e^{-i\omega_v t_2}-1) e^{-i\omega_v t_3} .
\end{align}
The phases accumulated within each waiting time by the vibrational state are given by
$a_{bra,1}=a_{bra,3}=0$ and
\begin{align}
a_{bra,2} & = -z_k^2 \sin (\omega_v t_2) . 
\end{align}

The expression of $r$, which determines the amplitude of $R_{4,jk}^{(v,3)}$ reads:
\begin{align}
    r & = z_j^2 (\cos\Lambda_{111}-1) + z_j^2 (\cos\Lambda_{010}-1)    
     \nonumber\\
    & + z_j z_k (\cos\Lambda_{001} + \cos\Lambda_{100} -\cos\Lambda_{011} - \cos\Lambda_{110} ) .
\end{align}
The phase of $R_{4,jk}^{(v,3)}$ is given by the following combination of sinusoidal terms:
\begin{align}
    \varphi & = - z_j^2\sin\Lambda_{111} + z_k^2\sin\Lambda_{010} \nonumber\\
    & + z_j z_k (\sin\Lambda_{001}-\sin\Lambda_{100}-\sin\Lambda_{011}+\sin\Lambda_{110}).
\end{align}

\paragraph{Excited state absorption, rephasing term.}
In this case [Fig. \ref{fig:dsfd2}(e)], the sequence of electronic state in the ket is given by
$k_1=0$, $k_2=k$ and $k_3=l$. Therefore, the sequence of coherent states at the end of the three waiting times is specified by
$ \alpha_{ket,1}=0$ 
and
\begin{align}
    \alpha_{ket,3}& =z_k(e^{-i\omega_v t_2}-1)e^{-i\omega_v t_3} + z_{l} (e^{-i\omega_v t_3}-1),
\end{align}
while the expression of $\alpha_{ket,2}$ coincides with that given in Eq. (\ref{eq:A05}). 
The phases accumulated within the three waiting times are $a_{ket,1}=0$ and
\begin{align}
a_{ket,3} & = -z_{l}^2\sin(\omega_v t_3) - z_k z_{l} \{ \sin[\omega_v (t_2+t_3)] \nonumber\\ & 
    - \sin(\omega_v t_3) - \sin(\omega_v t_2) \},
\end{align}
with $a_{ket,2}$ that reads as in Eq. (\ref{eq:A06}).

The sequence of electronic states in the bra is given by
$b_1=b_2=b_3=j$.
This results in the following sequence of coherent states:
\begin{align}
    \alpha_{bra,3}& =z_j[e^{-i\omega_v (t_1+t_2+t_3)}-1] ,
\end{align}
while $\alpha_{bra,1}$ and $\alpha_{bra,2}$ are already given in Eqs. (\ref{eq:A01},\ref{eq:A08}).
The phases accumulated within each waiting time by the vibrational state are given by:
\begin{align}
a_{bra,3} & = -z_j^2 \{\sin [\omega_v (t_1+t_2+t_3)]-\sin [\omega_v (t_1+t_2)]\}, 
\end{align}
with $a_{bra,1}$ and $a_{bra,2}$ as in Eqs. (\ref{eq:A02},\ref{eq:A10}).

As a result, $r=-\frac{1}{2}|\alpha_{ket}-\alpha_{bra}|^2$, which determines the modulus of $R_{3,jkl}^{(v)}$, takes the form:
\begin{align}
    r & = -[z_j^2+z_{l}^2+z_k^2-z_{l}(z_j+z_k)] \nonumber\\
    & + z_{l k} z_{l j} \cos\Lambda_{001} + z_k z_{kl} \cos\Lambda_{010}  + z_j z_k\cos\Lambda_{100} \nonumber\\
    & + z_k z_{l j} \cos\Lambda_{011} - z_j z_{kl} \cos\Lambda_{110} - z_j z_{l j} \cos\Lambda_{111} .
\end{align}
The phase of the response function, resulting from the difference between the phase factors accumulated by the ket and the bra, and from the inner product $\langle\alpha_{bra}|\alpha_{ket}\rangle$, reads:
\begin{align}
    \varphi & = -z_{l k} z_{l j} \sin\Lambda_{001} - z_k z_{kl} \sin\Lambda_{010}  + z_j z_k\sin\Lambda_{100} \nonumber\\
    &- z_k z_{l j} \sin\Lambda_{011} - z_j z_{kl} \sin\Lambda_{110} - z_j z_{l j} \sin\Lambda_{111} .
\end{align}

\paragraph{Excited state absorption, non-rephasing term.}
In this case [Fig. \ref{fig:dsfd2}(f)], the sequence of electronic state in the ket is given by
$k_1=k_2=j$ and $k_3=l$. Therefore, the sequence of coherent states at the end of the three waiting times is specified by the same $ \alpha_{ket,1}$ and $ \alpha_{ket,1}$ as the ones reported in Eqs. (\ref{eq:A03},\ref{eq:A11})
and by
\begin{align}
    \alpha_{ket,3}& = z_j [e^{-i\omega_v (t_1+t_2)}-1]e^{-i\omega_v t_3} + z_{l} (e^{-i\omega_v t_3}-1).
\end{align}
The phases accumulated within the three waiting times are given in Eqs. (\ref{eq:A04},\ref{eq:A12}) and by
\begin{align}
a_{ket,3} & = -z_{l}^2\sin(\omega_v t_3) - z_j z_{l} \{ \sin[\omega_v (t_1+t_2+t_3)] \nonumber\\ & 
    - \sin(\omega_v t_3) - \sin[\omega_v (t_1+t_2) ]\}.
\end{align}

The sequence of electronic states in the bra is given by
$b_1=0$ and $b_2=b_3=k$.
This results in the following sequence of coherent states:
\begin{align}
    \alpha_{bra,2}& =z_k(e^{-i\omega_v t_2}-1)\\
    \alpha_{bra,3}& =z_j[e^{-i\omega_v (t_2+t_3)}-1] ,
\end{align}
while $\alpha_{bra,1}=0$.
The phases accumulated within each waiting time by the vibrational state are given by:
\begin{align}
a_{bra,2} & = -z_k^2 \sin (\omega_v t_2) \\
a_{bra,3} & = -z_k^2 \{\sin [\omega_v (t_2+t_3)]-\sin (\omega_v t_2)\} 
\end{align}
with $a_{bra,1}=0$.

The distance between the wave packets in the ket and bra states is quantified by $r$, which is given by:
\begin{align}
    r & = -[z_j^2+z_k^2+z_{l}^2-z_{l}(z_j+z_k)] \nonumber\\
    & + z_{l k} z_{l j} \cos\Lambda_{001} + z_k z_{l j} \cos\Lambda_{010}  + z_j z_k\cos\Lambda_{100} \nonumber\\
    & - z_k z_{l k} \cos\Lambda_{011} + z_j z_{jl} \cos\Lambda_{110} + z_j z_{kl} \cos\Lambda_{111} .
\end{align}
Finally, the phase of the response function is given by the following combination of sinusoidal functions:
\begin{align}
    \varphi & = -z_{l k} z_{l j} \sin\Lambda_{001} + z_k z_{l j} \sin\Lambda_{010}  - z_j z_k\sin\Lambda_{100} \nonumber\\
    & - z_k z_{l k} \sin\Lambda_{011} - z_j z_{jl} \sin\Lambda_{110} - z_j z_{kl} \sin\Lambda_{111} .
\end{align}

\paragraph{Double quantum coherence, first term.}
In this case [Fig. \ref{fig:dsfd2}(g)], the electronic part of the ket evolves according to the sequence
$k_1=j$, and $k_2=k_3={l}$.
The evolution of the vibrational part is thus given by:
\begin{align}
    \alpha_{ket,2}&=z_j(e^{-i\omega_v t_1}-1)e^{-i\omega_v t_2} + z_{l} (e^{-i\omega_v t_2}-1)\\
    \alpha_{ket,3}&=z_j(e^{-i\omega_v t_1}-1)e^{-i\omega_v (t_2+t_3)} + z_{l} [e^{-i\omega_v (t_2+t_3)}-1] ,
\end{align}
while $\alpha_{ket,1}$ is given by Eq. (\ref{eq:A03}).

The phases accumulated at the end of the three waiting times are:
\begin{align}
    a_{ket,2}&= -z_{l}^2 \sin (\omega_v t_2) - z_j z_{l} \{ \sin[\omega_v (t_1+t_2)] \nonumber\\
    & - \sin(\omega_v t_2) - \sin(\omega_v t_1)\}\\
    a_{ket,3}&= -z_{l}^2 \{\sin [\omega_v (t_2+t_3)]-\sin (\omega_v t_2)\} \nonumber\\ 
    & - z_j z_{l} \{ \sin[\omega_v (t_1+t_2+t_3)] -\sin[\omega_v (t_1+t_2)] \nonumber\\
    & - \sin[\omega_v (t_2+t_3)]+ \sin(\omega_v t_2) \} ,
\end{align}
with $a_{ket,1}$ expressed in Eq. (\ref{eq:A04}).

The electronic state in the bra undergoes the following evolution:
$b_1=b_2=0$, $b_3=k$. 
This results in the following sequence of coherent states
$\alpha_{bra,1}=\alpha_{bra,2}=0$ and
\begin{align}
    \alpha_{bra,3}& =z_k(e^{-i\omega_v t_3}-1).
\end{align}
The phases accumulated within each waiting time by the vibrational state are given by
$a_{bra,1}=a_{bra,2}=0$ and
\begin{align}
a_{bra,3} & = -z_k^2 \sin (\omega_v t_3) .
\end{align}

The resulting amplitude of the response function, depending on the distance between the two above wave packets, is an exponential function of:
\begin{align}
    r & = -[z_j^2+z_{l}^2+z_k^2-z_{l} (z_j+z_k)] \nonumber\\
    &  + z_k z_{kl} \cos\Lambda_{001} +z_k z_{l j} \cos\Lambda_{010}  - z_j z_{l j} \cos\Lambda_{100}  \nonumber\\ 
    & + z_{l j} z_{l k} \cos\Lambda_{011} +z_j z_k \cos\Lambda_{110} + z_j z_{l k} \cos\Lambda_{111}  . 
\end{align}
The phase, resulting from the difference between the phase factors accumulated by the ket and bra and from the overlap $\langle\alpha_{bra}|\alpha_{ket}\rangle$, is given by the expression:
\begin{align}
    \varphi & =  z_k z_{kl}\sin\Lambda_{001}-z_k z_{l j} \sin\Lambda_{010}  + z_j z_{l j} \sin\Lambda_{100} \nonumber\\
    &  - z_{l j} z_{l k} \sin\Lambda_{011}  -z_j z_k \sin\Lambda_{110} - z_j z_{l k} \sin\Lambda_{111}  . 
\end{align}

\paragraph{Double quantum coherence, second term.}
In this case [Fig. \ref{fig:dsfd2}(h)], the electronic part of the ket evolves according to the sequence
$k_1=j$, $k_2={l}$, and $k_3=k$.
The evolution of the vibrational part is thus given by:
\begin{align}
    \alpha_{ket,2}&=z_j(e^{-i\omega_v t_1}-1)e^{-i\omega_v t_2} + z_{l} (e^{-i\omega_v t_2}-1)\\
    \alpha_{ket,3}&=[z_j(e^{-i\omega_v t_1}-1)e^{-i\omega_v t_2} + z_{l} (e^{-i\omega_v t_2}-1)]e^{-i\omega_v t_3} \nonumber\\ &+ z_k (e^{-i\omega_v t_3}-1),
\end{align}
with $\alpha_{ket,1}$ given by Eq. (\ref{eq:A03}).

The phases accumulated at the end of the three waiting times are:
\begin{align}
    a_{ket,2}&= -z_{l}^2 \sin (\omega_v t_2) - z_j z_{l} \{ \sin[\omega_v (t_1+t_2)] \nonumber\\
    & - \sin(\omega_v t_2) - \sin(\omega_v t_1)\}\\
    a_{ket,3}&=-z_k^2\sin(\omega_v t_3) -z_j z_k \{ \sin[\omega_v (t_1+t_2+t_3)] \nonumber\\ & -\sin[\omega_v (t_1+t_2)] -\sin[\omega_v (t_2+t_3)] + \sin(\omega_v t_2) \} \nonumber\\
    & - z_{l} z_k \{ \sin[\omega_v (t_2+t_3)] -\sin\omega_v (t_2) -\sin(\omega_v t_3) \} ,
\end{align}
while $a_{ket,1}$ is given in Eq. (\ref{eq:A04}).

The electronic state in the bra doesn't undergo any evolution
($b_1=b_2=b_3=0$). 
Therefore, the vibrational state is also frozen
($\alpha_{bra,1}=\alpha_{bra,2}=\alpha_{bra,3}=0$),
and no phase is accumulated during the three waiting times
($a_{bra,1}=a_{bra,2}=a_{bra,3}=0$).

The resulting amplitude of $R_{8,jkl}^{(v,3)}$ is an exponential function of:
\begin{align}
    r & = -[z_j^2+z_{l}^2+z_k^2- z_{l}(z_j +z_k)] \nonumber\\
    &  + z_k z_{kl}\cos\Lambda_{001}+z_{l k} z_{l j} \cos\Lambda_{010}  + z_j z_{jl} \cos\Lambda_{100} \nonumber\\ 
    &  +z_k z_{l j} \cos\Lambda_{011} + z_j z_{l k} \cos\Lambda_{110} + z_j z_k \cos\Lambda_{111} . 
\end{align}

\section{Initialization in a generic \\ coherent state \label{app:D}}
If the initial vibrational state of corresponds to a generic coherent state $|\alpha_0\rangle$, the above equations have to be generalized by adding further terms. This can be done simply by means of the following replacements:
\begin{align}
    {\y \alpha_{\chi,j}\ \longrightarrow}\ \alpha_{\chi,j}' = \alpha_{\chi,j} + \beta \equiv \alpha_{\chi,j} + \alpha_0 \exp\left(\sum_{k=1}^j t_k\right),
\end{align}
where $\chi = ket, bra$. The change in the vibrational state resulting from the initialization in a generic coherent state is thus independent on the pathway. In addition, the phase factors undergo a change, which is instead pathway dependent. In fact, one has that
\begin{align}
    a_{ket,j}' & = a_{ket,j} + z_{k_j} {\rm Im} \left[\alpha_0 (e^{-i\omega_v t_j}-1)e^{-i\omega_v \sum_{k=1}^{j-1} t_k}\right] \\
    a_{bra,j}' & = a_{bra,j} + z_{b_j} {\rm Im} \left[\alpha_0 (e^{-i\omega_v t_j}-1)e^{-i\omega_v \sum_{k=1}^{j-1} t_k}\right] .
\end{align}

The additional phase in the overlap $\langle\phi_{bra}|\phi_{ket}\rangle$ resulting from the fact that $\alpha_0\neq 0$ can be derived by considering the two contributions [see Eq. (\ref{eq:ph}), whose generalization to the case $M>3$ is straightforward]. The first one, resulting from the overlap $\langle\alpha_{bra}'|\alpha_{ket}'\rangle$, is given by
\begin{align}
    (\Delta\varphi)_1 = {\rm Im}(\beta\alpha_{bra}^*+\beta^*\alpha_{ket})={\rm Im} [\beta^*(\alpha_{ket}-\alpha_{bra})].
\end{align}
The second contribution comes from the difference in the phases accumulated with the application of the displacement operators, $a_{ket}-a_{bra}$, and reads
\begin{align}
    (\Delta\varphi)_2 & = -\sum_{j=1}^M (z_{k_j}-z_{b_j}) {\rm Im}[\alpha^*_0(e^{i\omega_v t_j}-1)\, e^{i\omega_v\sum_{k=1}^{j-1}t_k}] , 
\end{align}
where $|k_j\rangle$ and $|b_j\rangle$ are the electronic ket and bra states during the $j$-th waiting time.
Replacing in the two above equations the expressions of $a_{ket}$, $a_{bra}$, $\alpha_{ket}$, and $\alpha_{bra}$ given in the present Appendix, one can verify that $(\Delta\varphi)_1=(\Delta\varphi)_2$, and thus 
\begin{align}
(\Delta\varphi)=2(\Delta\varphi)_1= 2 {\rm Im} [\alpha_0^*(\alpha_{ket}-\alpha_{bra})e^{i\omega_v\sum_{k=1}^{M}t_k}] .
\end{align}

As a final step, we proceed to the derivation of the response function corresponding to the thermal state. In view of the above equation, the initialization to a coherent state that differs from the ground state results in a prefactor $e^{i\Delta\varphi}$, where $\Delta\varphi\equiv 2\,{\rm Im}(\alpha_0^* Q)$ with $|Q|^2=|\alpha_{ket}-\alpha_{bra}|^2=-2r$. The thermal state can be expressed in the coherent state representation as \cite{Scully97a}
\begin{align}
    \rho_{\rm Th} = \int\,d^2\alpha_0\, \frac{e^{-|\alpha_0|^2/\langle n \rangle}}{\pi\langle n \rangle} | \alpha_0\rangle \langle\alpha_0 |,
\end{align}
where $\langle n \rangle=(e^{\hbar\omega_v/k_BT}-1)^{\y -1}$ is the average phonon number. As a result, the finite temperature response function is given by the zero-temperature one, times a factor
\begin{align}
    \int\,d^2\alpha_0\, \frac{e^{-|\alpha_0|^2/\langle n \rangle}}{\pi\langle n \rangle}\, e^{\alpha_0^* Q -\alpha_0 Q^*} = e^{-\langle n \rangle|Q|^2} .
\end{align}
Being the zero-temperature response functions $R_{T=0}=e^{-r/2}e^{i\varphi}$, their finite-temperature counterparts are obtained by multiplying the real part in the exponent by a factor $(1+2\langle n \rangle) = \coth (\hbar\omega_v/2k_B T)$. 

\section{General expression \\ of the response function\label{app:C}}

As a starting point, the $M$-th order response function can be written as a propagator of the vibrational state, and in particular as:
\begin{align}\label{eq:60}
   R^{(v,M)}& = \langle 0 | (\mathcal{D}_{0j_1}\,\mathcal{V}_1\,\mathcal{D}_{j_10})\,\dots\,(\mathcal{D}_{0j_M}\, \mathcal{V}_M\, \mathcal{D}_{j_M0}) | 0 \rangle,
\end{align}
where 
$\mathcal{D}_{jk} \equiv \mathcal{D}(z_{jk}) $, $z_{jk}\equiv z_j-z_k$ (therefore, $z_{j0}=-z_{0j}=z_j$) 
and 
$\mathcal{V}_k$ are given by the operators $\exp (i\,s\, \omega_v\,a^\dagger a\,\sum_j t_j) $. The sum in the exponent is performed on variable numbers of consecutive waiting times, each one corresponding to a time interval between consecutive interactions with the field on the left ($s=-1$) or on the right ($s=+1$) side of the diagram. This follows directly from the possibility of rewriting the time evolution operator $e^{-it H_{v,k}}$ in terms of the undisplaced-oscillator operator $e^{-it\omega_v a^\dagger a}$ and of the displacement operators $\mathcal{D}(\pm z_k)$ (see Appendix \ref{app:A}). The sums in the exponents that define $\mathcal{V}_1$ ($\mathcal{V}_M$) include all the waiting times between the first and second interactions of the bra (ket) with the field, those in $\mathcal{V}_2$ ($\mathcal{V}_{M-1}$) include the times between the second and third interactions; and so on.

Hereafter, we demonstrate the equivalence between the expressions of the response function given in Eq. (\ref{eq:60}) and Eq. (\ref{eq:65}). The former one can in fact be rewritten as
\begin{align}\label{eq:70}
   R^{(v)} & = \langle 0 | (\mathcal{V}_0\,\mathcal{D}_{j_0j_1})\, (\mathcal{V}_1\,\mathcal{D}_{j_1j_2})\,\dots (\mathcal{V}_M\,\mathcal{D}_{j_Mj_{M+1}}) | 0 \rangle , 
\end{align}
where $\mathcal{V}_0 = \mathcal{I}$ and $j_{M+1}=j_0=0$. This simply results from the fact that $\mathcal{D}_{j_k 0}\,\mathcal{D}_{0j_{k+1}}=\mathcal{D}_{j_kj_{k+1}}$.

The vibrational state that is obtained after applying to the vacuum state $|0\rangle$ the first $q$ operators on the right-hand side,
\begin{align}
    |\alpha_q\rangle = (\mathcal{V}_{M-q+1}\,\mathcal{D}_{j_{M-q+1}j_{M-q+2}})\,\dots (\mathcal{V}_M\,\mathcal{D}_{j_Mj_{M+1}}) | 0 \rangle ,
\end{align}
is a coherent state identified by the complex number
\begin{align}
    \alpha_q = \sum_{k=1}^q z_{j_{M+k-q},j_{M+k-q+1}} \prod_{l=1}^k u_{l+M-q} .
\end{align}
This equation is obtained by noting that $\alpha_{q}=(\alpha_{q-1}+z_{j_{M-q+1}j_{M-q+1}})\, v_{M+1-q}$. The response function can be expressed as the overlap between the vacuum state of the bra and  the coherent state defined by the complex number
\begin{align}
    \alpha_{M+1} = \sum_{k=1}^{M+1} z_{j_{k-1},j_{k}} \prod_{l=1}^k v_{l-1} .
\end{align}
The modulus of such overlap is given by $e^{-\eta/2}$, where
\begin{align}
\eta\equiv |\alpha_{M+1}|^2 &= \sum_{k=1}^{M+1} z^2_{j_{k-1},j_{k}} + 2\sum_{k=1}^{M+1}\sum_{k'=1}^{k-1} z_{j_{k-1},j_{k}} \nonumber\\ &
\times z_{j_{k'-1},j_{k'}} {\rm Re} \left(\prod_{l=k'+1}^k v_{l-1}\right).
\end{align}
The application of a displacement operator to a coherent state also implies the appearence of a phase factor (see Appendix \ref{app:A}). Being $-z\,$Im$(\alpha)$ the phase change induced by applying $\mathcal{D}(z)$ to $|\alpha\rangle$, the overall phase factor accumulated after the application of the $M+1$ displacement operators in Eq. (\ref{eq:70}) is $-$Im$(\xi)$, where
\begin{align}
\xi & \equiv \sum_{q=1}^{M} \alpha_q z_{j_{M-q},j_{M-q+1}} & \nonumber\\ &  = \sum_{k=1}^{M+1}\sum_{k'=1}^{k-1} z_{j_{k-1},j_{k}} z_{j_{k'-1},j_{k'}} \prod_{l=k'+1}^k v_{l-1}.
\end{align}
Combining together the two equations above, one obtains the expression of the exponent that defines the response function:
\begin{align}
\ln [R^{(v,M)}] & = -\frac{1}{2} \eta - {\rm Im} \xi = -\frac{1}{2}\sum_{k=1}^{M+1} z^2_{j_{k-1},j_{k}} \nonumber\\ & - \sum_{k=1}^{M+1}\sum_{k'=1}^{k-1} z_{j_{k-1},j_{k}} z_{j_{k'-1},j_{k'}} \left(\prod_{l=k'+1}^k v_{l-1}\right).
\end{align}

One can finally establish a one to one correspondence between the $M(M+1)/2$ products of consecutive functions $f_{kl}\equiv \prod_{j=k}^{l} v_j$ (with $l>k$) that appear in the above equation, and the $M(M+1)/2$ functions $h_{mn}\equiv 1-\chi_{mn}= e^{-i\omega_v\sum_{j=m}^{n}t_j}$ or $h_{mn}^*$ (with $n>m$). {\y We start by introducing the notation that will be used in the demonstration.
The numbers} $M_b$ and $M_k=M-M_b+1$ are the numbers of interactions of the field that affect respectively the bra and the ket (arrows on the right and on the left of the Feynman diagram). {\y Therefore,} $v_l=e^{i\omega_v (t_{i_l}+\dots+t_{f_l})}$ for $l\le M_b$ and $v_l=e^{-i\omega_v (t_{i_l}+\dots+t_{f_l})}$ for $l>M_b$ (each function $v_l$ thus involves $f_l-i_l+1$ waiting times). Besides, if $l<M_b$, then $v_l$ and $v_{l+1}$ correspond to consecutive time intervals, with $i_{l+1}=f_l+1$; if $l>M_b$, then again $v_l$ and $v_{l+1}$ correspond to consecutive time intervals, but with $i_{l+1}=f_l-1$; finally, if $l=M_b$, then $f_l=i_{l+1}=M$. {\y (}Let's refer to the pathway in Fig. \ref{fig:n5}(b) to clear the notation with an example. Here there are $M_b=2$ arrows on the right of the diagram and $M_k=4$ on the left. The values of the indices $i_l$ and $f_l$ are: $i_1=1$ and $f_1=2$, $i_2=3$ and $f_2=5$, $i_3=f_3=5$, $i_4=f_4=4$, $i_5=3$ and $f_5=2${\y .)}

We {\y next show} that each function $f_{kl}$ corresponds to a function $h_{mn}$. In fact, if $l\le M_b$, then $f_{kl} = e^{i\omega_v(t_{i_k}+\dots+t_{f_l})}$ ({\it i.e.} the exponent includes all the consecutive waiting times from $t_{i_k}$ to $t_{f_l}$), and thus coincides with $h_{i_kf_l}^*$. If instead $k > M_b$, then $f_{kl} = e^{-i\omega_v(t_{i_l}+\dots+t_{f_k})}$ ({\it i.e.} the exponent includes all the consecutive waiting times from $t_{i_l}$ to $t_{f_k}$), and thus coincides with $h_{i_l f_k}$. In all the other cases, namely for $k\le M_b$ and $l>M_b$, one has that $f_{kl} = e^{i\omega_v[(t_{i_k}+\dots+t_{M})-(t_{i_l}+\dots+t_M)} $: this coincides with $h_{i_k\, i_l-1}^*$ if $i_k<i_l$ and with $h_{i_l\,i_k-1}$ if $i_k>i_l$ (whereas one cannot have that $i_k=i_l$, because this would imply an arrow at the same time on the two sides of the diagram). 

We {\y conclude by showing} that each function $h_{mn}$ corresponds to a function $f_{kl}$. Each of the waiting times is delimited between two consecutive interactions with the field. If the interaction before $t_m$ and after $t_n$ are both on the right (left) side of the diagram, then there must exist a pair of functions $v_k$ and $v_l$ such that $m=i_k$ and $n=f_l>m$ ($m=i_l$ and $n=f_k>m$), so that $f_{kl}$ coincides with $h_{mn}$. If the arrow before $t_m$ is on the right (left) and that after $t_n$ on the left (right), then there must be a pair of functions $v_k$ and $v_l$ such that $m=i_k$ and $n=i_l-1>m$ ($m=i_l$ and $n=i_k-1>m$), so that $f_{kl}$ coincides with $h_{mn}$. 

Examples of the correspondence between the functions $h_{mn}$ and $f_{kl}$ are given in Section V for the two Feynman diagrams reported in Fig. \ref{fig:n5}.

\section{Coupling to a bath \\ of vibrational modes\label{app:Z}}

In the multimode case, the overall response function is given by the product of the ones corresponding to each mode [Eq. (\ref{eq:mmode})]. If each mode $\xi$ is in a thermal state, the single-mode response function is given by Eq. (\ref{eq:thermal}). Writing the overall and the single-mode response functions as $R^{(v,M)}_T=\exp(f_{T})$ and $R^{(v,M,\xi)}_T=\exp(f_{T}^{\xi})$, one has that $f_{T}=\sum_{\xi=1}^B f_{T}^{\xi} = \sum_{\xi=1}^B [\coth(\hbar\omega_v/2k_BT)\,{\rm Re}(f_\xi)+i\,{\rm Im} (f_\xi)] $. The functions $f_\xi$ represent the zero-temperature limits of the $f_{T}^{\xi}$, and are given by the sum of terms $z_{jk,\xi}\,z_{j'k',\xi}\chi_{mn}=z_{jk,\xi}\,z_{j'k',\xi}[1-\exp(i\omega_\xi t_{mn})]$, being $\omega_\xi$ the frequency of the vibrational mode $\xi$. Therefore, the exponent $f_{T}$ of the response function is given by the sum (over the modes and over the functions $\chi_{mn}$, multiplied by the corresponding displacements) of terms
\begin{gather}
    z_{jk,\xi}\,z_{j'k',\xi} \{i \sin(\omega_\xi t_{mn})\nonumber\\ +\coth(\hbar\omega_v/2k_BT)[1-\cos(\omega_\xi t_{mn})]\} .
\end{gather}
Passing from a discrete set of vibrational modes to a continuum, one can eventually write the exponent  $f_{T}$ as a sum of the functions given in Eq. (\ref{eq:gf}).

\end{document}